\documentclass[useAMS,usenatbib]{mn2e}
 
\usepackage{amssymb}     
\usepackage{graphicx}
\usepackage{enumerate}
\usepackage{lscape}

\title[Fundamental properties of FR II radio galaxies via Monte Carlo simulations]{Fundamental properties of Fanaroff-Riley II radio galaxies investigated via Monte Carlo simulations} 
 
\author[A.D. Kapi\'{n}ska, P. Uttley \& C.R. Kaiser]{A. D. Kapi\'{n}ska$^{1,2,}$\thanks{E-mail: anna.kapinska@port.ac.uk}, P. Uttley$^{1,3}$ \& C.R. Kaiser$^{1,4}$ \\ 
$^1$ School of Physics \& Astronomy, University of Southampton, Southampton SO17 1BJ, U.K.\\ 
$^2$ Institute of Cosmology \& Gravitation, University of Portsmouth, Portsmouth PO1 3FX, U.K.\\
$^3$ Astronomical Institute `Anton Pannekoek', University of Amsterdam, Science Park 904, 1098 XH, Amsterdam, Netherlands\\ 
$^4$ Blumenstra\ss e 14e, Holzkirchen 83607, Germany} 
 
\begin{document} 
 
\date{Accepted ... Received ..; in original form ...} 
 
\pagerange{\pageref{firstpage}--\pageref{lastpage}} \pubyear{...} 
 
\maketitle 
 
\label{firstpage} 
 
\begin{abstract} 
Radio galaxies and quasars are among the largest and most powerful single objects known and are believed to have had a significant impact on the evolving Universe and its large scale structure. We explore the intrinsic and extrinsic properties of the population of Fanaroff-Riley II (FR~II) objects, that is their kinetic luminosities, lifetimes, and the central densities of their environments. In particular, the radio and kinetic luminosity functions of these powerful radio sources are investigated using the complete, flux limited radio catalogues of 3CRR and Best et al.. We construct multidimensional Monte Carlo simulations using semi-analytical models of FR~II source time evolution to create artificial samples of radio galaxies. Unlike previous studies, we compare radio luminosity functions found with both the observed and simulated data to explore the best-fitting fundamental source parameters. The new Monte Carlo method we present here allows us to: (i) set better limits on the predicted fundamental parameters of which confidence intervals estimated over broad ranges are presented, and (ii) generate the most plausible underlying parent populations of these radio sources. Moreover, as has not been done before, we allow the source physical properties (kinetic luminosities, lifetimes and central densities) to co-evolve with redshift, and we find that all the investigated parameters most likely undergo cosmological evolution. Strikingly, we find that the break in the kinetic luminosity function must undergo redshift evolution of at least $(1+z)^3$. The fundamental parameters are strongly degenerate, and independent constraints are necessary to draw more precise conclusions. We use the estimated kinetic luminosity functions to set constraints on the duty cycles of these powerful radio sources.  A comparison of the duty cycles of powerful FR~IIs with those determined from radiative luminosities of AGN of comparable black hole mass suggests a transition in behaviour from high to low redshifts, corresponding to either a drop in the typical black hole mass of powerful FR~IIs at low redshifts, or a transition to a kinetically-dominated, radiatively-inefficient FR~II population.

\end{abstract} 
 
\begin{keywords} 
galaxies: active -- galaxies: radio galaxies, quasars -- galaxies: evolution -- galaxies: jets -- galaxies: luminosity functions -- methods: numerical, statistical. 
\end{keywords}

\section{Introduction}

Radio galaxies and radio loud quasars are believed to have a significant impact on the evolving Universe and its large scale structure \cite[][]{g2,r2,s7}. They are often found in galaxy groups and clusters \citep[][]{l7,h3,a6,d2,z1,b12}. Since their jets inject a significant amount of energy into the surrounding medium, stored in the so-called radio lobes, they can provide useful information in the study of the density and evolution of the intergalactic and intracluster medium.  
The jet activity is also believed to regulate the growth of massive galaxies \citep[][]{b14,m9,s14,c12, s15}. Therefore the estimation of the kinetic power and specification of the environments of these powerful radio sources at various cosmological epochs is an important task. The intracluster medium can be explored with the use of X-ray observations up to $z\sim 1$ \citep[e.g.][]{r3,p2}. On the other hand radio galaxies are found at $z\sim3$ and beyond \citep[e.g.][]{c11,v1,j3} and hence they can provide a valuable insight into the high-redshift Universe. Moreover, complete catalogues of radio galaxies and quasars indicate that these sources were of higher number density during the so-called `quasar era' \citep{d3}, and since they can be observed at high redshifts the knowledge of the fundamental aspects of these sources as a population is of importance for cosmological studies.

\cite{f2} divided the extragalactic large scale radio sources into two main classes, namely low luminosity FR~I and more luminous FR~II types. Sources of the two classes are morphologically different, FR~Is are core-jet bright, edge-darkened objects and contain presumably turbulent jets, while FR~IIs are limb-brightened, often symmetrical objects with well defined features such as jets, hotspots and radio lobes; the latter are often referred to as classical double radio sources. The question of whether they originate from the same parent population but are dependent on the environment they reside in, or whether they are intrinsically different, is still open \citep[e.g.][]{b18, b17,m13, g1,kb1,k8,w6}.

Because of their relatively simple structure (contrary to the turbulent nature of FR~Is), some sophisticated semi-analytical models of FR~IIs' growth have been developed.  These models can predict source observables, that is radio lobe luminosity ($L_{\nu}$, where $\nu$ is the observed frequency) and linear source size ($D$), from underlying physical properties such as kinetic luminosity ($Q$), source age ($t$), and the density of the intergalactic medium ($\rho$). This has led to a number of studies that investigated whole populations of these powerful sources and their evolution \citep[][]{d1,brw,k1,b1,b2,w1}. The basic methodology of these studies was to construct virtual populations of FR~II sources choosing underlying properties for the population and running them through semi-analytical models to predict the  distribution of source observables, the linear size and radio lobe luminosity. The virtual populations were further compared to observed data from complete samples of radio sources. However, those studies focused on the commonly used radio power -- linear size ($P_{\nu} - D$) distribution diagram introduced by \cite{s5}. One of the major problems with such an analysis is  that $P_{\nu} - D$ diagram is difficult to interpret in terms of most commonly used techniques to study source populations such as the luminosity function. Moreover, many studies \citep[e.g.][]{k5,s6,brw} focused predominantly on the relationship between the observables and their trends with cosmological epochs. These observables are determined by the fundamental source properties, and hence they carry convolved effects of possible cosmological evolution of the underlying physical properties, as well as the influence of observational biases. Although, there have been attempts to investigate the fundamental source parameter space \citep{k1,m1,w1}, sometimes very strong assumptions have been adopted.

In this paper we use the standard Monte Carlo method, as summarized above, to generate virtual populations of FR~II sources. However, contrary to previous studies, we use distributions of radio lobe luminosity that can easily be transformed into radio luminosity functions, rather than the $P_{\nu} - D$ diagram consistently used in previous semi-analytical population studies, to compare the simulated population and the observed data. Also, we attempt to make as few assumptions as possible about the underlying physical source properties to obtain more general results (although we are still restricted by computing time). To do so we repeat the Monte Carlo simulation multiple times following grid minimisation that searches broad ranges of the possible source underlying properties. This allows us to create confidence intervals of the estimated parameters. In addition, as has not been done before, we allow for co-evolution of the physical source properties; this will enable us to determine the dominant type of evolution, if any. Further, we investigate the influence of various assumptions, such as the density profile characteristics or the jet particle content, on the best-fitting properties of source populations.  The method we develop here allows us to constrain the parent population of the sources while minimising the inevitable effects of the radio surveys' flux limits.

The paper is structured as follows: the observed samples are presented and discussed in \S\ref{sec:obs-data}, and in \S\ref{sec:theory} we summarize the theoretical models used. The construction of the simulated samples and statistical methods used are discussed in detail in \S\ref{sec:mc-modelling}. The results are presented and discussed in \S\ref{sec:results} and \S\ref{sec:discussion} respectively, and a summary is given in \S\ref{sec:summary}.

 
\begin{figure} 
\includegraphics[width=92mm]{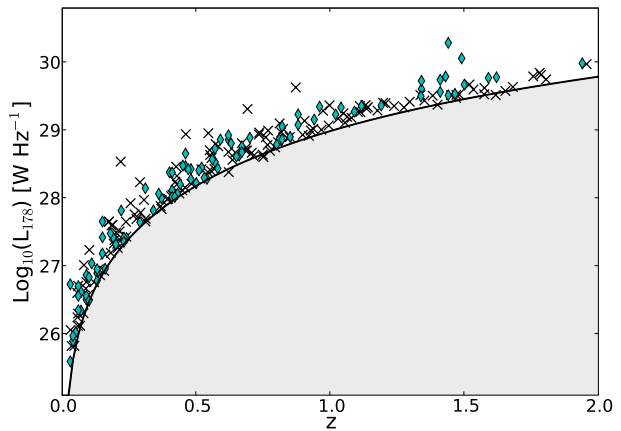} 
\caption{Radio luminosity vs redshift plane for the 3CRR (crosses) and BRL (diamonds) samples used in this work (sources of FR~II morphology with fluxes above the limit {\mbox{$S=10.9$~Jy}} at 178~MHz (solid line) and with $z<2.0$ are shown).} 
\label{rys:lum-red-plane} 
\end{figure} 
 

\section{Observational data sets} 
\label{sec:obs-data}

We use complete flux-limited radio samples which contain all extragalactic radio sources in a given sky area and above the sensitivity limit specified for each survey. Currently, we concentrate only on sources of the FR~II morphology due to the availability of semi-analytical models of their time evolution; however, the approach presented here may be extended to cover the whole radio source population once theoretical models for FR~Is are developed. We use both radio galaxies and radio-loud quasars as it is assumed, following the unification models, that the only difference between these types of sources is just their viewing angle \citep[e.g.][]{b13}. Additionally, due to the sparse population of sources at  high redshifts, and hence poor representation, we limit our analysis to sources with redshifts up to $z = 2$. Due to the nature of the theoretical models that assume that most of the radio luminosity comes from the radio lobes of the sources, one should use samples observed at low radio frequencies ($\sim {\rm few} \times 100$~MHz) to avoid compact radio emission that dominates at GHz frequencies. There are few commonly used low-frequency radio catalogues. In this paper, we present the work done with two such radio samples, namely the well known Third Cambridge Revised Revised Catalogue \citep[3CRR]{l1}, and the complete radio sample of \citet[][hereafter BRL]{b3}.

\begin{figure} 
\begin{center} 
\includegraphics[width=70mm, angle=270]{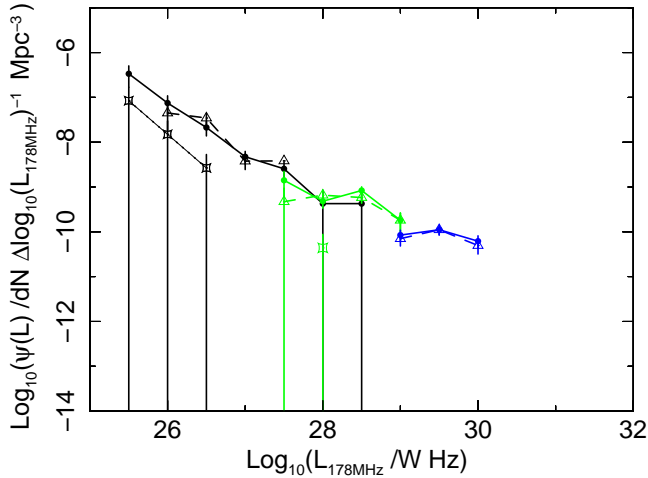} 
\caption{The observed radio luminosity functions of the analysed data (3CRR and BRL catalogues) for each of the considered redshift ranges, where $z_1$ is drawn in black, $z_2$ in blue, and $z_3$ in green. Each of the redshift bins is further divided in size bins, where solid lines (filled circles) correspond to smallest sources, dashed (open triangles) to medium size sources, and dotted (open square lozenges) to giant sources (see \S\ref{sec:obs-data} for the exact values).} 
\label{rys:obs-RLF} 
\end{center} 
\end{figure} 

The 3CRR radio sample is very shallow (its flux limit equals $10.9$~Jy defined at $178$~MHz), but covers a large area of the sky ($4.23$~sr); it contains some of the most powerful radio galaxies. 146 sources of FR~II morphology are in the sample; however, we excluded 3C~231 (M82), as well as, following \citet[][hereafter BRW]{brw}, two additional sources, 3C~345 and 3C~454.5 (due to the Doppler boosting). Of the remaining sources 23 are of FR~I type, and all are found at very low redshifts, not exceeding $z \approx 0.25$.

The BRL sample is defined at an observing frequency of $408$~MHz with flux limit of $5$~Jy. The sample has been created to complement other radio samples, and specifically with the 3CRR catalogue, as it provides coverage of the entire sky above $-30^{\circ}$ in declination. Its flux limit translates to $\sim 9.7$~Jy at $178$~MHz assuming a typical radio spectral index $\alpha = 0.8$\footnote{The relation between the flux of a source ($S_{\nu}$) at a frequency $\nu$, and its radio spectral index $\alpha$, is defined as $S_{\nu} \propto \nu^{-\alpha}$.}. The sample is 100~per cent spectroscopically complete \cite[][]{b21,b22}. In total, the sample contains 178 sources of which 124 are of FR~II morphology.

 
\begin{table} 
  \centering 
  \caption{Demography of the observational data from the 3CRR and BRL radio samples. Quoted numbers of sources valid after certain criteria are met ($S_{178MHz}>10.9$~Jy, $D>10$~kpc). FR~I type sources are listed for reference only (for details see \protect\S\ref{sec:obs-data}).} 
  \label{tab:obs-data} 
  \begin{tabular}{lcccc} 
    \hline 
    \hline 
    Redshift & \multicolumn{4}{c}{No. of sources}\\ 
    range    & \multicolumn{2}{c}{ 3CRR }        & \multicolumn{2}{c}{ BRL }   \\ 
     $(z)$   &  FR~I        & FR~II              &  FR~I       & FR~II         \\ 
    \hline 
    $z_1 \in [ 0.0 , 0.3]$ & 23 & 40                 & 7       & 31\\ 
    $z_2 \in ( 0.3 , 0.8]$ &    & 45                 &         & 35\\ 
    $z_3 \in ( 0.8 , 2.0]$ &    & 50                 &         & 26\\ 
    \hline 
    \hline 
  \end{tabular} 
 
\end{table} 
 

The two samples are analysed together and both are brought to the common observed frequency of $178$~MHz. In the case of both samples the measured radio spectral index of each source is used\footnote{The spectral indices of the sources contained in the 3CRR sample is measured between 178 and 750~MHz, while in the BRL sample between 408 and 1400~MHz.}. The redshift distribution of all sources included in our analysis is presented in Figure~\ref{rys:lum-red-plane}.  Further, the population is divided into size and luminosity bins in given redshift ranges. Due to the number of available sources from the catalogues we are only able to consider three redshift ranges, $z_1 \leq 0.3, 0.3 < z_2 \leq 0.8$ and $0.8 < z_3 \leq 2.0$, the borders of which are chosen to ensure similar source numbers per redshift bin. Incidentally, these redshift ranges span over similar relative light travel time intervals, i.e. $t_{\Delta z_1} = 3.370$~Gyr, $t_{\Delta z_2} = 3.361$~Gyr and  $t_{\Delta z_3} = 3.364$~Gyr. 
These subgroups of different redshifts are considered separately unless otherwise stated. Sources are grouped in size bins in the following manner: in $z_1$ the size ranges are $10 - 245$~kpc, $245$~kpc~$- 1$~Mpc and $>1$~Mpc, in $z_2$ these are $10 - 208$~kpc, $208$~kpc~$- 1$~Mpc and $>1$~Mpc, and in $z_3$  $10 - 100$~kpc, $100$~kpc~$- 1$~Mpc and $>1$~Mpc.  The size ranges were chosen to contain similar number of sources in the case of the smallest and medium sized sources, while the last bin tracks giant radio galaxies more accurately (giant radio galaxies are defined as sources of total linear sizes of more than 1~Mpc). For each of these size bins a radio lobe luminosity distribution  is constructed. The distribution is constructed at at an observing frequency of 178~MHz from ${\rm log_{10}}(L_{178MHz})=22.0$ to ${\rm log_{10}}(L_{178MHz})=35.0$ in steps of $\Delta {\rm log_{10}}(L_{178MHz})=0.5$. Such a division of the population allows us to track the relative number of the sources with a given size, and later to construct radio luminosity functions of the sources at the given redshift ranges and of the given sizes. The counts of sources used are summarized in Table~\ref{tab:obs-data}, and their observed radio luminosity functions are presented in Figure~\ref{rys:obs-RLF}.

As pointed out by BRW the use of only one radio sample in any population study may result in inevitable biases while interpreting results.  This is due to the strong luminosity -- redshift ($L_{\nu} - z$) dependence (Figure~\ref{rys:lum-red-plane}). It is important, therefore, to either include a few radio samples that will supplement each other in the radio lobe luminosity -- redshift plane, or to be very cautious with interpreting results. For the extensive discussion on this and also frequency related issues we refer our reader to the original BRW paper. Commonly used catalogues include 6CE \citep{e1} and 7CRS \citep[based on the Seventh Cambridge Redshift Survey Catalogue, ][]{m3, g5, w1} besides 3CRR \citep[e.g. BRW, ][]{w2}. The BRL sample has also been used along with the other three by \citet{w1}. However, here, we do not attempt to reconstruct the previous population studies of the [$P-D-z$] plane, but rather analyse binned distributions of the sources' radio lobe luminosities, hence we present work on only two samples. Because of considerably different flux limits of various samples, other ones such as 7CRS will be analysed separately to 3CRR and BRL.

We assume a flat Universe with the Hubble constant of $H_0 = 71$ km~s$^{-1}$~Mpc$^{-1}$, and $\Omega_{\rm M} = 0.7$ and $\Omega_{\Lambda} = 0.3$ throughout the paper.

\section[]{Theoretical models of radio galaxy and quasar evolution} 
\label{sec:theory}

A basic picture of the FR~II type radio source growth, where the relativistic plasma is ejected in two opposite directions forming collimated outflows, is nowadays widely accepted. These outflows expand and interact with the surrounding medium until they terminate in a strong shock and create so-called radio lobes where the excess transferred energy is stored. The idea first proposed by \citet{b6} and \citet{s2} has been developed, over the past  15 years, into a few more sophisticated models of the evolution of FR~II sources. In particular, there are three models which have gained the most attention, \citet[][part 1, hereafter KA]{ka} and subsequently \citet[][part 2, hereafter KDA]{kda}, BRW, and \citet[][hereafter MK]{mk}. It is not our intent to compare the existing models in this paper; however, we will briefly summarize their main features.

The BRW and MK models follow the dynamical evolution of radio sources as described by KA, which we will focus on. The density profile of the environment into which the source expands, is approximated by the generalised King (\citeyear{k2}) profile

\begin{equation} 
\rho = \rho_{\rm o} \left ( \frac{r}{a_{\rm o}}\right )^{-\beta} \,\,{\rm for \,\,} r > a_{\rm o}, 
\label{eqn:density} 
\end{equation} 
where $r$ is the radial distance from the AGN core, and $\rho_{\rm o}$ is a constant central density within the core radius  $a_{\rm o}$. The index $\beta$ is usually constrained by the observations, although some restrictions may apply in the case of certain assumptions \citep[e.g. the self-similarity assumption requires $\beta < 2$, see][]{f1}. Note, however, that the model depends on the combination of $(\rho_{\rm o} a_{\rm o}^{\beta})$ and not the parameters separately.  The approximation is valid at distances of at least a few core radii.

Further, as \citet{f1} and KA show, the source expansion problem can be solved purely with a dimensional analysis where the source growth depends purely on the jet of the source with constant power ($Q$, hereafter referred to as radio source kinematic luminosity), its age ($t$) and the environment in which it self-similarly expands. The linear length of the lobe ($D_{\rm l}$) of the source is hence defined as

\begin{equation} 
D_{\rm l}  = c_1 \left ( \frac{Q}{\rho_oa_o^{\beta}}\right )^{\frac{1}{5-\beta}} t^{\frac{3}{5-\beta}}, 
\label{eqn:size} 
\end{equation} 
where the $c_1$ parameter depends solely on $\beta$ and the source aspect ratio $R_{\rm T}$ (i.e. the ratio of the source length to its width, see \S\ref{sec:aspect-ratio}). The lobe pressure ($p_{\rm l}$) of a radio source of a given age is further found with

\begin{equation} 
p_{\rm l}  \propto (\rho_o a_o^{\beta})^{1/3} Q^{2/3} D_{\rm l}^{(-4-\beta)/3}. 
\label{eqn:pressure} 
\end{equation} 
 
To calculate the radio lobe luminosity the KDA model incorporates the energy losses of the relativistic electrons, using the adiabatic expansion, synchrotron and inverse Compton losses. The radio lobe luminosity at a radio frequency $\nu$ (note that $L_{\nu} = 4\pi P_{\nu}$) of the source becomes therefore 
\begin{equation} 
L_{\nu} \propto Q p^{(m+1)/4} t \int_{x_{\rm min}}^1 C\left(x\right){\rm d}x, 
\label{eqn:luminosity} 
\end{equation} 
where $m$ denotes the injection spectral index of the energy distribution of the relativistic electrons (\S\ref{sec:spectral-index}), and  the term $C\left(x\right)$ depends on the energy losses and $x = t_i / t$ with $t_i$ being time at which a particle was injected into the radio lobe. For the full derivation and discussion see KDA and \cite[][\citeyear{kb2}]{kb1}.

The BRW and MK models differ from KDA in their assumptions about the luminosity evolution of the sources determined by the way the relativistic particles are injected from the jet to the radio lobe, and the particle transport. In particular, BRW argue for the importance of the hotspot region pointing out the over-simplicity of the KA/KDA model and its assumption of source self-similar expansion. Additionally, BRW assume that spectral index of the radio lobes at a specific time is driven by breaks in frequency in the head region. Further, MK expands the KDA and BRW models incorporating the particle transport mechanisms. It is worth noting that assumptions of BRW and MK lead to much steeper evolutionary tracks of the sources in the radio power -- linear size ($P_{\nu}-D$) plane as compared with the predictions of the KDA model.

More recently \citet{b1,b2} attempted to develop a more accurate model by testing and modifying the three leading theoretical models. They report that none of the existing models can give fully acceptable fits to all of the properties deduced from the radio surveys (especially the synchrotron spectral index $\alpha$); however, according to their findings KA/KDA give the best overall results as compared to the other two models. In this work we employ the original KA/KDA model.

\section{Monte Carlo simulations of  the complete virtual radio samples} 
\label{sec:mc-modelling} 
 
In the traditional and most common way, the radio source evolution has been investigated through the so-called $P_{\nu}-D$ diagram \citep{s5}, which in recent studies has been extended into radio lobe spectral index ($\alpha$) and redshift ($z$) parameter space. Here, we use an alternative approach and instead of investigating [$P_{\nu}-D-z- \alpha$] space we use radio luminosity functions (RLFs) in the analysis. An advantage of doing so is the direct link to the results of other source population studies which are most commonly expressed in terms of luminosity or mass functions.

Due to degeneracy between the fundamental properties it is impossible to infer the source kinetic luminosity, the central ambient density and the source age directly from the observables, and hence Monte Carlo simulations are necessary in order to generate simulated RLFs. In this section we present our Monte Carlo method (\S\ref{sec:mc-sim}). The parameter distribution functions and other fixed assumptions of the model  for constructing the radio source population are discussed in \S\ref{sec:distributions}. The statistical methods used to determine confidence intervals and goodness-of-fit test are discussed in \S\ref{sec:stats}.

\subsection{Monte Carlo simulation} 
\label{sec:mc-sim}

To construct a single virtual sample we generate, in total, {\mbox{ $\sim\rm few \times 10^6$}} virtual radio sources. The subsequent prescription is followed. 
 
\begin{enumerate}[1.] 
 
\item For each source of the virtual sample we assign a set of physical parameters summarized in Table~\ref{tab:model-params}. Each of the parameters is either drawn from its respective distribution or is the same for each source; the details of these physical `input' parameters are presented in \S\ref{sec:distributions}.

\item Based on the theoretical model of KA/KDA  the source is evolved and its linear size $D$ is calculated.

\item The linear size of the source is being checked for its reliability, that is whether the source expansion exceeds an assumed fraction of the light speed (maximum allowed head advance speed, see \S\ref{sec:adv-speed}). If it does, the source is rejected. Otherwise, the source is accepted and its linear size is corrected for the projection angle (\S\ref{sec:Dproj}).

\item The radio lobe luminosity $L_{\nu}$  of the source at a frequency $\nu=178$~MHz and its randomly generated redshift is calculated.

\item The radio luminosity is subsequently randomized by adding a Gaussian variable of standard deviation  $\Delta L_{\rm err}=0.08 \times L_{\nu}$ to account for any instrumental/systematic errors. Moreover, since the source is evolved in its rest-frame, a correction of  $(1+z)^{\alpha}$ transforming the calculated radio luminosity to the common frequency must be included. The common frequency of $178~$MHz is used. These corrections are discussed in detail in \S\ref{sec:spectral-index}.

\item Steps 1 - 5 are repeated {\mbox{ $\sim\rm few \times 10^6$}} times.

\item The radio lobe luminosity histograms for each redshift range and linear size bin are generated as discussed in \S\ref{sec:obs-data}. The binning of the virtual sample histograms matches exactly the one used for the real observed data sample. However, to ensure that the probability density functions of kinetic luminosities (Eqn.~\ref{eqn:schech-stdn} and Eqn.~\ref{eqn:schech-wlt}) are represented by reasonable number of sources at the functions' exponentially falling ends, we initially generate the virtual population assuming that their kinetic luminosities are drawn from a uniform distribution, and later a weighting factor is applied to each source (see \S\ref{sec:jetpow}).

\item At this stage the histograms represent the total number of sources in the simulated sample since  we have not determined yet the fluxes of the generated  sources. In this sense, those are the `true' numbers of the entire source population. However, to be able to compare the simulated sample to the observed data the selection effect arising from limited flux sensitivity must be taken into account. We achieve it by using radio luminosity functions. RLFs are an ideal tool when dealing with coupled selection effects such as limited sensitivity of radio surveys. They are defined as number density of sources per unit comoving volume per unit luminosity. The initial histogram of the entire simulated population is therefore transformed into RLF by using the survey volume $V_{\rm survey}$, 
\begin{equation} 
\phi_{{\rm sim}, i} = \frac{A}{4\pi} \frac{n_{L_i}}{ V_{\rm survey}}, 
\end{equation} 
where $A$ is the area of the sky that the survey covers, and $n_{L_i}$ represents the data counts in luminosity bins $L_i$. Further, the RLF is transformed back into a histogram of number of expected sources for each linear size and luminosity bin using $V_{\rm lim}$ -- a maximum co-moving volume in which a source in a given luminosity bin and with the flux limit of the survey would be included in the sample \citep{s10}, 
\begin{equation} 
n_{L_i, S_{\rm lim}} = \frac{4\pi}{A} \phi_{{\rm sim,}i}  V_{\rm lim}. 
\label{eqn:rlf-schmidt} 
\end{equation} 
The two transformations simplify to 
\begin{equation} 
n_{L_i, S_{\rm lim}} = n_{L_i} \frac{ V_{\rm lim}}{V_{\rm survey}}. 
\label{eqn:sim-data-counts} 
\end{equation} 
Note that flux densities of the simulated  radio galaxies ($S_{{\rm sim,} i}$) are not examined directly, and hence no radio sources are rejected based on the limited sensitivity bias (i.e. requirement that $S_{{\rm sim,} i} > S_{\rm lim}$). Instead, a correction factor ($V_{\rm lim} / V_{\rm survey}$), which indicates the probability of radio source occurrence at a given flux limit, is employed. The effect of the limited flux sensitivity is an average for each considered bin. To ensure greater accuracy we perform the flux correction on much finer bin widths than those of the final histograms. In particular, in this step we use $20$ times finer bins than used in the final histograms. These are later summed to match the initially constructed distributions.

\item So far, we have not discussed the number of progenitors becoming active and turning into FR~II sources. To ensure a reasonably good statistics, our simulated sample is generated with a much larger number of virtual radio sources than the observed data sets contain. Instead of introducing a corresponding distribution function we use the full simulated sample. However, since both the simulated and observed samples must be of a similar size, that is of a similar number of sources considered, we need to renormalize the data counts, which are found with Eqn.~\ref{eqn:sim-data-counts}, in the virtual sample. We use the binned maximum likelihood method (MLM, see \S\ref{sec:mlm}) together with Brent's method \citep{b19,p1} to do so. The normalization is set as a free parameter, but represents an average for the considered redshift range.

\item Finally, the goodness-of-fit test is performed. The statistics used, normalization of the sample and the goodness-of-fit test are discussed in \S\ref{sec:stats}. 
 
\end{enumerate}

\subsection{Monte Carlo simulation input parameters} 
\label{sec:distributions}

\subsubsection{Kinetic luminosity distribution} 
\label{sec:jetpow}

The kinetic luminosity of each generated source is drawn randomly from a distribution function that acts as a probability density function. The form of the distribution function of sources' kinetic luminosities is not known, and various forms have been assumed in previous works ranging from a simple uniform distribution between minimum ($Q_{\rm min}$) and maximum ($Q_{\rm max}$) kinetic luminosity \citep[e.g.][]{w1}, through  power-law scaling as $Q^{x}$ (e.g. BRW) to more complex functions as used by Willott et al. (\citeyear{w2}, discussed below).

We consider two models for the initial distribution functions of the kinetic luminosities. In Model S the kinetic luminosity distribution function is assumed to be modelled by the so-called Schechter function \citep{s1} of a form 
\begin{equation} 
\psi(Q)~ {\rm d}Q = \psi^* \left (\frac{Q}{Q_{\rm B}} \right )^{-\alpha_{\rm s}} {\rm exp} \left (\frac{-Q}{Q_{\rm B}} \right ) ~{\rm d}Q 
\label{eqn:schech-stdn} 
\end{equation} 
where the slope of the function for the kinetic luminosity values below the kinetic luminosity break ($Q_{\rm B}$) is described by the exponent $\alpha_{\rm s}$, and drops exponentially for higher $Q$. $\psi^*$ is a normalization constant, which in our case is  neglected at this stage as Eqn.~\ref{eqn:schech-stdn} is used as a probability density function.

Model W follows \citet{w2}, who introduce modified version of the above Schechter function of a form  
\begin{equation} 
\psi(Q) ~{\rm d}Q = \psi^* \left (\frac{Q}{Q_{\rm B}} \right )^{-\alpha_{\rm s}} {\rm exp} \left (\frac{-Q_{\rm B}}{Q} \right ) ~{\rm d}Q, 
\label{eqn:schech-wlt} 
\end{equation} 
and where  the exponent $\alpha_{\rm s}$ describes the function for kinetic luminosities higher than $Q_{\rm B}$, while for lower kinetic luminosities the function drops exponentially, and $\psi^*$ is used as in the Model S. In their original paper \citet{w2} use a combination of Eqn.~\ref{eqn:schech-stdn} and Eqn.~\ref{eqn:schech-wlt} to describe the whole population that consists of high and low radio luminosity sources. Eqn.~\ref{eqn:schech-wlt} was introduced to specifically model the high radio luminosity subpopulation of radio sources. Since we do not consider FR~I type sources in our study we will not follow the original method of \citet{w2}, and Eqn.~\ref{eqn:schech-wlt} alone is used to describe the considered subpopulation of powerful radio sources.

Both functions describe the distribution of kinetic luminosities between $Q_{\rm min}$ and $Q_{\rm max}$, which are set in such a way that that contribution to the RLFs of sources with kinetic luminosities outside this range is negligible. To ensure these probability density functions are represented by reasonable number of sources, what will assure good statistics at the functions exponential ends, the kinetic luminosities are initially generated from a uniform distribution between $Q_{\rm min}$ and $Q_{\rm max}$. Each of these kinetic luminosities is assigned a probability of its occurrence (a weighting factor) according to the considered probability function (Eqn.~\ref{eqn:schech-stdn} and Eqn.~\ref{eqn:schech-wlt}). This weighting factor is applied to each source while constructing histograms at a later stage.

 
\begin{table*} 
  \centering 
  \caption{Overview of a source physical parameters and default distributions from which they are drawn. Details on the assumed distribution or value of the respective parameters are discussed in sections as given in column 4.} 
  \label{tab:model-params} 
  \begin{tabular}{llll} 
    \hline 
    \hline 
    Parameter  & Assumed      & Description  & Discussed in \\ 
               & distribution &      \\ 
    \hline\\ 
    \multicolumn{4}{c}{\sc Physical Parameters}\\\\
    $z$       & distribution  & source redshift      & \S\ref{sec:redshift}\\ 
    $Q$       & distribution  & source kinetic luminosity     & \S\ref{sec:jetpow}\\ 
     
    $\rho_o$  & distribution  & mean central density in which source expands & \S\ref{sec:dens}\\ 
    $a_o$     & value         & core radius & \S\ref{sec:dens}\\ 
    $\beta$   & distribution  & power law index of the radial density distribution & \S\ref{sec:dens}\\ 
 
    $t$         & distribution  & source current age & \S\ref{sec:age}\\  
    $t_{\rm max}$& distribution  & source maximum age & \S\ref{sec:age}\\  
     
    $\alpha$  & calculated value & radio spectral index & \S\ref{sec:spectral-index}\\  
    $m$       & distribution  & power-law exponent of the relativistic particles' energy distribution& \S\ref{sec:spectral-index}\\ 
    $\gamma_{\rm min}$ & value & minimum Lorentz factor of relativistic particles& \S\ref{sec:spectral-index}\\ 
    $\gamma_{\rm max}$ & value & maximum Lorentz factor of relativistic particles& \S\ref{sec:spectral-index}\\ 
 
    $R_{\rm T}$ & distribution & aspect ratio     & \S\ref{sec:aspect-ratio}\\    
    $\vartheta$& distribution & projection angle & \S\ref{sec:Dproj}\\ 
 
    $\Gamma_{\rm x}$& value    & adiabatic index of the IGM & \S\ref{sec:adiab-indices}\\ 
    $\Gamma_{\rm c}$& value    & adiabatic index of the radio lobes & \S\ref{sec:adiab-indices}\\ 
    $\Gamma_{\rm b}$& value    & adiabatic index of the magnetic field energy density & \S\ref{sec:adiab-indices}\\ 
    $k'$           & value    & ratio of thermal to electron energy densities in the jet& \S\ref{sec:particle-content}\\ 
    $v_{\rm max}$   & value    & maximum allowed head advance speed & \S\ref{sec:adv-speed}\\ 
    \\
    \multicolumn{4}{c}{\sc Distributions' Parameters}\\  \\
    $Q_{\rm B}$       & value & kinetic luminosity break                         & \S\ref{sec:jetpow}\\ 
    $\alpha_{\rm s}$  & value & exponent of the kinetic luminosity distribution  & \S\ref{sec:jetpow}\\ 
    $n_{\rm q}$       & value & strength of the kinetic luminosity break redshift evolution         & \S\ref{sec:cosmo-combined}\\
    $\rho_{\rm m}$    & value & mean of log-normal distribution of radio sources' central densities & \S\ref{sec:dens}\\ 
    $\sigma_{\rho_o}$ & value & standard deviation of log-normal distribution of radio sources' central densities & \S\ref{sec:dens}\\ 
    $n_{\rm r}$       & value & strength of the central density redshift evolution                  & \S\ref{sec:cosmo-combined}\\
    $t_{\rm max_{\rm m}}$& value & mean of log-normal distribution of radio sources' maximum ages     & \S\ref{sec:age}\\  
    $\sigma_{t_{\rm max}}$& value & standard deviation of log-normal distribution of radio sources' maximum ages     & \S\ref{sec:age}\\  
    $n_{\rm t}$       & value & strength of the maximum source's lifetime redshift evolution& \S\ref{sec:cosmo-combined}\\
    \hline 
    \hline 
  \end{tabular} 
\end{table*}


\subsubsection{Ambient density distribution} 
\label{sec:dens}

The central density value ($\rho_o$) for each generated source is randomly drawn from a log-normal distribution with the mean value ${\rm log}_{10}(\rho_{\rm m})$ and standard deviation of $\sigma_{{\rm log}_{10}(\rho_o)} = 0.15 $. The standard deviation is not introduced as a free parameter; however, we test how strong an effect it has on the results (see discussion in \S\ref{sec:assump-disc}).  
 
Since the type of the surrounding environment (clusters of galaxies or field galaxies) is tightly linked to the core radius of the source, these should never be discussed separately. Here, however, we set one value of $a_o$ for all the simulated sources (2~kpc). The choice of $a_o$  may determine the most likely environments found in our simulation; this issue is discussed in detail in \S\ref{sec:results} and \S\ref{sec:res-density}.  The exponent $\beta$ of the density profile  (Eqn.~\ref{eqn:density}) is randomly chosen from a uniform distribution between {\mbox{$\beta_{\rm min} = 1.2$}} and $\beta_{\rm max} = 2.0$. Although many authors use a constant value of $\beta$ (e.g. $\beta= 1.5$ is used by \citealt{d1}, BRW, \citealt{w4}; $\beta= 1.9$ and $\beta= 2.0$ by KDA, \citealt{w1}), some observational evidence suggest that the parameter may vary between sources (e.g. Alshino et al. 2010).

\subsubsection{Age distribution of the simulated sources} 
\label{sec:age} 
 
It is assumed that radio sources live up to a certain maximum age ($t_{\rm max}$) after which they instantly die.  Although this may seem to be an oversimplification as any relic radio galaxies are going to be neglected, from previous work of KDA and BRW for instance, one can see that the luminosity tracks of radio sources steepen rapidly at the late stages of the source's life, quickly dropping below detectable flux levels.   
 
Here the age of  a source is  randomly drawn from  a uniform distribution between $t=0$ and $t_{\rm max}$. The expectation that all radio sources have the same maximum age seems to be unrealistic, hence we introduce a log-normal spread of the maximum ages around the mean $t_{\rm max_{\rm m}}$, i.e. $\sigma_{\rm log_{10}(t_{max})} = 0.05$. We investigate the effect of the width of this distribution in \S\ref{sec:assump-disc-tmaxstdev}.

\subsubsection{Injection and radio spectral indices} 
\label{sec:spectral-index}

The energy distribution of the relativistic electrons initially follows a power-law relation with exponent $m$, {\mbox{$N(E) dE \propto E^{-m} dE$.}}  The exponent $m$ assigned to a source is drawn from a uniform distribution with the minimum and maximum value allowed by the theoretical model of source growth used in this study, that is $m_{\rm min} = 2$ and $m_{\rm max} = 3$. However, theoretical studies of the particle acceleration in the relativistic shocks  suggest a universal value \citep[$m \sim 2.2-2.4$, e.g.][]{k7,s11}, while studies based on observations of gamma-ray burst afterglows favour a Gaussian distribution of this parameter \citep[][]{c10}. Also, \citet{m7} conclude that the distribution of $m$ is rather narrow with $m\in[2.0,2.4]$ according to their modelling, and \citet{m12} report on good fits of the spectra by a power-law with index $m\cong 2.0 -2.3$. 
 
Moreover, the power-law of the relativistic electron energy distribution  extends between $\gamma_{\rm min}$ and $\gamma_{\rm max}$, i.e. between the Lorentz factors of the least and most energetic electrons. We assume $\gamma_{\rm min}= 1$ and $\gamma_{\rm max}= 10^{10}$. KDA stresses that $\gamma_{\rm min} \ll \gamma_{\rm max}$. We used much higher maximum Lorentz factor than KDA who used $\gamma_{\rm max}=10^5$. Lorentz factors of $\gamma_{\rm max}\sim 10^5 - 10^6$ have also been measured by \citet{m10} and \citet{m8}, while BRW favour $\gamma_{\rm max}\cong 10^{14}$. We decided to use the intermediate value as a default one. Further,  \citet{b5} report a new estimate for the low-energy cut-off of the energy distribution of relativistic electrons in FR~II type sources. Based on the observations of Cygnus A they estimate a rather high value of  $\gamma_{\rm min} = 10^4$. On the other hand, investigations of hot spots done by \citet{m10} suggest that  the minimum Lorentz factors, below which the synchrotron losses are unimportant, are typically $\gamma_{\rm min} \sim 10^2$, but may reach values of $1 - 10^4$. Values of $10^2$ for the minimum energy cut-off are also supported by, for example, \citet{b1} and \citet{m12}. Since there is a large discrepancy, especially in the estimations of the Lorentz factors, we investigate the possible effects of these different assumptions on the simulated source populations in \S\ref{sec:assump-disc-m-lorentz}.

To find the radio spectral index $\alpha$ one needs at least two data points, that is two radio lobe luminosities of the same source measured at two different frequencies.  A simple power-law is employed between $L_{\nu_{178}}$ and  $L_{\nu_{\rm x}}$ to find $\alpha$, which is finally used to correct the $L_{\nu_{178}}$; the radio lobe luminosity estimated with the model, $L_{\nu_{178}}$, is in the source rest-frame and one needs to convert it to the common frequency using $(1+z)^{\alpha}$. For consistency and to mimic behaviour of the 3CRR sample we chose $\nu_{\rm x} = 750 $~MHz \citep[after][]{k1}. We note here, however, that BRW, \cite{j1} and \cite{j2} argue that this most commonly used relation (i.e. simple power-law) may be too simplistic, and instead, curved radio spectra should be considered.

\subsubsection{Aspect ratios ($R_{\rm T}$) and self-similarity} 
\label{sec:aspect-ratio}

One of the consequences of the KA dynamical model of the source expansion is its self-similar growth. Instead of measuring or introducing the jet opening angle, the aspect ratio ($R_{\rm T}$) is used. In the KA model the aspect ratio is linked to the jet opening angle ($\theta$) by $R_{\rm T} \propto 1 / \theta $ based on the assumption that the lobe pressure is balanced by the external gas ram pressure. $R_{\rm T}$ stays constant through the source lifetime. The BRW model, on the other hand, assumes that $R_{\rm T}$ increases with the source growth, and their results suggest some dependence on the source kinetic luminosity and/or its age, while \cite{m1,m2} suggest that the self-similarity may not hold for old sources.

\citet{m11} present a detailed investigation of aspect ratios of 3C sources of $z<1.0$. The observed $R_{\rm T}$ values are found to fall in a range of $1<R_{\rm T}< 8$, with a median within $\langle R_{\rm T}\rangle \in 1.6 - 2.6$ (which depends on spectral class). Mullin et al. (2008) report that higher aspect ratios seem to occur for larger sources ($>100$~kpc) which may suggest a non self-similar source growth, or a self-similar growth occurring only in the early stages of a source life.  However, the aspect ratio has a predominant influence on the radio source linear size (hence its age). In particular, the smaller the aspect ratio gets, the wider the opening angle of the jet becomes, and thus a higher pressure of the lobe is needed to account for the faster sideways expansion of the lobe. This in turn will lead to smaller head advance speeds of the jet, as well as larger synchrotron losses, and will result in smaller linear sizes as compared to the sources with larger $R_{\rm T}$. This may be one of the reasons why larger sources seem to have larger aspect ratios. Similarly, BRW suggested that the aspect ratio might be higher for more powerful sources. One must notice, however, that for a given age and environment of a radio source higher kinetic energy will lead to larger linear size, and, according to our argument above, larger aspect ratios may be expected.

 Since the distribution of $R_{\rm T}$ is not yet well constrained;  we decided to assume a uniform distribution of the aspect ratio with {\mbox{$6.0 > R_{\rm T} > 1.3$}}, which translates to the range of jet opening angles of $ 13.7^{\circ}<\theta<63^{\circ}$ (values based on \citealt{l3,d1}; Machalski et al. 2004a; Mullin et al. 2008), and to follow the original KDA model which implies self-similar growth of a radio source. A distribution of aspect ratios allows for a variation in growing rates of radio sources and hence is more realistic than a single value.

\subsubsection{Jet particle content} 
\label{sec:particle-content}

The ratio ($k'$) of the energy densities of thermal particles to the energy densities of the electrons at the time they are injected into the cocoon  is assumed to be $k'=0$. Note that this definition of $k'$ differs slightly from typical assumptions ($k$) in such a way that relativistic and non-relativistic electrons are already included even if $k'=0$ (this already implies that the typically assumed ratio is $k\neq0$).

There is much debate on the particle content of the radio galaxy jets. Some argue that the FR~II jets are lightweight (electron-positron dominated), while FR~I jets are heavy, that is they may possess a significant proton content \cite[e.g.][]{c9,w3,k6,d4,c7,c8}. Moreover, \citet{h4} based on the investigation of Cygnus A report that FR~II radio galaxies may be more proton dominated ($k\sim 1-4$) if they reside in very rich environments.  KDA showed that heavy FR~II jets will require significantly higher jet powers to reproduce the same radio luminosity as the electron-positron dominated jets, hence they concluded that $k'$ must be close to 0. Initially, in our simulation we followed the conservative KDA assumption, but the effect of changing $k'$ (to allow a proton content in the jets) on the whole population of simulated sources is investigated in \S\ref{sec:assump-disc-k}.

\subsubsection{Maximum head advance speeds} 
\label{sec:adv-speed} 
 
It is at first assumed that the head advance speed of a source ($v_{\rm max}$) at the time of observation may not exceed $0.4c$, all sources that surpass this limit are rejected at the stage of generating the population. Such an upper limit is consistent with the speeds inferred from the synchrotron spectral ageing of high luminosity double radio sources \citep[e.g.][]{l5, b9}, and supported by the estimation of dynamical ages of FR~II sources by \citet{m7} who report $v_{\rm max}\leq 0.3c$. However, there have been discussions on the possible overestimation of the lobe advance speed upper limits; results obtained through the analysis of lobe asymmetries and the steepening of radio spectra suggest that the head advance speeds do not exceed $0.15c$ \citep[e.g.][]{a4}, or even 0.05c \citep{s3}. We discuss the effect of different assumptions of $v_{\rm max}$ on the results in \S\ref{sec:assump-disc-advspeed}.

\subsubsection{Adiabatic indices of radio lobes, magnetic field, and external medium} 
\label{sec:adiab-indices} 
 
After KA/KDA, we assume the adiabatic indices of the radio lobes, magnetic fields and the external medium to be $\Gamma_{\rm l}= \Gamma_{\rm b} = 4/3$ and $\Gamma_{\rm x} = 5/3$ respectively, that is we assume a non-relativistic equation of state for the closest external medium of the radio source, and relativistic particles inside the radio lobes.

 
\begin{figure} 
\includegraphics[width=64mm, angle=270]{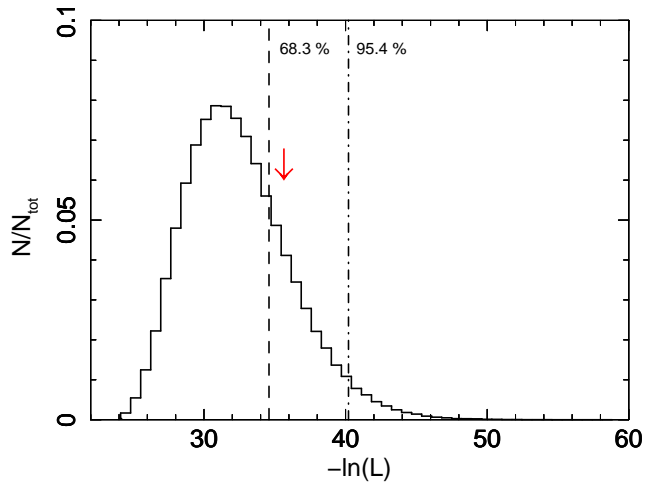} 
\caption{The goodness-of-fit test. The synthetic log-likelihood distribution of the $2 \times 10^5$ synthetic data sets generated from the model average histogram (consult \S\ref{sec:stats} for term explanation). The red arrow indicates the position of the actual data set log-likelihood. The dashed line (black) marks the $68.3$ per cent containment (from the minimum value of $-{\rm ln}(\mathcal{L})$, note that the distribution is one-sided), and the dot-dashed one (black) the $95.4$ per cent containment. The histogram of the synthetic log-likelihood distribution of the lowest redshift data of Model S is shown. The data is consistent with the model at the 90 per cent confidence level ($p-$value=0.234).} 
\label{rys:gof-test} 
\end{figure} 
 

 
\begin{table*} 
  \centering 
 
  \caption{Searched ranges and steps of the distribution parameters in the grid minimisation.} 
  \label{tab:grid-min-ranges} 
  \begin{tabular}{lclll} 
    \hline 
    \hline 
     Parameter      & Model &   Searched range       &  Step   & Unit            \\ 
    \hline 
                    &       & {\sc Independent-{\it$z$}}\\ \\
    $Q_{\rm B}$      & S     & $[$\phantom{$-$}$36.90 ,$\phantom{$ -.$}$44.10]$  &  0.15  &log$_{10}$(W)   \\
    $\alpha_{\rm s}$ & S     & $[-13.00,$\phantom{$ -0.$}$1.70]$               &  0.1   &   \\
   
    $Q_{\rm B}$      & W     & $[$\phantom{$-$}$35.60 ,$\phantom{$ -.$}$ 42.00]$ &  0.15  &log$_{10}$(W)   \\
    $\alpha_{\rm s}$ & W     & $[$\phantom{$0$}$-0.45,$\phantom{$ -.0$}$9.50]$                &  0.15  &    \\
   
    $\rho_o$        & S, W  & $(-26.00, -16.80] $                           &  0.2  &log$_{10}$(kg$\,$m${^{-3}}$)    \\
   
    $t_{\rm max}$    & S, W  & $[$\phantom{$ -0$}$5.18,$\phantom{$ -.$}$ 10.00]$   &  0.15 &log$_{10}$(yr)    \\\\
   
                    &       & {\sc Combined-{\it$z$}}\\ \\
    $Q_{\rm B}(z=0)$  & S    & $[$\phantom{$-$}$35.00 , $\phantom{$ -.$}$40.70]$  &  0.3  & log$_{10}$(W)  \\
    $\alpha_{\rm s}$  & S    & $[$\phantom{$0$}$-1.20 , $\phantom{$ -0.$}$1.60]$    &  0.2      \\

    $Q_{\rm B}(z=0)$  & W    & $[$\phantom{$-$}$34.70 ,$\phantom{$ -.$}$ 39.50]$ &  0.3  & log$_{10}$(W)   \\
    $\alpha_{\rm s}$  & W    & $[$\phantom{$-0$}$0.00 ,$\phantom{$ -0.$}$ 4.00]$    &  0.2      \\

    $n_{\rm q}$       & S, W & $[$\phantom{$-0$}$0.00 ,$\phantom{$ -.$}$ 12.00]$   &  0.5      \\
    $\rho_o(z=0)$    & S, W & $(-26.00,-19.40 ]$ &  0.3 & log$_{10}$(kg$\,$m${^{-3}}$)    \\
    $n_{\rm r}$       & S, W & $[$\phantom{$-0$}$0.00 ,$\phantom{$ -.$}$12.00]$   &  0.5      \\
    $t_{\rm max}(z=0)$& S, W & $[$\phantom{$-0$}$6.58 ,$\phantom{$ -.0$}$ 9.58]$   &  0.3  & log$_{10}$(yr)   \\
    $n_{\rm t}$       & S, W & $[$\phantom{$-0$}$0.50 ,$\phantom{$ 0.$}$ -5.00]$    &  0.5      \\
    \hline 
    \hline 
  \end{tabular} 
 
\end{table*}


\subsubsection{Projection effects} 
\label{sec:Dproj} 
 
The linear size calculated based on the randomly chosen and set physical parameters of each source in the generated population (see steps 2. and 3. in \S\ref{sec:mc-sim}) is the source true size, and is further randomly oriented in the sky as it is observed at some projection angle $\vartheta$. It is assumed that the projection angle is distributed uniformly in $[1- {\rm cos}(\vartheta)]$ plane. Therefore, the size becomes $D_{\rm proj} = D_{\rm true} {\rm sin}(\vartheta)$. For simplicity we will refer to this projected size as $D$, and the true size is not considered from now on.  
 
Also, note that Eqn.~\ref{eqn:density} is an approximation and is valid for distances $r$ greater than few core radii, hence we do not consider sources smaller than 10~kpc in our samples, either the simulated or the observed ones\footnote{9 sources in total, resulting from this requirement, were excluded from  the observed samples.}.

\subsubsection{Redshift distribution} 
\label{sec:redshift}

It is assumed that sources are uniformly distributed in space, between minimum ($V_{\rm min}$) and maximum ($V_{\rm max}$) comoving volume depending on the considered redshift range. From this a radio source redshift ($z$) is found. Redshifts from {\mbox{$z=0$}} to {\mbox{$z=2$}} in steps of $\Delta z = 0.001$ are considered in this work. Note that evolution of source number density is not initially modelled in the construction of the simulated population; however we apply a normalization (scaling), which is a free parameter in each redshift range (step 9. in \S\ref{sec:mc-sim}, \S\ref{sec:mlm}), before the maximum likelihood method and goodness-of-fit test are performed. Hence we obtain an average in radio source number density for each $z$ bin considered.

\subsection{Confidence intervals and Goodness-of-Fit test} 
\label{sec:stats}

\subsubsection{Binned Maximum Likelihood Method} 
\label{sec:mlm}

We use the binned maximum likelihood method (MLM) in the statistical tests \cite[e.g.][]{c2}. A log-likelihood is found in fitting each of the simulated samples to the observed data according to 
\begin{equation} 
{\rm ln \mathcal{L}} = \Sigma^N_{i=1}\left [x_i {\rm ln}\left(\frac{y_i}{n}\right) -\frac{y_i}{n} - {\rm ln}(x_i!) \right ] 
\label{eqn:log-like} 
\end{equation} 
where $N$ is the number of bins, each with an expected number of sources $y_i$ and with the observed number of counts $x_i$, and $n$ is the normalization constant discussed in step 9. in \S\ref{sec:mc-sim}.

\subsubsection{Goodness-of-Fit} 
\label{sec:gof} 
 
To perform the goodness-of-fit (GoF) test we use the Monte Carlo analysis. The final histogram of the simulated radio sources is in fact the mean of all possible realizations of the underlying `true' population, which we call the `model average histogram'. The Monte Carlo procedure undertaken here is to randomize each bin of the model average histogram within the Poisson regime to obtain a `synthetic data set', a single realization of the `true population'. Further, we apply the binned MLM method to find the log-likelihood between the newly created synthetic data set and the model average histogram. We repeat the procedure multiple times ($\sim 2 \times 10^5$) to obtain a distribution of the possible log-likelihoods of the true population, which we refer to as the `synthetic log-likelihood distribution'. Finally, the log-likelihood of the observed data histogram is compared to the generated synthetic log-likelihood distribution. The $1\sigma$ level of consistency requires the actual data set log-likelihood to be placed within the 68.3 percentile of the log-likelihood distribution, i.e. between 0 and 68.3 per cent since our distribution is one-sided (Figure~\ref{rys:gof-test}). We quote the GoF test results in terms of $p-$value, which is the probability that the test statistic is at least as extreme as the one observed and assuming the null hypothesis is true. $1\sigma$ is equal to $0.317$ in terms of $p-$value. The null hypothesis is rejected if the $p-$value is less than the significance level $\alpha_{\rm sig}$; we assume  $\alpha_{\rm sig}=0.1=10$ per cent. The choice of the number of synthetic histograms allows us to check the agreement up to $5\sigma$ level ($p-{\rm value}=6\times10^{-7}$). It is important, however, to remember that the confidence levels obtained in such a way are nominal only as the synthetic log-likelihood distribution is not Gaussian in our case.

Although this technique tests whether a model fits data well, it is incapable of distinguishing the best-fitting models. To compare these best-fitting models one needs to perform the so-called likelihood ratio test the statistic of which is 
 
\begin{equation} 
d = -2 ~[ ~{\rm log}(\mathcal{L}(H_{\rm m0})) - {\rm log}(\mathcal{L}(H_{\rm m1}))~] 
\label{eqn:f-test} 
\end{equation} 
and where $\mathcal{L}$ denotes log-likelihood, $H_{\rm m0}$ is the null model and $H_{\rm m1}$ an alternative one. The $p$-value of the $d$ statistics may be obtained to find which model should be preferred. The models must be nested, however, and if this requirement is not fulfilled, the likelihood ratio test is invalid. For the models $H_{\rm m0}$ and $H_{\rm m1}$ to be nested it is necessary that $H_{\rm m1}$ contains the same parameters as $H_{\rm m0}$, and has at least one additional parameter.


\subsubsection{Confidence intervals} 
\label{sec:intervals}

The source physical parameters that are the main focus in this study are $t_{\rm max_{\rm m}}$ (\S\ref{sec:age}), $\rho_{\rm m}$ (\S\ref{sec:dens}), and $Q_{\rm B}$ and $ \alpha_{\rm s}$ (\S\ref{sec:jetpow}). To obtain the confidence intervals of these parameters we perform grid minimisation searching their  broad ranges (Table \ref{tab:grid-min-ranges}), and follow the method of  \cite{c2}. For each parameter set in our searched grid there is a corresponding log-likelihood. From these, the global $\mathcal{L}_{\rm max}$ is found, indicating the best fit to the observed data (Eqn.\ref{eqn:log-like}). Furthermore, as pointed out by \cite{c2},  one may find the difference between $\mathcal{L}_{\rm max}$ and $\mathcal{L}$ for all the other sets of parameters in the evaluated grid, the so-called $\Delta C$ statistic, which is defined as 
\begin{equation} 
\Delta C = [-2 ~{\rm ln} (\mathcal{L}_{\rm max}) + 2 ~{\rm ln} (\mathcal{L}) ]. 
\end{equation} 
$\mathcal{L}$ is the log-likelihood of the sub-grid, which is extracted from the global grid by setting each point of a parameter which we are focused on as non-varying and the corresponding set of log-likelihoods of these points are listed based on the  all the other parameters that vary.

The $\Delta C$ statistic is distributed as $\chi^2$ with $n_{\mathcal{L}_{\rm max}} - n_{\mathcal{L}}$ degrees of freedom (dof), where $n_{\mathcal{L}_{\rm max}}$ denotes degrees of freedom associated with the global $\mathcal{L}_{\rm max}$ and $n_{\mathcal{L}}$ denotes the dof of restricted sub-grid (dof $= 2$ in our plots). The confidence intervals are, therefore, defined such that contours encircle parameters for which their log-likelihood is above a certain value ($\mathcal{L} > \mathcal{L}_0$), and levels of $\chi^2$ distribution may be used. Note that this may lead to disconnected regions in the case the likelihood function is highly irregular. For an in-depth description of constructing contour plots in cases such as this one we refer the reader to \cite{c1} and \cite{l2}.

\section{Results} 
\label{sec:results} 
 
In this section we present and discuss the results of our Monte Carlo simulations. The fitted histograms and  the radio luminosity functions of the analysed observed and simulated samples are presented in \S\ref{sec:rlf}. The evidence for the cosmological evolution of the physical parameters is discussed in \S\ref{sec:cosmo-evid}. An important note on the parameter degeneracy is discussed in \S\ref{sec:rho-tmax-degen}. Finally, \S\ref{sec:assump-disc} contains discussion on the possible effects of the model assumptions  on the results.

\subsection{Radio luminosity functions and fitted data counts} 
\label{sec:rlf}

\begin{figure*} 
\includegraphics[angle=270,width=180mm]{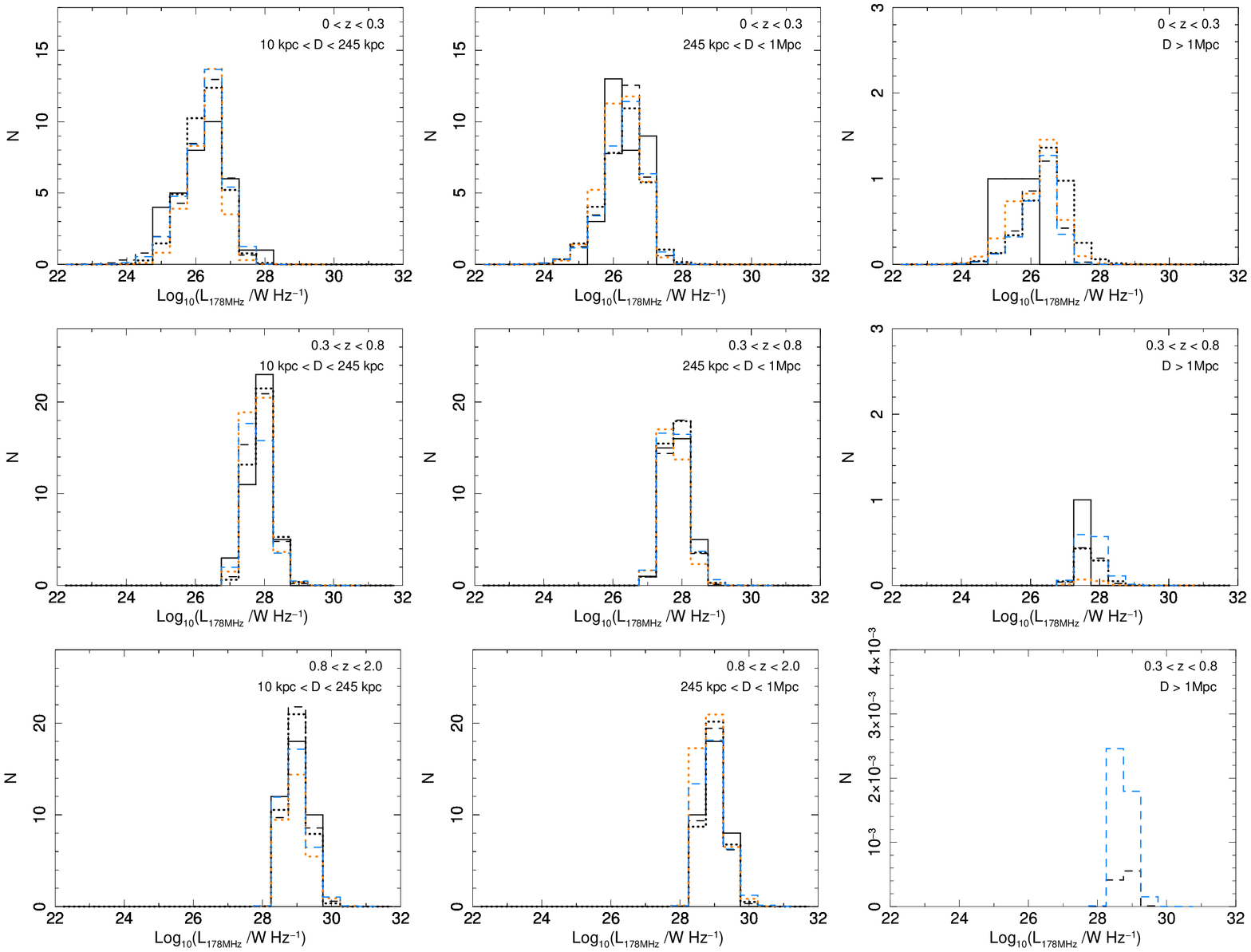} 
\caption{Histograms of the observed radio sample (black solid line) and simulated ones created with the best-fitting parameters of the independent-$z$ fits of model S (black dashed line) and model W (black dotted line), and the combined-$z$ fits of model S (blue dashed line)  and model W (orange dotted line), in $z_1$, $z_2$ and $z_3$. In each redshift range there are three separate size bins as described within the subplots. Each redshift bin is simulated independently, while the FR~IIs linear sizes within each redshift range are simulated simultaneously; that is, a good fit to all linear sizes simultaneously at the same redshift is required. The simulated source populations created with the best-fitting parameters are consistent with the data at the $90$ per cent confidence level based on the $\Delta C$ statistics (for exact $p-$values see Table \ref{tab:best-fits}).}
\label{rys:hist} 
\end{figure*}

\begin{figure*} 
\includegraphics[angle=270,width=180mm]{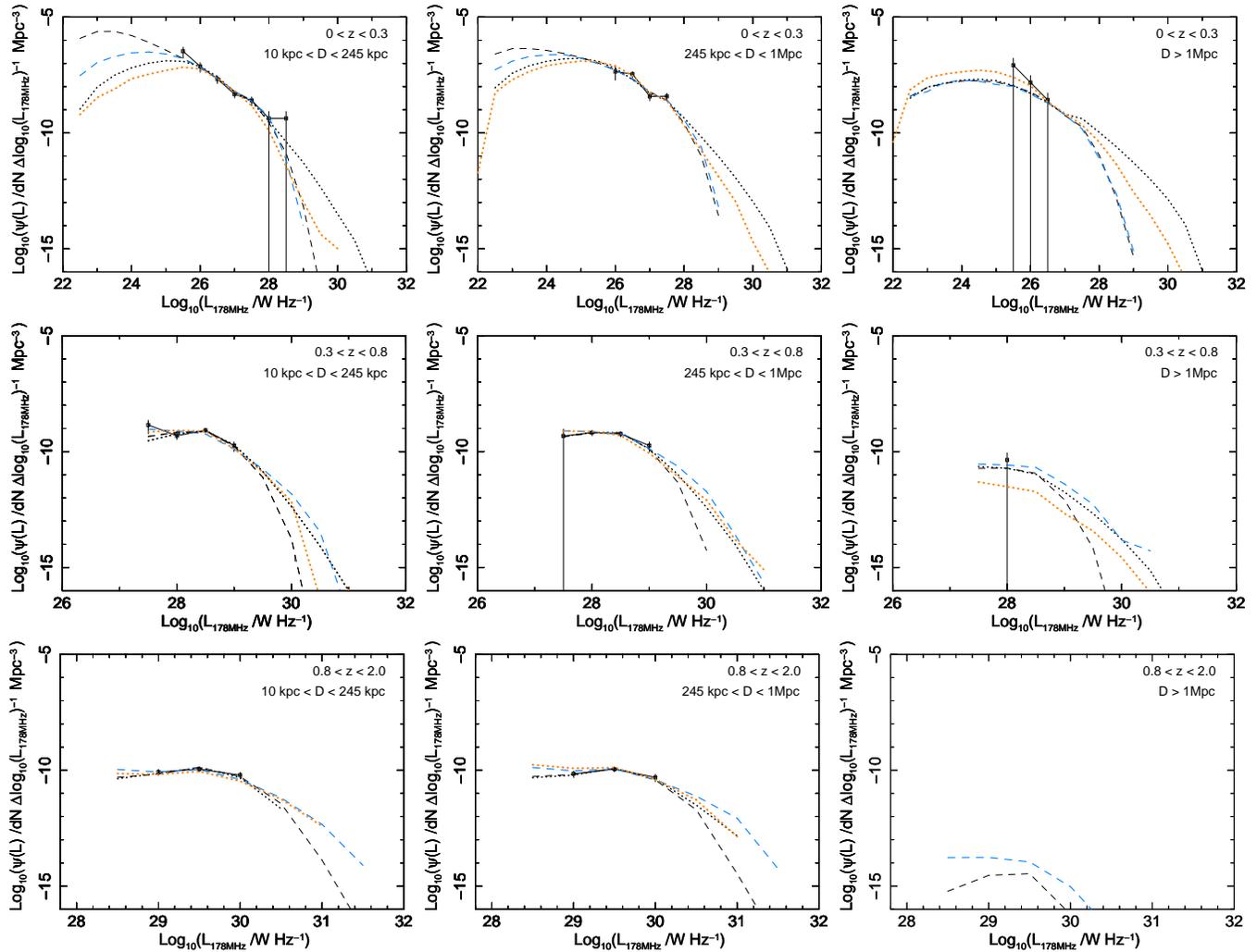} 
\caption{Radio luminosity functions of FR~II type sources from 3CRR and BRL catalogues (solid) and the simulated populations generated with the best-fitting parameters for each redshift and model tested (independent-$z$ fits of Model S drawn as black dashed line, independent-$z$ fits of Model W as black dotted line, combined-$z$ fits of model S as blue dashed line, and combined-$z$ fits of model W as orange dotted line). Each redshift bin is simulated independently, while the FR~IIs linear sizes within each redshift range are simulated simultaneously; that is, a good fit to all linear sizes simultaneously at the same redshift  is required. The simulated source populations created with the best-fitting parameters are consistent with the data at the $90$ per cent confidence level based on $\Delta C$ statistics (for exact $p-$values see Table \ref{tab:best-fits}).}
\label{rys:RLF} 
\end{figure*}

The best-fitting model histograms are presented in Figure~\ref{rys:hist}. The corresponding RLFs are shown in Figure~\ref{rys:RLF}. The quoted uncertainties on the RLFs are the Poissonian errors only. The simulated RLFs appear to be consistent with the observed data. Moreover, we were able to reconstruct the luminosity distributions of the observed samples for their given linear size subsamples as well as the redshift ranges. The match in the linear size distribution is of particular interest, as it has been noted previously that not all previous work succeeded in reconstructing the linear size distributions and often too many large sources have been created \citep[see e.g.][]{b1,b2}. It may be possible that such an effect might not be due to the specific models used, but rather is due to restricting the model parameters, which in our case were kept free. We have not tested analytical models other than KA/KDA in this work.

\subsection{Evidence for the cosmological evolution of the intrinsic and extrinsic source parameters} 
\label{sec:cosmo-evid}

In the simplest case we have tested, it was assumed that the parameters that are our main focus (that is $t_{\rm max_{\rm m}}, \rho_{\rm m}, Q_{\rm B}$ and $\alpha_{\rm s}$) do not evolve with redshift within each redshift bin. We analyse the redshift ranges independently, hence separate sets of best-fitting parameters for each redshift bin are found; if there is no redshift evolution of the intrinsic and extrinsic parameters, the results for each redshift range should not be significantly different. These are referred to as the independent-redshift fits and are presented in \S\ref{sec:cosmo-independent}. Subsequently, we have attempted to investigate the strength of the cosmological evolution of the source physical parameters by allowing for continuous redshift evolution of $t_{\rm max_{\rm m}}, \rho_{\rm m}$ and $Q_{\rm B}$ within each redshift bin, and fitting all the redshift bins simultaneously using the same parameters to describe the evolution. We refer to this case as the combined-redshift fits, and the results are presented in \S\ref{sec:cosmo-combined}.

\subsubsection{Independent-$z$ fits} 
\label{sec:cosmo-independent}

The confidence intervals of the best-fitting parameters for the simple Model S in the respective redshift bins are shown in Figure~\ref{rys:cont-sch-z1} -- Figure~\ref{rys:cont-sch-zall}, and similarly for the Model W in Figure~\ref{rys:cont-wlt-z1} -- Figure~\ref{rys:cont-wlt-zall}. The best fits and the GoF test results are listed in Table~\ref{tab:best-fits}, although we once again stress that the best-fitting parameters maximize the likelihood but are not the standard means of their respective distributions and the inspection of the associated confidence intervals is very important.

The confidence intervals span a wide range of the possible values for each of the parameters resulting in relatively large uncertainty in the best fit. One must bear in mind that these confidence intervals are based on the underlying five degrees of freedom allowed in the simulation, and any visible degeneracy or shift may be influenced by a change of other parameters. Indeed, some degeneracies are seen,  and are especially strong in the case of the maximum source age and the corresponding central density; however, the degeneracies seem to be also present for the $Q_{\rm B} - t_{\rm max_{\rm m}}$ and $Q_{\rm B} - \rho_{\rm m}$ results. The degeneracy issue is discussed separately in \S\ref{sec:rho-tmax-degen}.

The confidence intervals do not converge for the high kinetic luminosity break values ($>10^{42}$~W) in the case of Model S, and for the low kinetic luminosity breaks ($<10^{37}$~W) in the case of Model W. It is accompanied by negative slopes of approximately $0.5 - 1.5$ in both cases. At this stage the function becomes a power-law. Despite the fact that the kinetic luminosity break indicates values of $>10^{42}$~W, the number density of jets with kinetic luminosities close to this break is so low that the actual $Q_{\rm max}$ contributing to the simulated population is $\sim10^{41}$~W. Similarly, in the Model W the kinetic luminosity break shifts to very low values and only the high kinetic luminosity end of the distribution, which at this point is approximated by a power-law, contributes to the observed data counts.  In previous studies, authors have assumed such power-law distributions. Assuming that  
\begin{equation} 
\psi(Q) {\rm d}Q = Q^{-x}{\rm d}Q,  
\end{equation} 
slopes of $x=2.6$ (BRW), $x=3.3$ and $x=3.6$ \citep{b1}, $x=1.6$ \citep{kb1}, and $x\sim 0.9 - 1.3$ \citep{w1} have been found. Here, we observe slopes of $x\sim0.5 - 1.5$ (Model S and W) in the power-law extremum. We can conclude here, therefore, that the hypothesis that kinetic luminosities follow an unbroken power-law distribution cannot be ruled out at a confidence level of more than $95.4$ per cent.

To ease the investigation of the possible cosmological evolution of all the searched parameters Figure~\ref{rys:cont-sch-zall} and Figure~\ref{rys:cont-wlt-zall} show superimposed 90 per cent intervals of all redshifts. We find that the kinetic luminosity break shifts to higher values for higher redshift ranges in the case of both  Model S and Model W. Furthermore, Figure~\ref{rys:cont-sch-zall} clearly shows that there is no common solution found for the maximum source age and the central density for all three redshift ranges. Consequently, the hypothesis that there is no evolution of $t_{\rm max_{\rm m}}$ and $\rho_{\rm m}$ is ruled out with probability of $> 99$ per cent. One may notice, however, that it is possible to find $Q_{\rm B}$ to be constant for all redshifts (Model S only), but this would imply unrealistically strong evolution of $\rho_{\rm m}$. To avoid such a strong evolution more than one parameter would have to undergo an evolution with redshift.  
 As an additional test, we have attempted to fit the three redshift bins simultaneously with one set of parameters that would be valid for all redshift sub-samples. No satisfying results have been found for the latter test case, the consistency of the hypothesis with the data was ruled out at the nominal $5\sigma$ level.

\subsubsection{Combined-$z$ fits} 
\label{sec:cosmo-combined}

We have attempted to investigate the strength of the cosmological evolution of the intrinsic and extrinsic source parameters. The results of the independent-$z$ fit suggest evolution of more than one parameter, and bearing in mind the possible degeneracies (see discussion in  \S\ref{sec:rho-tmax-degen}), we tried to restrict as few parameters as possible. We allowed simultaneous evolution of the following parameters: (i) the kinetic luminosity break was allowed to evolve with redshift as $Q_{\rm B}(z) = Q_{\rm B}(z=0)(1+z)^{n_{\rm q}}$, (ii) the central density of the environment as $\rho_{\rm m}(z) = \rho_{\rm m}(z=0)(1+z)^{n_{\rm r}}$, and (iii) the maximum age of the sources was assumed to undergo evolution according to  $t_{\rm max_{\rm m}}(z) = t_{\rm max_{\rm m}}(z=0)(1+z)^{n_{\rm t}}$. However, due to limited computing time and power, we were forced to compromise on the resolution of the grid of the parameters searched over. For instance, the $n_{\rm q}, n_{\rm r}$ and $n_{\rm t}$ exponents were searched in steps of 0.5 only, and $Q_{\rm B}, \rho_{\rm m}$ and $t_{\rm max_{\rm m}}$ in steps of 0.3 in a logarithmic scale.

The results are displayed  in Figure~\ref{rys:schech-evol} (Model S) and Figure~\ref{rys:wlt-evol} (Model W). The best agreement between the simulated populations and the observed samples is listed in Table~\ref{tab:best-fits}. As discussed in \S\ref{sec:rho-tmax-degen} the degeneracy between parameters means that the $n_{\rm q}, n_{\rm r}$ and $n_{\rm t}$ cannot be easily constrained, and additional constraints on the parameters are necessary. An interesting result emerges for the kinetic luminosity break cosmological evolution, for which the trend with redshift seems to be unavoidable ($n_{\rm q}>3$) as our results show. We discuss these results in detail in \S\ref{sec:discussion}.

\subsection{Central density -- maximum age degeneracy} 
\label{sec:rho-tmax-degen} 
 
The degeneracy between central density and maximum age needs some extra attention as it seems to be often overlooked. The calculated radio lobe luminosity of a radio source depends on its kinetic luminosity, central density and the age of the source. However, one must be aware that restricting one parameter will yield an adjustment in another. In particular,  the maximum age of the sources and the central density in which sources evolve seem to be the most strongly correlated. From Eqn.~\ref{eqn:size} -- \ref{eqn:luminosity} one can deduce that 
%
\begin{eqnarray}
L_{\nu} \propto Q^{\frac{12+2m-\beta(3+m)}{2(5-\beta)}} (\rho_o a_o^{\beta})^{\frac{3+2m}{2(5-\beta)}}
               t^{\frac{6-4m-\beta(3+m)}{2(5-\beta)}} \int_{x_{\rm min}}^1 C\left(x\right)d\left(x\right). 
\label{eqn:tmax-degener} 
\end{eqnarray} 
%
Assuming $m=2.5$ and $\beta=1.5$ (most commonly assumed values) and ignoring for the moment the energy losses term, one will see that  
\begin{equation} 
L_{\nu} \propto Q^{1.25} \rho_o^{1.15} t^{-1.75}. 
\label{eqn:tmax-degener-assump} 
\end{equation} 
Clearly, if we keep radio lobe luminosity, kinetic luminosity and all the other values the same for a source, and change only its central density and the age, the two latter parameters will compensate for each other.  
For example, a change in the central density by a factor of $10$ (reaching less dense environments) will yield a change in source age by a factor of $7$ (implying younger ages) to maintain the same radio lobe and kinetic luminosities. Of course this is an approximation only as such a change will also yield changes in the source linear size and in the loss processes term $C(x)$ of Eqn.~\ref{eqn:luminosity} since both depend on the stage of the source life. Note that Eqn.~\ref{eqn:tmax-degener-assump} cannot be used as an approximation for simple $L_{\nu}$ calculation because the energy losses cannot be neglected. Eqn.~\ref{eqn:tmax-degener-assump} is used here solely for descriptive purposes.

\begin{figure*} 
\includegraphics[angle=270,width=136mm]{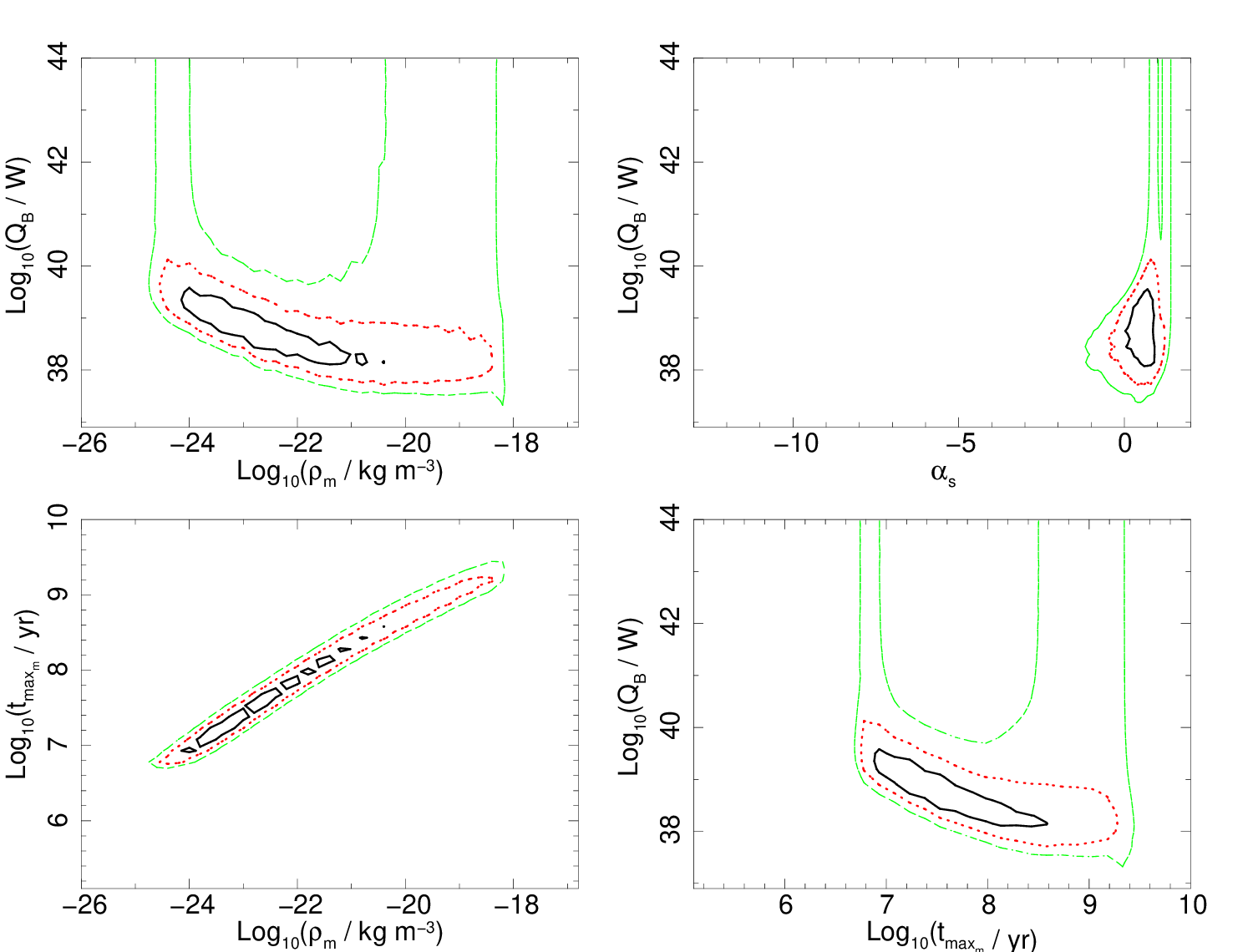} 
\caption{Joint confidence intervals for the independent-$z$ fits of Model S in $z_1$ redshift range are shown ($z_1\leq0.3$). 68.3 per cent (solid, black), 95.4 per cent (dotted, red) and 99.7 per cent (dashed green) contours, based on $\Delta C$ statistics (see \S\ref{sec:stats}), are shown. The best-fitting parameters are consistent with the data at the 90 per cent confidence level  ($p-{\rm value} = 0.234$).} 
\label{rys:cont-sch-z1} 
\end{figure*}

\begin{figure*} 
\includegraphics[angle=270,width=136mm]{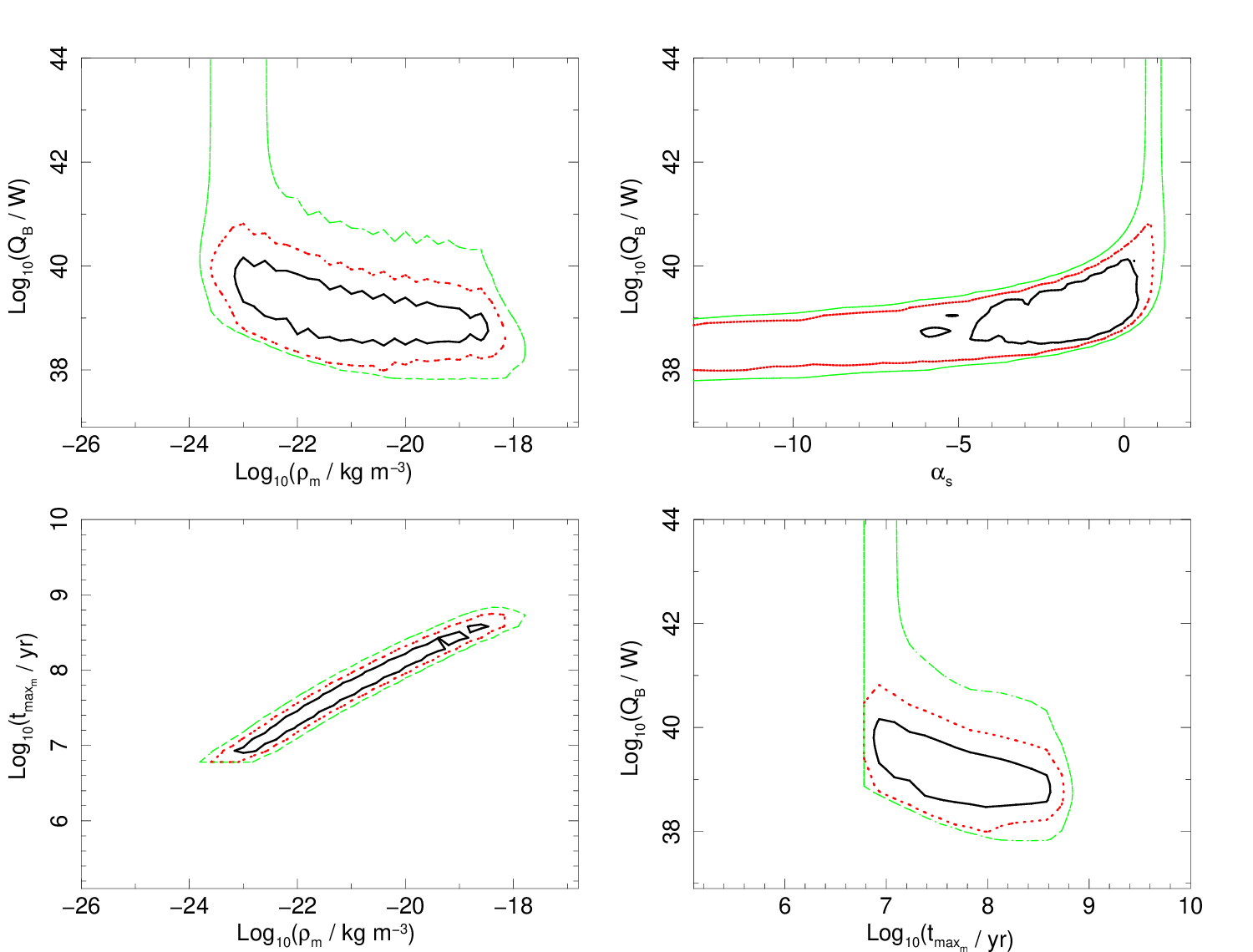} 
\caption{Joint confidence intervals for the independent-$z$ fits of Model S in $z_2$ redshift range are shown ($0.3<z_2\leq0.8$). 68.3 per cent (solid, black), 95.4 per cent (dotted, red) and 99.7 per cent (dashed green) contours, based on $\Delta C$ statistics (see \S\ref{sec:stats}), are shown. The best-fitting parameters are consistent with the data at the 90 per cent confidence level  ($p-{\rm value} = 0.639$).} 
\label{rys:cont-sch-z2} 
\end{figure*}

\begin{figure*} 
\includegraphics[angle=270,width=136mm]{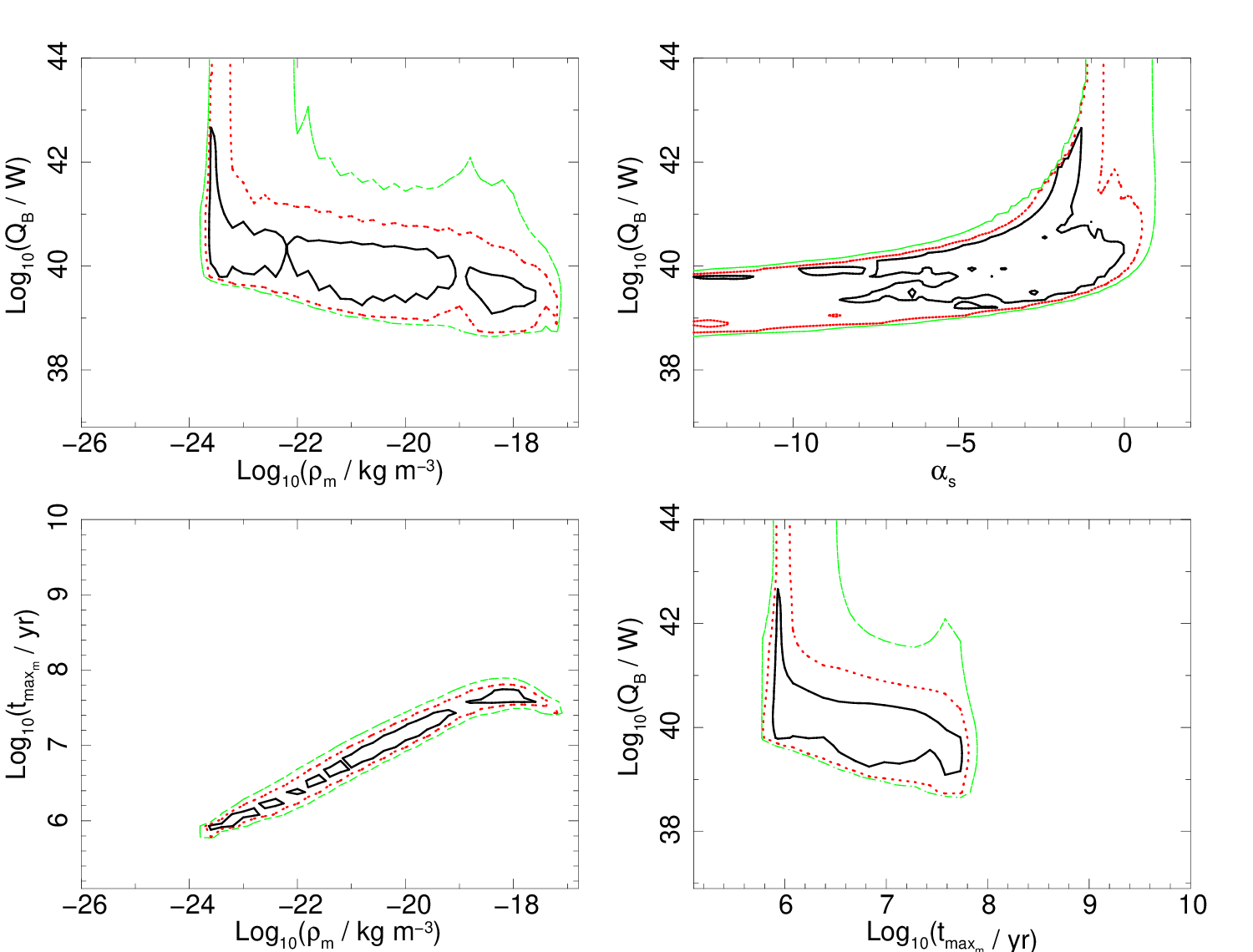} 
\caption{Joint confidence intervals for the independent-$z$ fits of Model S in $z_3$ redshift range are shown ($0.8<z_3\leq2.0$). 68.3 per cent (solid, black), 95.4 per cent (dotted, red) and 99.7 per cent (dashed green) contours, based on $\Delta C$ statistics (see \S\ref{sec:stats}), are shown. The best-fitting parameters are consistent with the data at the 90 per cent confidence level  ($p-{\rm value} = 0.925$).} 
\label{rys:cont-sch-z3} 
\end{figure*}

\begin{figure*} 
\includegraphics[angle=270,width=136mm]{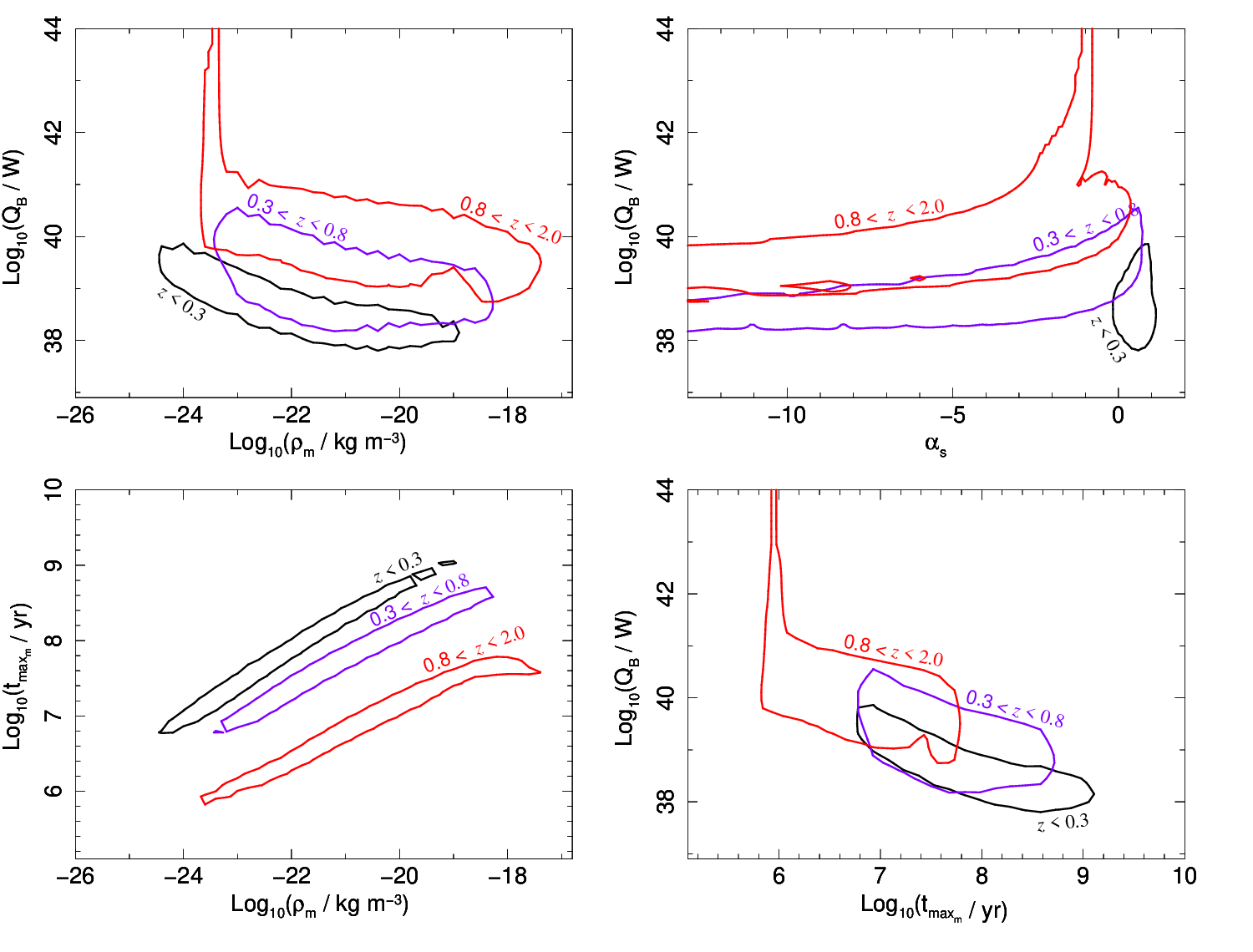} 
\caption{Overlaid 90 per cent confidence intervals  based on $\Delta C$ statistics (\S\ref{sec:stats}) for the three redshift ranges considered ($z_1$ drawn in black, $z_2$ in blue, $z_3$ in red) of the independent-$z$ fits of Model S. } 
\label{rys:cont-sch-zall} 
\end{figure*}

\begin{figure*} 
\includegraphics[angle=270,width=136mm]{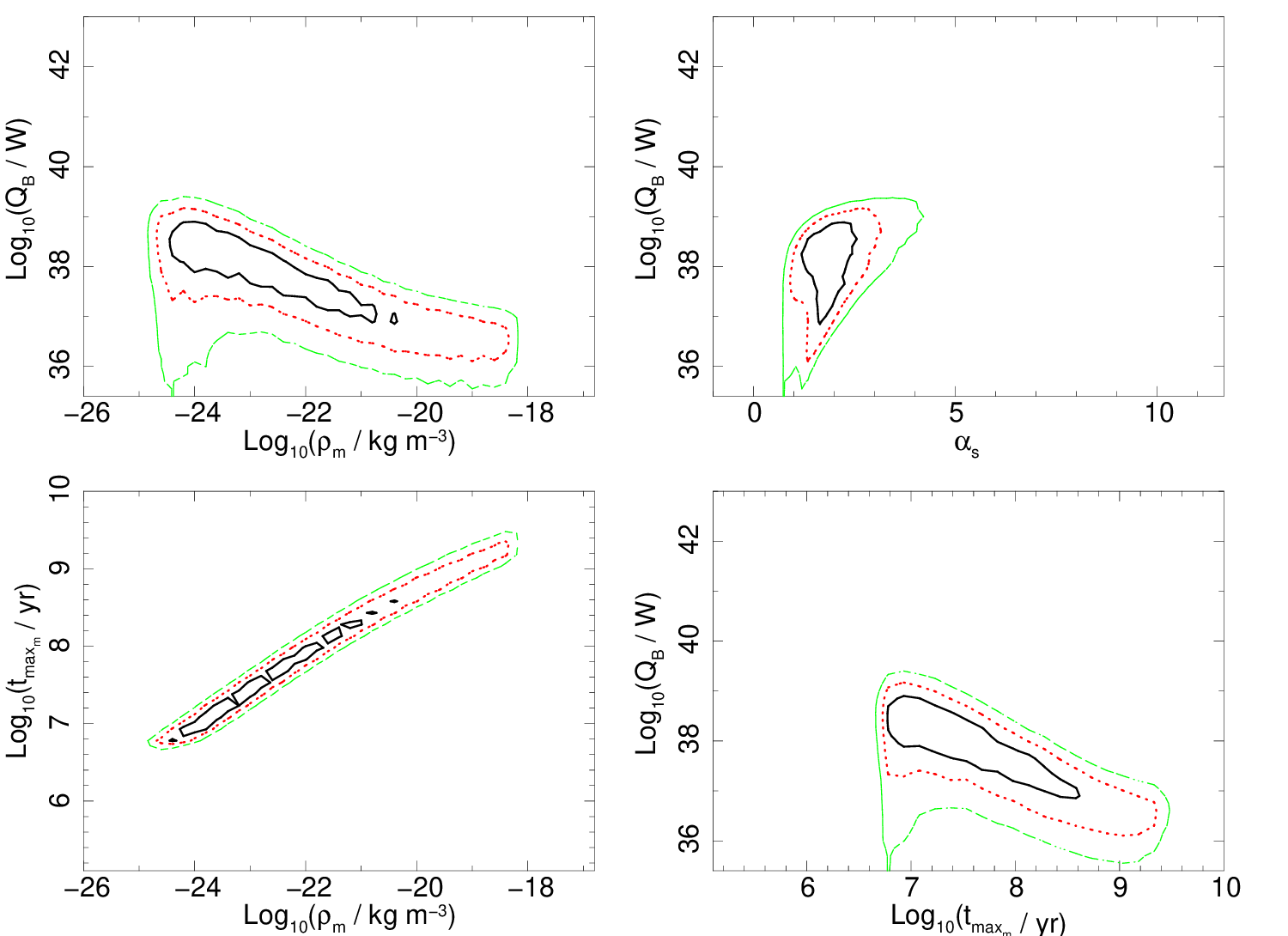} 
\caption{Joint confidence intervals for the independent-$z$ fits of Model W in $z_1$ redshift range are shown ($z_3\leq0.3$). 68.3 per cent (solid, black), 95.4 per cent (dotted, red) and 99.7 per cent (dashed green) contours, based on $\Delta C$ statistics (see \S\ref{sec:stats}), are shown. The best-fitting parameters are consistent with the data at the 90 per cent level  ($p-{\rm value} = 0.251$).} 
\label{rys:cont-wlt-z1} 
\end{figure*}

\begin{figure*} 
\includegraphics[angle=270,width=136mm]{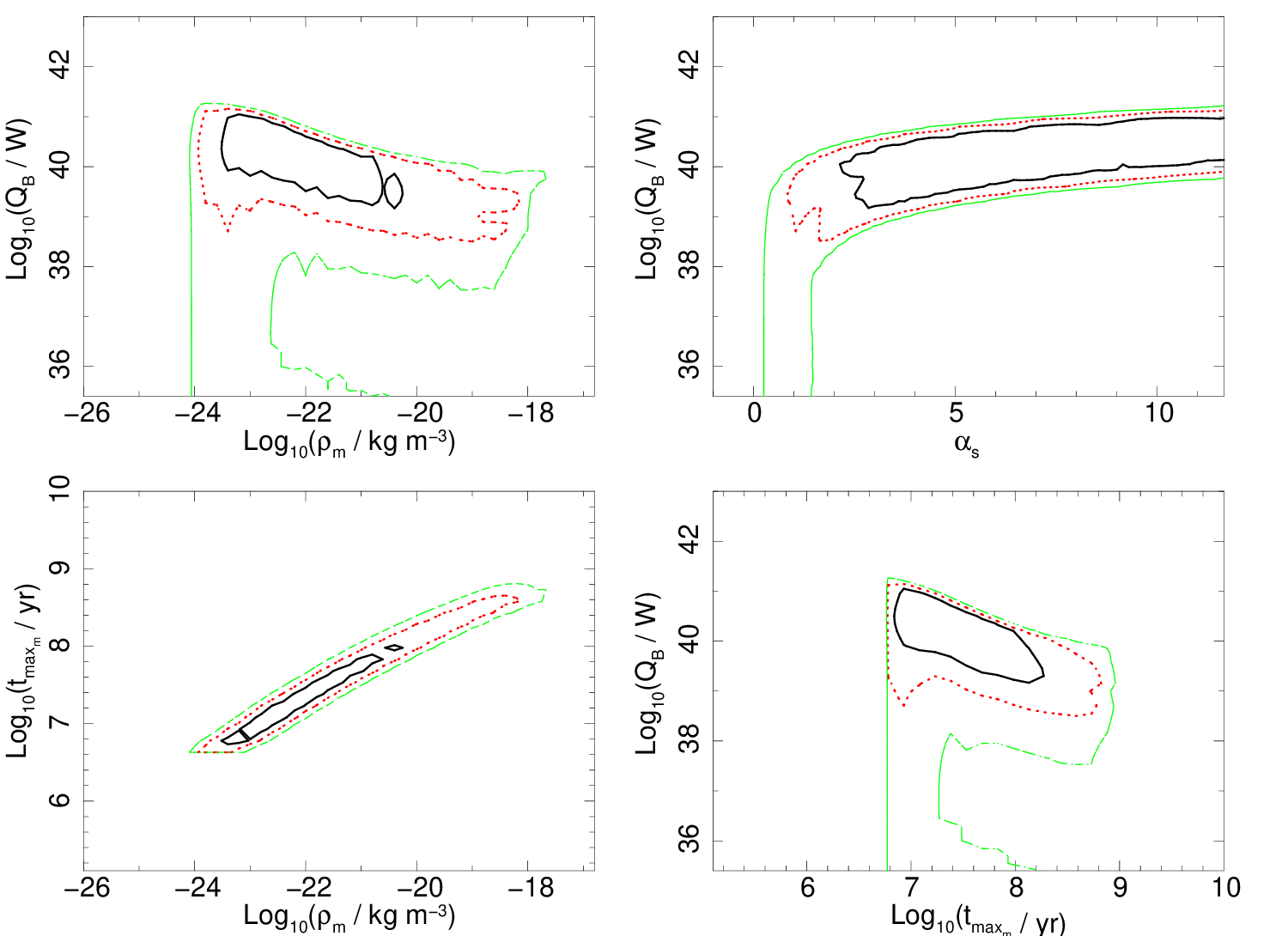} 
\caption{Joint confidence intervals for the independent-$z$ fits of Model W in $z_2$ redshift range are shown ($0.3<z_3\leq0.8$). 68.3 per cent (solid, black), 95.4 per cent (dotted, red) and 99.7 per cent (dashed green) contours, based on $\Delta C$ statistics (see \S\ref{sec:stats}), are shown. The best-fitting parameters are consistent with the data at the 90 per cent level  ($p-{\rm value} = 0.584$).} 
\label{rys:cont-wlt-z2} 
\end{figure*}

\begin{figure*} 
\includegraphics[angle=270,width=136mm]{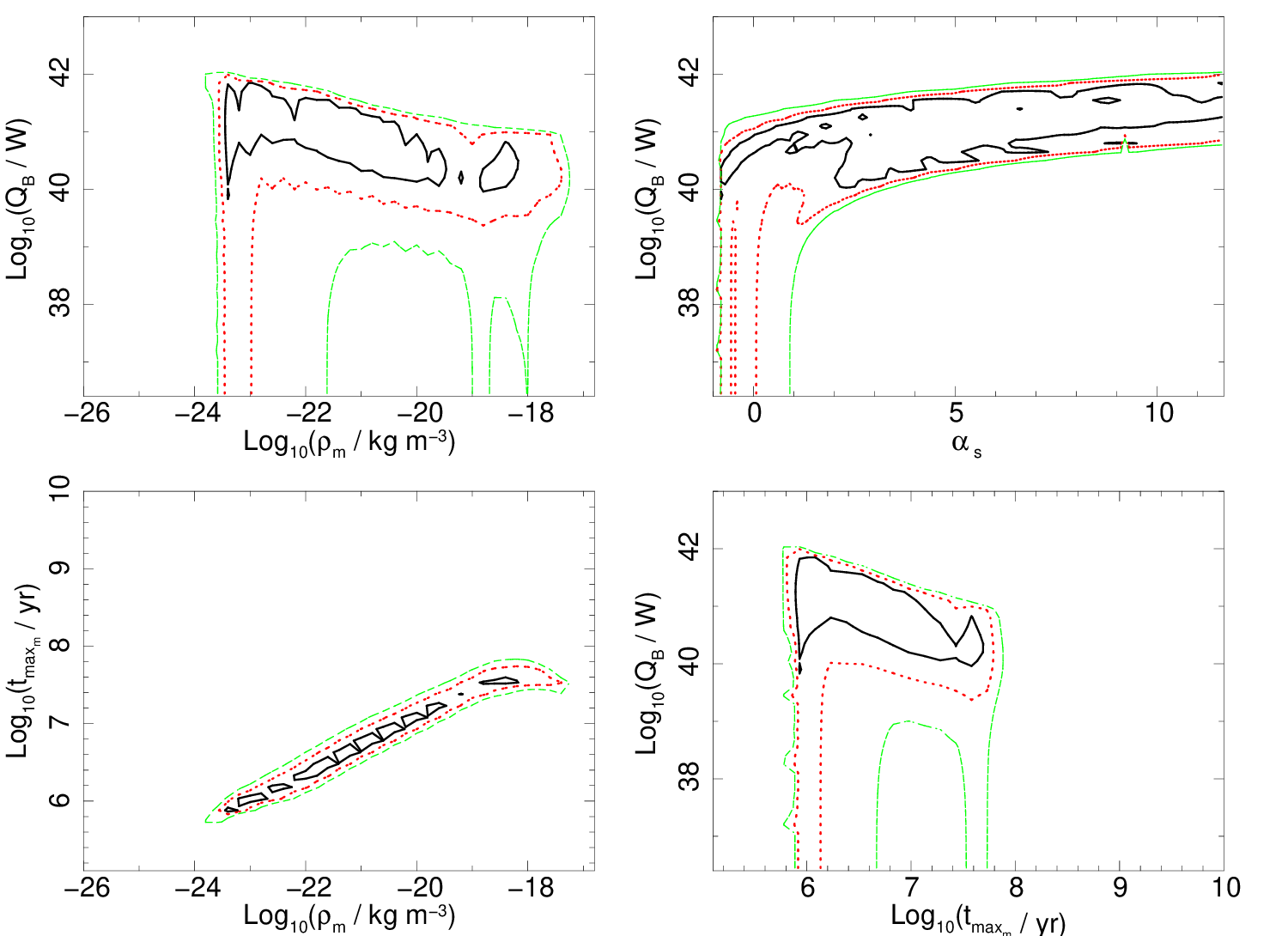} 
\caption{Joint confidence intervals for the independent-$z$ fits of Model W in $z_3$ redshift range are shown ($0.8<z_3\leq2.0$). 68.3 per cent (solid, black), 95.4 per cent (dotted, red) and 99.7 per cent (dashed green) contours, based on $\Delta C$ statistics (see \S\ref{sec:stats}), are shown. The best-fitting parameters are consistent with the data at the 90 per cent level  ($p-{\rm value} = 0.954$).} 
\label{rys:cont-wlt-z3} 
\end{figure*}

\begin{figure*} 
\includegraphics[angle=270,width=136mm]{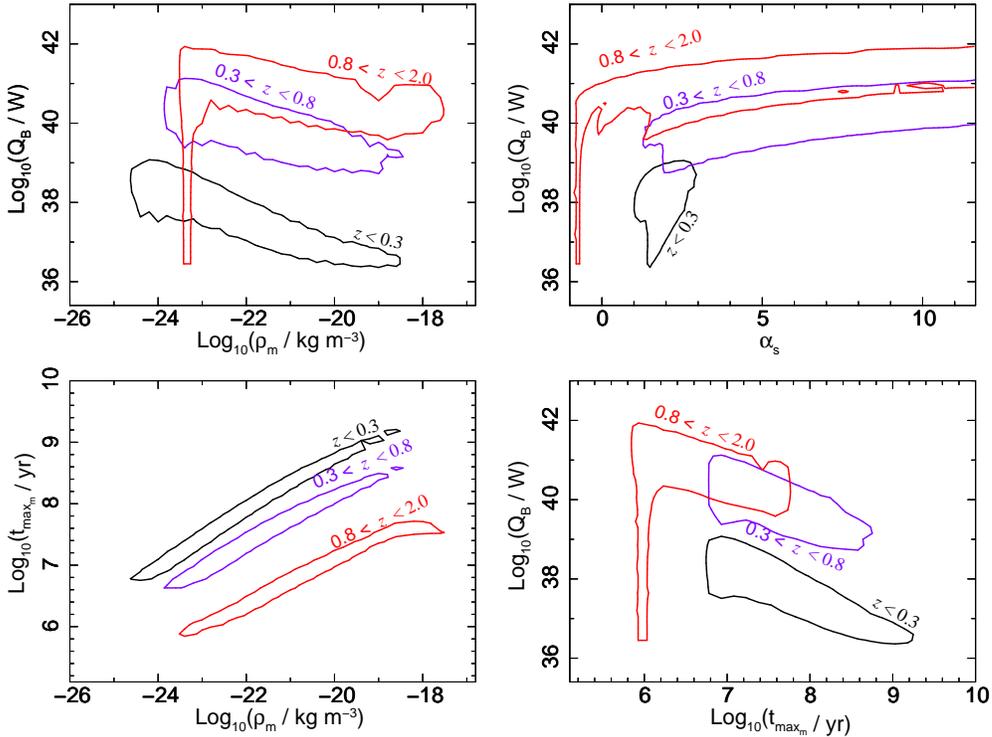} 
\caption{Overlaid 90 per cent confidence intervals  based on $\Delta C$ statistics (\S\ref{sec:stats}) for the three redshift ranges considered ($z_1$ drawn in black, $z_2$ in blue, $z_3$ in red) of the independent-$z$ fits of Model W. } 
\label{rys:cont-wlt-zall} 
\end{figure*}

\begin{figure*} 
\includegraphics[angle=270,width=170mm]{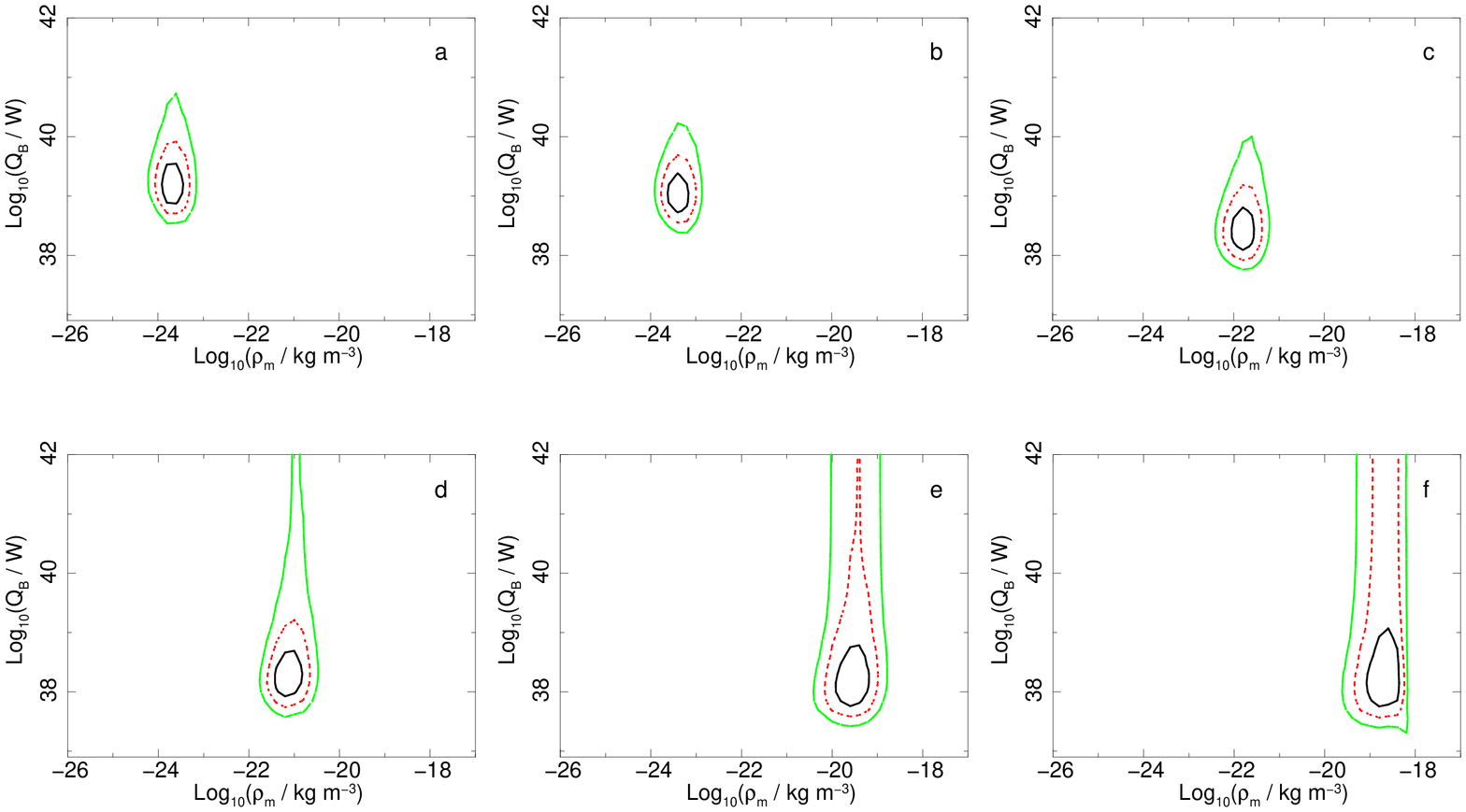} 
\caption{From top right: (a) the joint confidence intervals of the best-fitting $Q_{\rm B}$ and $\rho_{\rm m}$ for population of sources with fixed maximum lifetime of $t_{\rm max_{\rm m}} = 8.5 \times 10^6$ yr is presented, (b) with $t_{\rm max_{\rm m}} = 1.2 \times 10^7$ yr, (c) with $t_{\rm max_{\rm m}} = 6.8 \times 10^7$ yr, (d) with $t_{\rm max_{\rm m}} = 1.3 \times 10^8$ yr, (e) with $t_{\rm max_{\rm m}} = 5.4 \times 10^8$ yr, and (f) with the maximum source age of $t_{\rm max_{\rm m}} = 1.1 \times 10^9$ yr. In each case $t_{\rm max_{\rm m}}$ is drawn from log-normal distribution with $\sigma_{\rm t_{\rm max_{\rm m}}} = 0.05$ (see \S\ref{sec:age}). The independent-$z$ fits of the Model S  in the lowest redshift range ($z_1\leq0.3$) is presented.  For each sub-plot the solid (black) line corresponds to 68.3 per cent of the best-fitting parameters for the particular choice of fixed $t_{\rm max}$, the dotted (red) line corresponds to 95.4 per cent, and the dashed (green) contours contain 99.7 per cent of the best fits based on the $\Delta C$ statistics. It is clearly visible that the assumption on  the maximum allowed age of the sources will strongly determine the density of the environment in which sources evolve (predominantly), as well as their jet powers (less strongly).} 
\label{rys:tmax-correl} 
\end{figure*}

\begin{figure*} 
\includegraphics[angle=180,width=175mm]{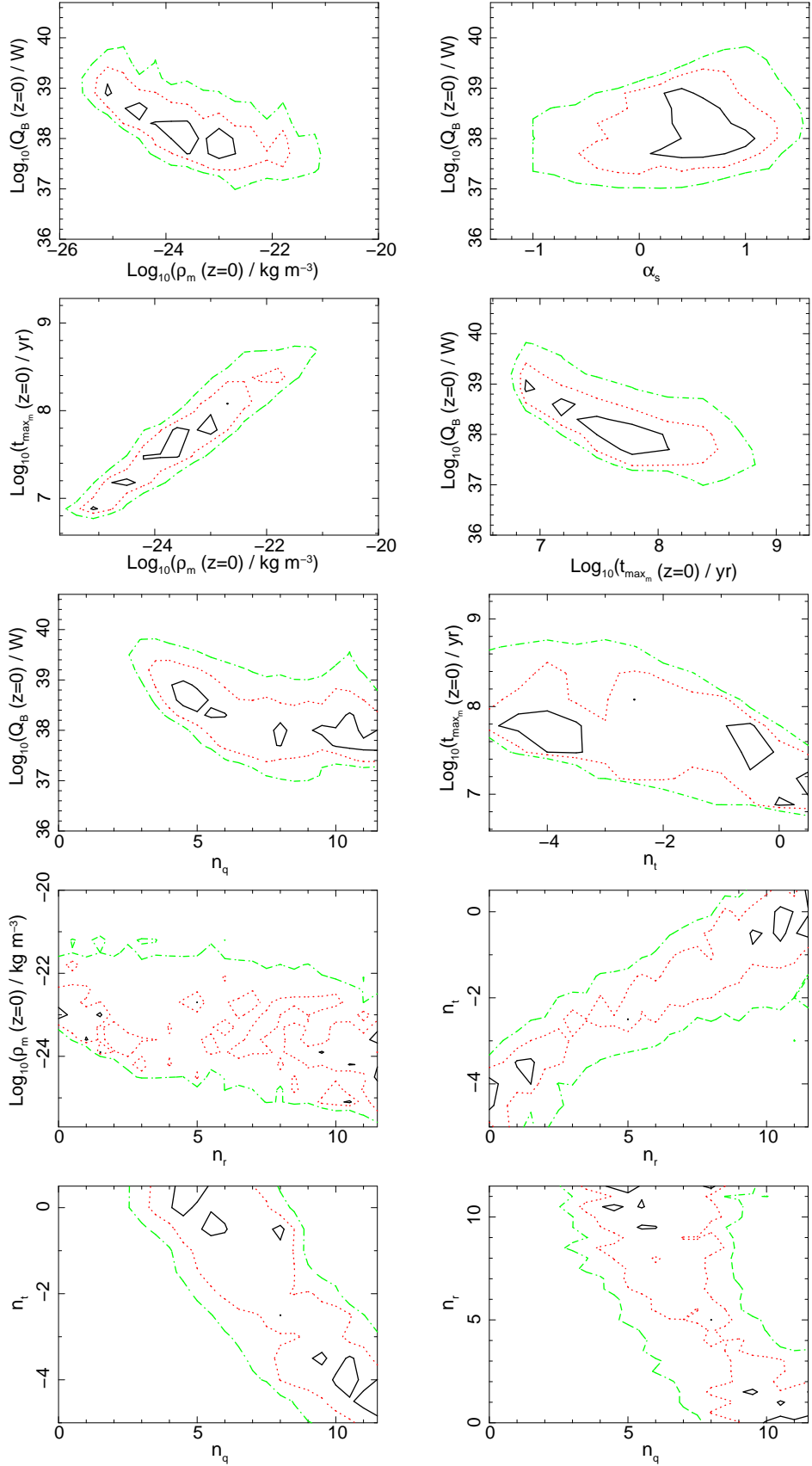} 
\caption{68.3 per cent (solid, black), 95.4 per cent (dotted, red) and 99.7 per cent (dash-dotted, green) joint confidence intervals (based on $\Delta C$ statistics) of all the searched parameters of the combined-$z$ fits of the Model S. Parameters $n_{\rm t}, n_{\rm q}$ and $n_{\rm r}$ quantify the strength of the redshift evolution of $t_{\max_{\rm m}}, Q_{\rm B}$ and $\rho_{\rm m}$ respectively. } 
\label{rys:schech-evol} 
\end{figure*}

\begin{figure*} 
\includegraphics[angle=180,width=133mm]{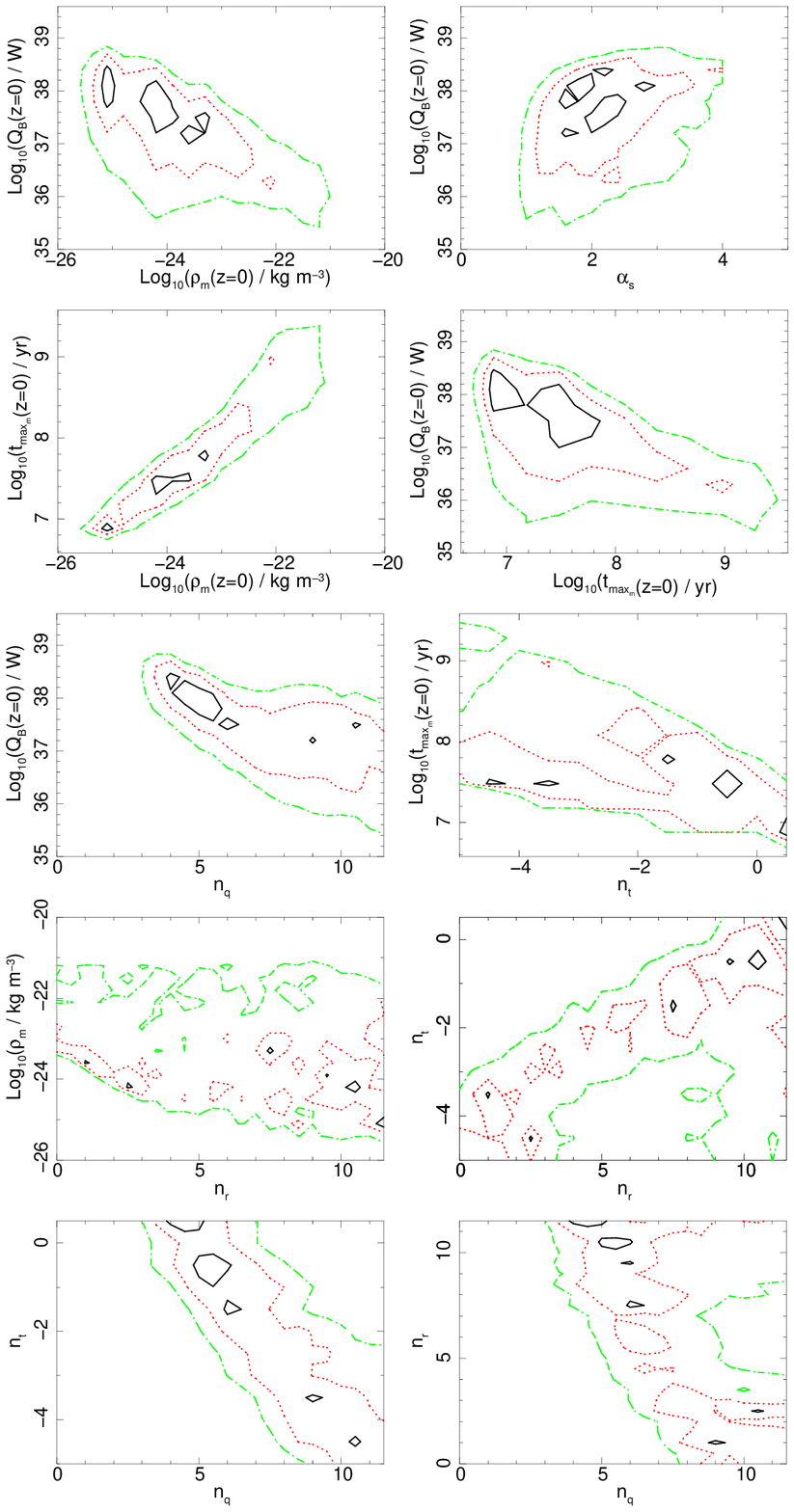} 
\caption{68.3 per cent (solid, black), 95.4 per cent (dotted, red) and 99.7 per cent (dash-dotted, green) joint confidence intervals (based on $\Delta C$ statistics) of all the searched parameters of the combined-$z$ fits of the Model W. Parameters $n_{\rm t}, n_{\rm q}$ and $n_{\rm r}$ quantify the strength of the redshift evolution of $t_{\max_{\rm m}}, Q_{\rm B}$ and $\rho_{\rm m}$ respectively. } 
\label{rys:wlt-evol} 
\end{figure*}

One may also argue that the kinetic luminosity may compensate for the change of the source age instead of the ambient density (Eqn.\ref{eqn:tmax-degener}). Our results suggest, however, that the $\rho_{\rm m} - t_{\rm max_{\rm m}}$ degeneracy is dominant. The effect of this degeneracy is seen in Figure~\ref{rys:cont-sch-z1} -- Figure~\ref{rys:cont-wlt-z3}, and additionally in Figure~\ref{rys:tmax-correl} where results in the $Q_{\rm B} - \rho_{\rm m}$ plane for source populations of different assumed maximum ages are displayed.


\begin{landscape} 
\begin{center}

\begin{table}
\begin{minipage}{20.5cm}
  \caption{The best-fitting parameters for the all tested cases of Model S and W for each redshift range. Due to occurring degeneracies (\S5.3) one  should always consult the corresponding confidence intervals (see Figure~6 -- Figure~13).  The following standard deviations of $\rm log_{10}(\rho_{\rm m})$ and $\rm log_{10}(t_{\rm max_{\rm m}})$ log-normal distributions are used: $\rm log_{10}(\sigma_{\rho_o})$ = 0.15 and $\rm log_{10}(\sigma_{t_{\rm max}})$ = 0.05.  90 per cent uncertainties are quoted. }
  \label{tab:best-fits}
  \begin{tabular}{llccrccccc}

    \hline
    \hline \\
    Model & $z$ & ${\rm log}_{10}(Q_{\rm B}/\rm W)$ & $n_{\rm q}$&$\alpha_{\rm s}$ & ${\rm log}_{10}(\rho_{\rm m} / \rm kg\, m^{-3})$& $n_{\rm r}$& ${\rm log}_{10}(t_{\rm max_{\rm m}}/ \rm yr)$&  $n_{\rm t}$& $p-$value\\ &&&&&&&&&\\
    \hline \\
    Model S              & $z_1$ & $39.15^{+0.30}_{-0.30} $ & -- & $0.6^{+0.3}_{-0.6}$ & $-23.4^{+0.6}_{-0.4}$ &  -- & $7.23^{+0.30}_{-0.15}$& -- & 0.234  \\
   independent-$z$ fits  & $z_2$ &  $39.00^{+0.60}_{-0.45} $ & -- &$-0.9^{+1.3}_{-7.7}$ & $-20.4^{+1.8}_{-2.0}$ &  -- & $7.83^{+0.60}_{-0.75}$& -- & 0.639  \\
                         & $z_3$ & $39.90^{+0.60}_{-0.45} $ & -- &$-1.7^{+1.7}_{-8.3}$ &$-20.0^{+2.0}_{-2.8}$   &  -- & $7.13^{+0.60}_{-1.05}$& -- & 0.925  \\   \\

    Model S              & all &  $38.0^{+(<0.3)\ddagger}_{-0.3}$ & $10.5^{+(<0.5)\ddagger}_{-0.5}$ & $0.6^{+(<0.2)\ddagger}_{-0.6} $ & $-23.0^{+(<0.3)\ddagger}_{-(<0.3)\ddagger} $ &$0.0^{+(<0.5)\ddagger}_{-(<0.5)\ddagger}$ & $7.8^{+(<0.3)\ddagger}_{-(<0.3)\ddagger} $& $-4.0^{+(<0.5)\ddagger}_{-(<0.5)\ddagger} $ &0.380\\
    combined-$z$ fits     & \\  \\

    Model W               & $z_1$ &  $38.25^{+0.45}_{-0.45}$& -- & $1.80^{+0.75}_{-0.60}$            & $-23.4^{+0.8}_{-1.0}$   &  -- & $7.23^{+0.45}_{-0.45}$ & -- &  0.251  \\
     independent-$z$ fits & $z_2$ & $40.60^{+0.30}_{-0.75}$ & -- & $7.30^{+5.60\dagger}_{-5.35}$      & $-22.6^{+0.8}_{-0.8}$   & -- &$7.08^{+0.30}_{-0.30}$  & -- &  0.674  \\
                          & $z_3$ &  $41.35^{+0.45}_{-0.9}$ & -- & $3.50^{+7.40\dagger}_{-3.95\dagger}$& $-23.0^{+2.2}_{-0.4}$   &  -- & $6.08^{+0.75}_{-0.15}$& -- &  0.885  \\ \\

    Model W               & all &   $37.8^{+(<0.3)\ddagger}_{+(<0.3)\ddagger}$ & $10.5^{+(<0.5)\ddagger}_{-0.5}$ & $2.8^{+(<0.2)\ddagger}_{+(<0.2)\dagger} $ & $-23.6^{+(<0.3)\ddagger}_{-(<0.3)\ddagger} $ &$1.0^{+(<0.5)\ddagger}_{-(<0.5)\ddagger}$ & $7.8^{+(<0.3)\ddagger}_{-(<0.3)\ddagger} $& $-5.0^{+(<0.5)\ddagger}_{-(<0.5)\ddagger} $ & 0.557\\
    combined-$z$ fits     &\\
     
    \hline
    \hline

  \end{tabular}

  \begin{flushleft}
  \medskip
      {{\bf Notes.} The resolution of the results is $\Delta\rm{log}_{10}(\rho_{\rm m})=0.2$, $\Delta\rm{log}_{10}(Q_{\rm B})=0.15$, $\Delta\rm{log}_{10}(t_{\rm max})=0.15$, and $\Delta\alpha_{\rm s}=0.1$ for the independent-$z$ fits, and  $\Delta \alpha_{\rm s} = $ $0.2$, $\Delta\rm{log}_{10}(\rho_{\rm m}) = \Delta\rm{log}_{10}(Q_{\rm B}) = \rm{log}_{10}(t_{\rm max}) = 0.3$, and  $\Delta n_{\rm t} = \Delta n_{\rm r} = \Delta n_{\rm q} = 0.5$ for the combined-$z$ fits.}\\

      {$\dagger$ For errors which may be extending beyond the searched ranges (see Table~3) value up to the range border is quoted.}\\

      {$\ddagger$ If errors are smaller than their respective resolution, the value of $<\Delta$ is quoted.}
  \end{flushleft}

\end{minipage}
\end{table}
\end{center} 
\end{landscape}


 
 
\subsection{Influence of the assumptions on the other model parameters} 
\label{sec:assump-disc}


Although, one can easily estimate how strongly a given physical model assumption influences a single source, such predictions may not be trivial when considering the whole population of radio sources. Here, we discuss the possible influence of the model assumptions introduced in \S\ref{sec:distributions} on the simulated source populations and our results. The corresponding confidence intervals for each case are plotted in the online Supplementary Material.

\subsubsection{Core radius and the shape of the central density distribution} 
\label{sec:assump-disc-denstdev} 
 
The central density is tightly linked to the core radius $a_o$, and, we stress again, the parameter ($\rho_o a_o^{\beta}$) should not be considered separately in terms of $\rho_o$ and $a_o$. In our simulation the core radius is kept fixed at a value of $2~\rm kpc$, and the corresponding values of the mean central density are searched through. Any change to the value of $a_o$ will affect the result of $\rho_o$ only, where for lower values of the core radius the mean central density will compensate shifting towards higher values, and for larger $a_o$ less dense environments will be preferred (assuming no change is made to $\beta$).

The initial log-normal distribution of central densities is assumed to be quite narrow, with $\sigma_{\rm log_{10}(\rho_{o})} = 0.15 $. We tested how much the results will be affected if we allow a broader central density spread. As we discussed in \S\ref{sec:rho-tmax-degen}, the change in density will have very strong effect predominantly on the maximum allowed age of the sources. Moreover, since both radio luminosities and linear sizes depend on the density of the medium, allowing for larger standard deviation of the $\rho_{\rm m}$ distribution  will affect the distributions of kinetic luminosities of the sources and their ages as these will have to compensate for the environment change to reproduce the observables (as compared to the initial narrow $\sigma_{\rm log_{10}(\rho_{o})}$). Indeed, the results show that the kinetic luminosity break shifts to slightly lower values as the central density standard deviation gets broader. The mean central density and the maximum age of the source of the best fits seem to oscillate around similar values for the tested cases, where $\sigma_{\rm log_{10}(\rho_{o})} = 0$ (delta function), $\sigma_{\rm log_{10}(\rho_{o})} = 0.15$, $\sigma_{\rm log_{10}(\rho_{o})} = 0.50$ and $\sigma_{\rm log_{10}(\rho_{o})} = 0.75$ [in units of $\rm log_{10}(\rho_{\rm o})$]. However, we see a significant change for $\sigma_{\rm log_{10}(\rho_{o})} = 1.00$, when the confidence intervals of all parameters become more defined, in particular $Q_{\rm B}$ drops to values of $10^{38}$~W and the maximum age of the sources oscillates around $\sim 1\times10^8$~yr in the case of $z_1$, while for $z_3$ there are only very specific regions of the parameter space allowed.  

\subsubsection{Distribution of the maximum source lifetimes} 
\label{sec:assump-disc-tmaxstdev}

The initial log-normal distribution of the maximum source lifetimes is assumed to be very narrow, with {\mbox{$\sigma_{\rm log_{10}(t_{\rm max})} = 0.05$.}} We tested how much the results will be affected if we allow for greater variety in the maximum lifetimes within one population of the sources. We re-run the simulation twice, with  $\sigma_{\rm log_{10}(t_{\rm max})} = 0.3 $, and with $\sigma_{\rm log_{10}(t_{\rm max})} = 0.6 $. Similarly to the case of the standard deviation of the central density distribution, we do not observe any significant changes in the confidence intervals of any of the searched parameters.

\subsubsection{Head advance speeds}
\label{sec:assump-disc-advspeed}

Since there is much discussion on the value of maximum head advance speed of the FR~II jets (see \S\ref{sec:adv-speed}) we re-run the simulation with $v_{\rm max} = 0.05c$, $v_{\rm max} = 0.15c$, and further with $v_{\rm max}=0.8c$, to investigate how strong an effect such an assumption may induce on the whole population.  
We notice that the lower   $v_{\rm max}$ is the more constrained confidence intervals become. Although there is no significant difference between the confidence intervals, stronger constraints, resulting from lower allowed $v_{\rm max}$, indicate that higher central densities, older source ages, and slightly less powerful sources (the latter seems to be the case only in the smallest redshift range) from the initial broad contours are indeed preferred.

Such results come from the fact that, in the non-relativistic case, the energy density in the jet head (hot spot) is $u_{\rm HS} = \rho_{\rm x} \nu_{\rm adv}^2$, where $\rho_{\rm x}$ is the external medium density, $\nu_{\rm adv}$  is the head advance speed, and $\nu_{\rm adv} = \nu_{\rm j} / [1 + (\rho_{\rm x}/\rho_{\rm j})^{1/2}]$ with the jet speed $\nu_{\rm j}$ and  jet density $\rho_{\rm j}$ (see Marti et al. 1997 for a relativistic extension of this calculation which leads to very similar results). This implies that $\nu_{\rm adv} \propto \rho_{\rm x}^{-1/2}$ \cite[see][]{s18}. Thus, the smaller the head advance speed is, the higher the ambient densities of the sources must become to reproduce their observed linear sizes and radio luminosities. If the external density changes then adjustment in the $t_{\rm max}$ and $Q_{\rm B}$ must develop to again ensure that the observed $L_{\nu}$ and $D$ are reproduced. This is indeed what is observed.

Note, however, that since from the KA model we know that the expansion speed decreases with the source age, the maximum head advance speed that we refer to here is in fact the maximum expansion speed a source may have at the time of observation. Since there is a higher probability of observing an old source than a young one, these $v_{\rm max}$ may not be the highest possible expansion speeds during a source lifetime as at the earlier stages of the source life its early-$\nu_{\rm adv}$ (non observed) may be higher than the decelerated late-$\nu_{\rm adv}$ (observed, at the later stages of the source life). The latter, only, is compared to $v_{\rm max}$.

\subsubsection{Energy distribution of the relativistic electrons: the injection index and the Lorentz factors} 
\label{sec:assump-disc-m-lorentz}

Initially, we assumed a uniform distribution of $m\in[2,3]$. We tested, however, the possible influence of the assumed $m$ distribution on our results, and we re-run the simulation twice assuming that $m$ is set to a single, universal value ($m=2.3$), and further that $m$ follows a Gaussian distribution with $m_{\rm mean}=2.4$ and $\sigma_{m} = 0.3$. Note that this normal distribution is asymmetric here since it is not allowed to extend beyond the minimum and maximum $m$ values. \cite{c10} finds $m_{\rm mean}=2.4$ and $\sigma_{m} = 0.6$; however, since $m$ is restricted by its minimum and maximum allowed value in our study, such a distribution would be nearly uniform for $m_{\rm min}=2$ and $m_{\rm max}=3$. For this reason we decided to assume a smaller standard deviation of the distribution in our tests. We find no difference between results found with either of the Gaussian, the uniform distributions or a single value of the injection index. Hence, we conclude that we are not sensitive enough to distinguish between the underlying distributions of $m$ at this point.

Furthermore, since there has been discussion on the low-energy cut-off of the energy distribution of relativistic electrons present in the radio jet we tested the following cases to investigate the possible changes to the results: (i) $\gamma_{\rm min} = 1$ and $\gamma_{\rm max} = 10^5$, (ii) $\gamma_{\rm min} = 10^2$ and $\gamma_{\rm max} = 10^{10}$, and (iii) $\gamma_{\rm min} = 10^4$ and $\gamma_{\rm max} = 10^{10}$. All of these were compared to our initially assumed case where $\gamma_{\rm min} = 1$ and $\gamma_{\rm max} = 10^{10}$. As expected, there is no difference between the initial assumptions and case (i), indicating that indeed the exact value of the maximum Lorentz factor is not crucial as long as $\gamma_{\rm min} \ll \gamma_{\rm max}$. In case (iii) we observe a significant change of the kinetic luminosity break confidence intervals, where for $\gamma_{\rm min}=10^4$  the kinetic luminosity break is smaller by  $\sim1.5$ decades,  as compared to the other cases. Such a drastic change in confidence intervals in case (ii) is not observed, despite $\gamma_{\min}$ being significantly larger than our initial case. Here, we conclude that change in the minimum Lorentz factor for the whole source population will significantly affect their kinetic luminosities only.

$\gamma_{\rm min}$, $\gamma_{\rm max}$ and $m$ directly influence the initial energy density distribution of the relativistic particles. The minimum Lorentz factor indicates how relativistic are the least energetic particles. The lower $\gamma_{\rm min}$ the more cold material is included. For cold plasma higher $Q$ is required in order to obtain the observed radio luminosity because a fraction of the kinetic luminosity will be lost for particle acceleration. However, if this material is not included, less power is required to reproduce the observed $L$, and hence significantly lower $Q_{\rm B}$ are observed in our results if we allow $\gamma_{\rm min}$ to be as high as $10^4$.  
The maximum Lorentz factor characterizes the cut off at the high energy end of the particle distribution; particles can only be accelerated up to a certain frequency above which loss rates take over \cite[see also][]{m19}. 
The index $m$ defines the slope  of the energy density distribution, and hence determines the relative fraction of low and high energetic particles, where the flatter the distribution is the more energetic particles are included. Although the change of $m$ may be significant for a single radio source, it does not seem to have a dramatic effect on the whole population of sources, and it is the low energy cut off that influences the average properties of the population.

\subsubsection{Jet particle content} 
\label{sec:assump-disc-k}

As discussed in \S\ref{sec:particle-content} there is much debate on the particle content of the radio  jets.  
Initially, in our simulation we followed the conservative KDA assumption ($k'=0$), but to test the influence of this assumption on the whole population of sources we assumed the particle content of radio jet to be drawn from a uniform distribution in a range $k'\in[0,10]$, and further also $k'\in[0,100]$. As an additional test we also investigated the case when $k'=100$. A significant change in the fitted source parameters were observed only in the two latter cases. While assuming $k'\in[0,100]$ (higher redshifts) or $k'=100$ the best-fitting kinetic luminosity breaks shift significantly to higher values by  $\sim 1.5$ decades. The other parameters do not seem to undergo any significant change.

Indeed, the addition of protons in the relativistic jets would require these particles, similarly to electrons, to be accelerated to the relativistic speeds. As already shown by e.g. \citet{b15}, relativistic protons store much more energy than electrons typically do, and this in turn will give rise to the radio source kinetic luminosity. The argument may be turned around: to maintain the expected kinetic luminosity, the observed radio lobe luminosity is much lower for proton dominated jets than for lightweight, electron-positron jets (see also KDA). However, many authors favour proton-electron jets. \cite{i1}, for instance, draw a hypothesis that some low-luminosity AGN have proton dominated jets, which are supported by extracting energy from black hole spin. \cite{s16} report that jets of radio-loud quasars are also most likely heavy due to the positron-electron kinetic energy being too small to support energetics of radio lobes. However, \cite{b16}, extending the work of \cite{c9}, show that in the case when jets are proton dominated the following criterion occurs: $\gamma_{\rm min}\cong10^2$. This is not required by positron-electron jets (see \S\ref{sec:assump-disc-m-lorentz}). Indeed, if one considers the net effect of the two assumptions tested, that is the jet particle content and the minimum Lorentz factor, one will notice that their effects may cancel out. Setting $\gamma_{\rm min}\cong 10^2$ or more one ensures that only initially highly relativistic particles are included in the jet material and these are later being re-accelerated in the jet front; it requires stronger momentum flux to affect the nonrelativitic protons than relativistic ones, hence the change in kinetic luminosity in these two cases.

We are unable to distinguish between purely electron-positron jets ($k'=0$) and those that contain modest numbers of protons ($k'\in[0,10]$). Very heavy FR~II jets ($k'\sim100$) consisting of an electron-proton plasma seem to be unlikely, unless the minimum Lorentz factor is set to large values ($\gamma_{\rm min}\cong 10^2 - 10^4$).

\section{Discussion} 
\label{sec:discussion}

\subsection{The lifetimes of radio galaxies} 
\label{sec:res-lifetimes} 
 
Our estimated maximum lifetimes of radio galaxies may seem to be rather low, especially given that sources of the observed ages of order of few $\times10^8$~yr are known (e.g. $\sim1.8\times10^8$yr of B0319-454 reported by \citealt{s8}, $\sim1.4\times10^8$yr in the case of B0313-683 reported by \citealt{s9}). However, our results are in  agreement with other estimates of observed ages and total lifetimes of, specifically, 3CRR radio galaxies evaluated by either numerical simulations or multi-frequency radio observations \citep[e.g.][]{a1, m7, w1, o2}. Our results are also consistent with work of \citet{b10} based on an independent low redshift sample, who determined the maximum lifetime of the galaxies to be $1.5\times 10^7$~yr. Further, the ages estimated by Machalski et al. (2004a) reach $1-2 \times 10^8$~yr, but only for giant radio sources, and for most normal radio galaxies they analysed, i.e. the ones with linear sizes smaller than 1~Mpc, ages few$ - 10$ times smaller were estimated. As a sanity check we find that for our hardwired maximum head advance speed of $0.4c$ a source of 1~Mpc in size may live for the maximum of $4\times10^6$~yr, while to increase the total lifetime to $1.6\times10^8$~yr, the source head advance speed would have to be of order of $0.01c$. Such a low head advance speed may be treated as a lower limit because the jet must be supersonic during the sources lifetime to maintain the FR~II morphology, and also classical double sources older than the inferred age are not observed. The model of \citet{f1} predicts that the source advance speed changes, i.e. decreases, with the age of the source. This would suggest that sources that initially have high maximum head advance speed, slow down as they get older and hence they reach older ages and smaller $v_{\rm max}$ are measured \citep[e.g.][]{m7}.

The radio samples we are currently using contain the most powerful FR~II radio galaxies and quasars, and the bias originating from the flux limit is particularly strong at high redshifts. This causes us to observe only the youngest and most powerful sources. \citet{b4} suggested an explanation of the noticeable decrease with redshift in source lifetimes. The term `Youth-Redshift degeneracy' has been coined to describe the effect. Due to the luminosity-redshift degeneracy originating  from the radio flux limits, at high redshifts one can observe only the most powerful objects. They seem to be also of smaller linear sizes as compared to the radio galaxies in the local Universe \cite[][]{k4,b11}. 
The leading theoretical models of radio galaxy time evolution (e.g. KDA, BRW) predict a short initial phase of the radio source growth where the radio lobe luminosity peaks (up to a $\sim$ few kpc), after which the radio lobe luminosity decreases as radio source ages and grows bigger. Therefore, to be able to produce linear sizes and radio lobe luminosities as observed at high redshifts, the sources must be younger than their low redshift counterparts.
Based on the 3CRR and BRL samples our results indeed suggest a trend of the total radio source lifetimes decreasing  with redshift. The best-fitting parameters found for the combined-$z$ fits of the Models S and W indicate $n_{\rm t} \in [-3.5 , -4.5]$, slightly stronger as compared to the  results obtained by \citet{w3} who found $n_{\rm t} \cong -2.5$ and \citet{w1} who report $n_{\rm t} = -2.4$. However again, due to the degeneracy between $t_{\rm max_{\rm m}}$ and $\rho_{\rm m}$ the strength of the possible evolution is not strongly constrained as it may be easily compensated by the redshift evolution of the central densities of the galaxy clusters or the kinetic luminosity break. Interestingly, a constant maximum lifetime for all redshifts is allowed, i.e. $n_{\rm t} \sim0$, but this would require the cosmological evolution of $\rho_{\rm m}$ to be unfeasibly strong (see \S\ref{sec:res-density}). Unfortunately, at the current stage we are unable to verify whether the decrease in the maximum lifetimes of these radio sources is genuine, or whether this is just an observational bias due to the flux limit. Deeper high redshift radio samples are needed to resolve this issue.


\begin{figure*} 
  \includegraphics[angle=270,width=176mm]{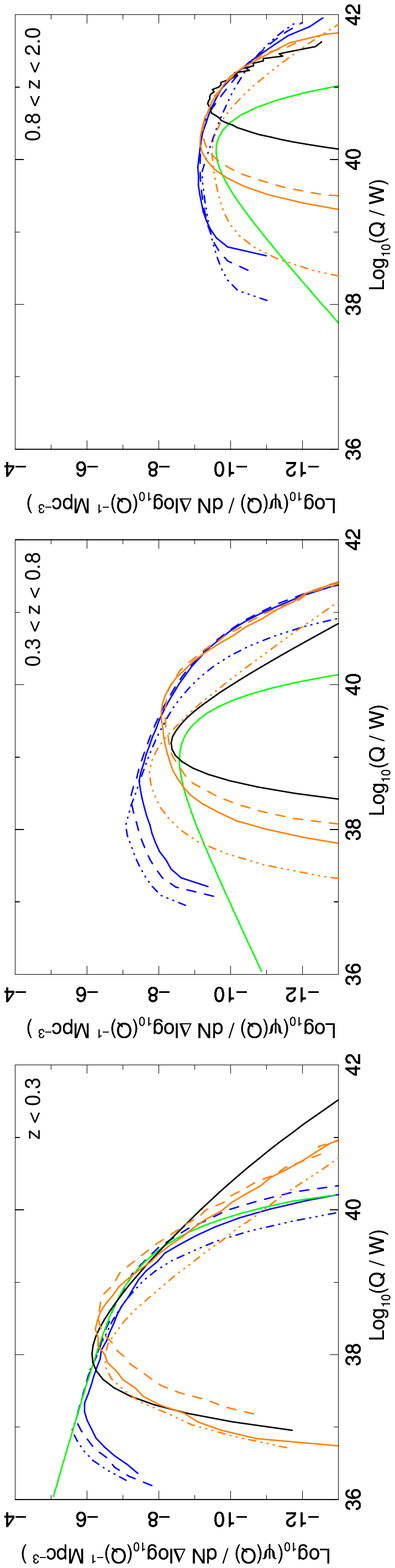} 
  \caption{Kinetic luminosity functions generated with the best-fitting parameter sets of the independent-$z$ fits of Model S (solid line, green) and Model W (solid line, black) for all redshift ranges, and combined-$z$ fits of Model S (blue) and Model W (orange). The combined-$z$ fits of Model S and Model W show their best fits (solid blue, solid orange respectively), and two restricted cases where $n_{\rm r}$ parameter is set to $n_{\rm r} = 2.5$ (dashed) and $n_{\rm r} = 4.0$ (dash-dot-dot-dotted).} 
\label{rys:klf} 
\end{figure*} 
 

\subsection{Ambient gas densities} 
\label{sec:res-density}

The central gas densities estimated here are consistent with recent X-ray studies \citep[e.g.][]{b12,c6,s4}. We observe that clearly higher densities are favoured for earlier epochs. Redshift evolution of the source environments is expected, at least, due to the expansion of the Universe. Invoking textbook physics it is known that  
\begin{equation} 
H^2(z) = H^2_0 (\Omega_{\rm M}(1+z)^3 +\Omega_{\rm k}(1+z)^2 + \Omega_{\rm \Lambda} ), 
\label{eqn:hubble} 
\end{equation} 
where $H$ is the Hubble parameter, $H_0$ is the Hubble constant, and the parameters $\Omega_{\rm M}$, $\Omega_{\rm \Lambda}$ and $\Omega_{\rm k}$ define the matter density, the vacuum density and the spatial curvature respectively. From the above equation one can deduce that the critical density evolves ($\rho_{\rm c}$) with redshift as 
\begin{equation} 
\rho_{\rm c}(z) = \frac{3H^2_0}{8\pi G} \times [\Omega_{\rm M}(1+z)^3 + \Omega_{\rm \Lambda} ] 
\end{equation} 
where $G$ is the gravitational constant and assuming a flat Universe ($\Omega_{\rm k}=0$). Since the radio galaxies are often found in clusters and galaxy groups, one may expect that their immediate environments will undergo similar evolution to the one predicted by models of cluster formation \cite[see e.g.][for a review]{a5}. Structure formation models predict that the mean dark matter density (the predominant constituent of galaxy clusters) scales as the critical density of the Universe, that is again $\rho_o(z) \propto \rho_{\rm c}(z) \propto H^2(z)$. The critical density will, therefore, change between redshifts $z_{\rm n2}$ and $z_{\rm n1}$ by a factor of $\rho_{\rm c}(z_{\rm n2}) / \rho_{\rm c}(z_{\rm n1})$. If $z_{\rm n1}=0$ and $z_{\rm n2}=2$ then the density would be expected to be higher at $z_{\rm n2}$ by a factor of $z_{\rm n2} /z_{\rm n1}\cong 8.8$, equivalent to $\sim0.94$ in logarithmic scale, and to the redshift dependence as approximately $(1+z)^2$.  Similar trends have been recently found by \cite{p2}, and as they point out it can be also deduced from previous observational studies \cite[e.g.][]{d5,d6}. On the other hand, \cite{o2} do not observe any density evolution up to $z\cong0.5$. 
Our best-fits for the combined-$z$ fits of the tested models (Table~\ref{tab:best-fits}) clearly show that $n_{\rm r}$ is degenerate and strongly dependent on  $n_{\rm t}$ and $n_{\rm q}$; the value expected by the galaxy cluster cosmological evolution models and the effect of the expanding Universe is possible for certain values of $n_{\rm t}$ and $n_{\rm q}$, but is not unique. Some may argue that environments of FR~II sources may depend on the cosmological epoch. That is, beside the argument of the evolution following the expansion of the Universe, at low redshifts weak clusters and galaxy groups are favoured by FR~II sources, while rich galaxy clusters are preferred at high redshifts \citep[][]{d2,z1,w5}. If this is the case the effect will be convolved in the $n_{\rm r}$ parameter resulting in evolution stronger than the expected $n_{\rm r} \sim 2.0 - 2.5$. The possible values of $n_{\rm r}$ and their consequences are discussed further in \S\ref{sec:klf}.

As has already been shown in \S\ref{sec:assump-disc} the spread of the log-normal distribution of the density has  a rather small effect on the results. This is highlighted by the fact that when a delta function is employed the final results are still in good agreement with the observed samples. We conclude that at the current stage we are unable to constrain the shape of the population ambient density distribution. Further, the mean central density value will depend on the assumed $a_o$ and $\beta$, and hence if our assumptions for these values are incorrect one may simply scale the fitted densities with the $(\rho_o a_o^{\beta})$ relation. We currently assume that there is no redshift evolution of the core radius or the density profile slope. If there is any intrinsic change this will be subsumed by the mean density value, possibly resulting in a stronger trend with redshift.

\subsection{Kinetic luminosities of powerful FR~II sources} 
\label{sec:klf}

The kinetic luminosity functions (KLFs) of the radio source populations generated with the  best-fitting parameters for all redshift ranges and models are presented in Figure~\ref{rys:klf}, while in Figure~\ref{rys:klf-frac-size} the kinetic luminosity distributions and their corresponding underlying parent populations from which the observed samples originate are plotted for each redshift range, size bin and model. There are two main conclusions. Firstly, the higher the redshift the more powerful the jets are that are included; at $z < 0.3$ the highest $Q$ are of order $10^{40}$~W (Model S), while in the highest redshift range, $z_3$, the most powerful sources seem to have jet luminosities reaching $10^{42}$~W (Figure~\ref{rys:klf}). Secondly, the larger the radio sources are the more powerful AGN on average they host (Figure~\ref{rys:klf-frac-size}). The second effect seems to be much milder than the dependence on redshift, but it is present; a similar pattern has been already observed by \citet[][]{g3}.

\subsubsection{Differences between results of Model S and W} 
 
As can be seen from  Figure~\ref{rys:klf} and  Figure~\ref{rys:klf-frac-size}, although the parent populations of Model S and W are different, the predicted flux-limited samples of the two models are very similar. More interestingly, one would expect kinetic luminosities of sources of FR~II type to be at least $10^{37}$~W. Indeed, although we simulated sources of kinetic luminosities of $10^{36}$~W, these do not contribute to the observed populations. Some consideration is also needed of the lowest redshift bin best fits for Model W, where the kinetic luminosity break reaches somewhat low values and the slope is flatter compared to the other redshift ranges. The kinetic luminosity distribution function of  Model W has been used by \citet{w2} as their initial kinetic luminosity distribution function but for the most powerful sources only. In their work the population of radio galaxies and radio-loud quasars was divided based on the emission line characteristics rather, that is high- and low-excitation sources (HEG and LEG respectively), than the historical FR~I and FR~II classes. HEG include powerful FR~II and the most powerful FR~I sources, while LEG include most FR~Is and low-luminosity  FR~II objects. As mentioned before Model W may not be able to reproduce the faint end of the distribution. Contrary to \citet{w2} we do include low luminosity FR~IIs and perhaps this is the cause of the observed shift of $Q_{\rm B}$ and $\alpha_{\rm s}$.

\subsubsection{Bending power-law as the initial distribution of kinetic luminosities}

The thin, dotted lines in Figure~\ref{rys:klf-frac-size} represent the possible parent populations generated according to the assumed distribution functions. The flux limit of the radio catalogues used is very high, and hence only the most powerful sources are observed; as estimated by our simulations their kinetic luminosities do not drop below $10^{37} - 10^{38}$~W. In such a strong flux limit regime either of the kinetic luminosity distribution functions (Models S and W) will shape the faint power end of the luminosity functions on the basis of a good fit to the high power end. However, with the lack of observed sources with $Q<10^{38}$~W the faint end of the function is highly unconstrained especially at higher redshifts where, given currently used radio samples, only the most powerful sources are considered. To investigate this issue we attempted to fit pre- and post- kinetic luminosity break slopes separately. To do this, we have tested the so-called curved power-law distribution  
\begin{equation} 
\psi(Q) ~{\rm d}Q = \psi^* \left [\left (\frac{Q}{Q_{\rm B}} \right )^{-\alpha_{\rm s1}} + \left (\frac{Q}{Q_{\rm B}} \right )^{-\alpha_{\rm s2}} \right ] ~{\rm d}Q. 
\label{eqn:schech-bend} 
\end{equation} 
where $\alpha_{\rm s1}$ describes the pre-break slope of the function, and $\alpha_{\rm s2}$ the post-break slope. To avoid introducing more free parameters we decided to set the high power end of the kinetic luminosity distribution function to $\alpha_{\rm s2}= 2.2$ \citep[after the radiative luminosity functions of][]{h2}, and allowed the pre-break slope to vary. The results, while assuming the curved power-law as the initial distribution function of source kinetic luminosities, are in good agreement with those found with the Model S. The faint end of the kinetic luminosity function is well constrained in the lowest redshift range, but since now it is not dependent on the high power end of the function it becomes highly unconstrained at higher redshifts. It is necessary, therefore,  to include lower luminosity radio samples in our analysis to investigate the faint end of the kinetic luminosity functions.

\subsubsection{Cosmic downsizing}

The concept of cosmic downsizing, as an anti-correlation between the stellar mass of galaxies and the epoch of their formation, was introduced by \citet{c5}. Since then, the term has been broadened to include the AGN evolution that has manifested similar anitihierarchical trends. The suggestion that the low-luminosity AGN X-ray luminosity functions seem to peak at lower redshifts than those of the luminous quasars \citep{u1,h1}, has been interpreted as implying black hole anitihierarchical growth, and hence seems to be analogous to the star-forming galaxy downsizing \citep[e.g.][]{m5}. Furthermore, since the growth of the black hole and its kinetic luminosity may be linked, \citet{m5} and \citet{c4} suggested that the kinetic energy density of these active galaxies undergoes a similar cosmological evolution such that the most powerful jets are found at earlier epochs. \citet{m5} based their findings on the analysis of radio core kinematic luminosity functions, while \citet{c4} analysed imprints of the extended radio emission on the X-ray emitting plasma. It should not be surprising, therefore, if such an effect is seen in our results.  The results of the independent-$z$ fits already suggest a redshift evolution of the kinetic luminosity break. Moreover, the results of the combined-$z$ fits of the fitted models confirm that this trend with redshift may be rather strong. Of course, the strength of this cosmological evolution depends on the assumed evolution of the sources' maximum lifetime and/or environment in which sources grow due to previously discussed degeneracies.  
One of the most plausible cases, where the galaxy cluster central environments evolve with the expanding Universe and according to the galaxy cluster evolutionary models, would require $n_{\rm r} \cong 2.0 - 2.5$, and this will lead to $n_{\rm q} = 9.5$ and $n_{\rm t} = - 3.0$. Given that $Q_{\rm B}(z=0) \cong 2\times 10^{38}$~W here, such a strong redshift evolution would imply $Q_{\rm B}(z=2.0) \cong 6.8\times 10^{42}$~W. This value is rather high, the highest kinetic luminosities of FR~II radio sources estimated in other works are of order of $10^{41}$~W.

To give an estimation on the black hole mass which would produce such a kinetic luminosity one needs to know the black hole radiative, $\epsilon(\hat{a})$, and jet, $\eta(\hat{a})$, efficiencies. Specifically, after \cite{2011MNRAS.414.1937M}, it is assumed here that the bolometric luminosity of an AGN accreting at the rate $\dot{m}$ is
\begin{equation}
L_{\rm bol} = \epsilon(\hat{a}) \dot{m} c^2,
\end{equation}
where $c$ is the speed of light. Furthermore, the AGN's kinetic luminosity depends also on the accretion rate $\dot{m}$ and is given by 
\begin{equation}
Q = \eta(\hat{a}) \dot{m} c^2.
\end{equation}
Since the Eddington ratio is defined as $\lambda_{\rm Edd} = L_{\rm bol} / L_{\rm Edd}$ one can easily derive the expression for black hole mass, which is
\begin{equation}
M_{\rm BH} = 0.8 \times 10^{-31} \frac{\epsilon(\hat{a})}{\eta(\hat{a})\lambda_{\rm Edd}} Q
\end{equation}
in the units of $M_{\odot}$. 
For non-spinning black holes, that is assuming that these radio sources power their jets via accretion at their Eddington limits ($\lambda_{\rm Edd} \approx 1.0$), and so the spin $\hat{a} =0$, the most powerful objects inferred here at $z\sim2.0$ ($Q_{\rm B} = 6.8\times 10^{42}$~W) would host black holes of masses of $\sim 2.7\times10^{14}$~M$_{\odot}$. For this case, the jet efficiency is found to be within $\eta(\hat{a})\in (10^{-5}, 10^{-3})$ depending on the theoretical model assumptions, and the radiative efficiency is $\epsilon(\hat{a}) \approx 0.05$ \cite[see references compiled by][]{2011MNRAS.414.1937M}.
Similar results are obtained for black holes of low spins (e.g. for $\hat{a}=0.4$ the efficiencies are  $\eta(\hat{a}) \approx 0.001$  and $\epsilon(\hat{a}) \approx 0.05$ which imply $\sim 2.7\times10^{13}$~M$_{\odot}$). However, assuming that the black holes are of high spins (analogically setting $\eta(\hat{a}) \approx 0.5$ and $\epsilon(\hat{a}) \approx 0.1$ which imply black hole spin of $\hat{a}\gtrsim 0.95$) the black holes' masses would need to be  $\sim 1.0\times10^{11}$~M$_{\odot}$ in order to produce quoted $Q_{\rm B}$. The most massive AGNs' black holes are found to not exceed $\sim 10^{10}$~M$_{\odot}$ \citep[e.g.][]{m16,n2,s12,s13}. Yet, equivalently to the independent-$z$ fits, at these high redshifts the $Q_{\rm B}$ values are of such small number density that the highest kinetic luminosities that contribute to the population do not reach $10^{42}$~W (Figure~\ref{rys:klf}). In such a case, assuming maximum kinetic luminosity of $Q=1.0 \times 10^{42}$~W the non-spinnig black hole as denoted earlier would need to have a mass of $\sim3.8\times10^{13}$~M$_{\odot}$, and the high spin black hole would need to have a mass of $\sim1.5\times10^{10}$~M$_{\odot}$.

\begin{figure*} 
\includegraphics[angle=270,width=176mm]{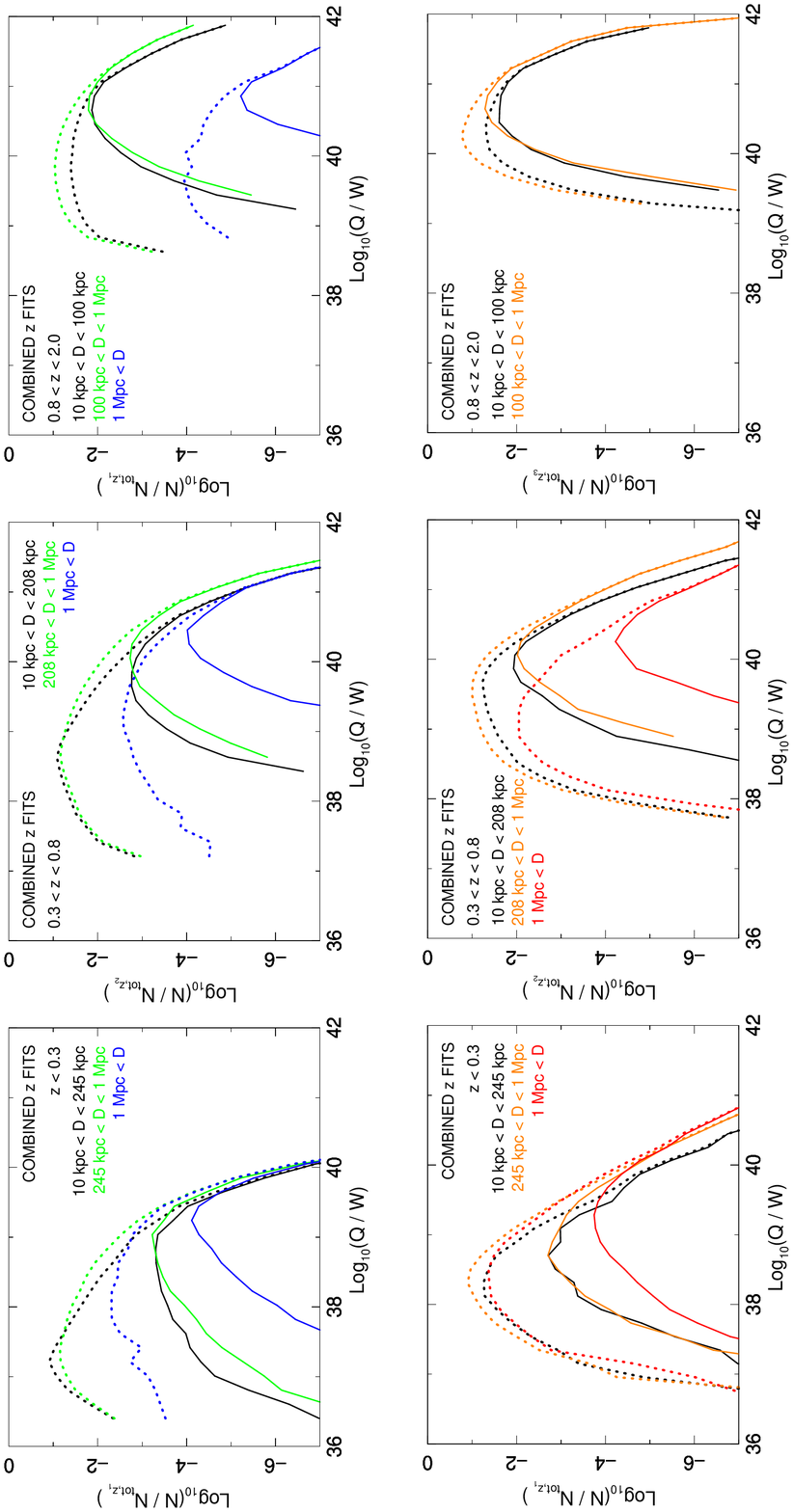}
\includegraphics[angle=270,width=176mm]{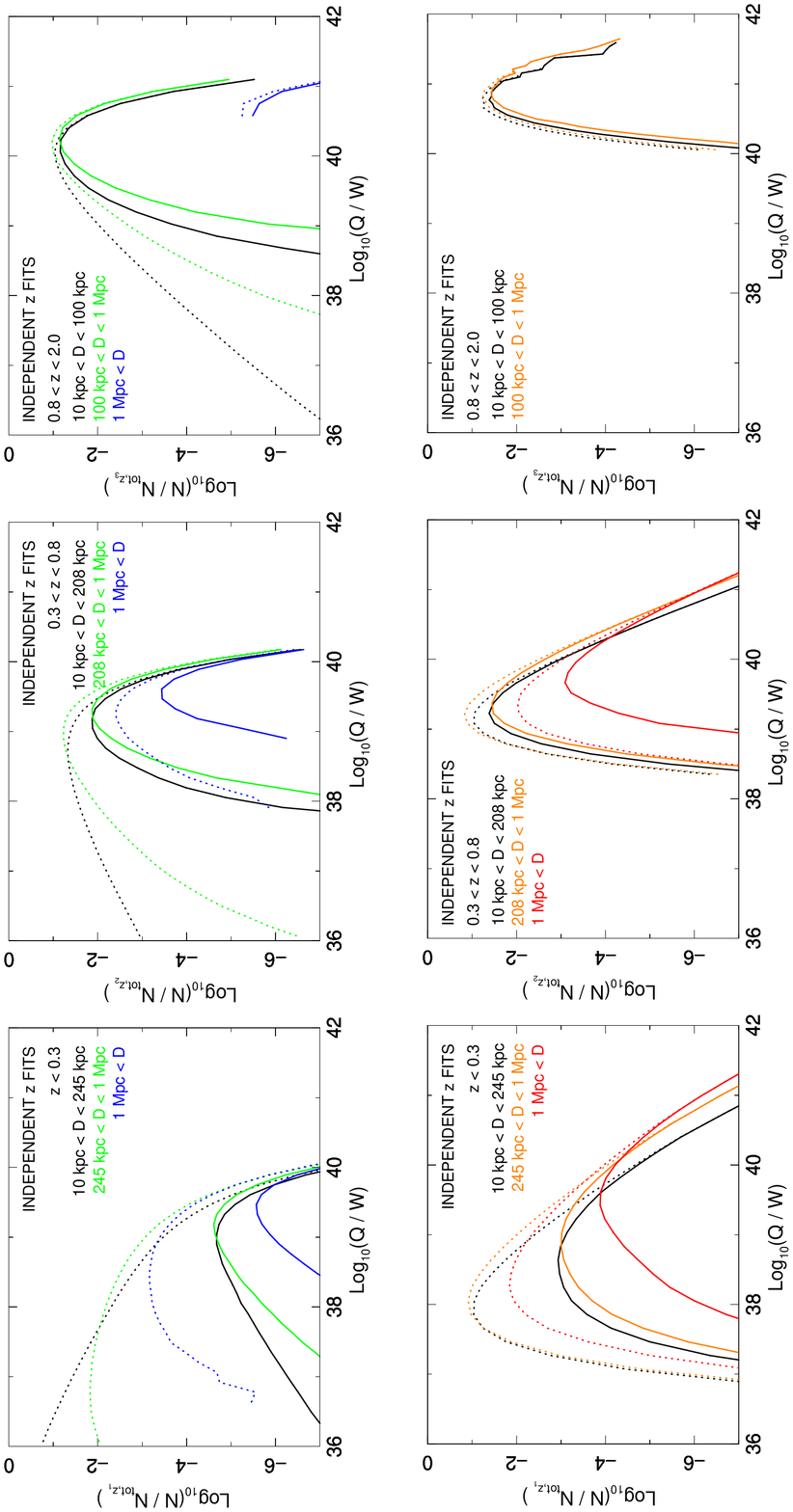} 
\caption{Functions showing kinetic luminosity distributions (normalized for each redshift range separately) of the observed, that is flux limited, sources (solid lines) and the corresponding underlying parent populations from which the observed samples originate (dotted lines). In each plot, for each redshift range and model considered, kinetic luminosity distributions for separate linear size bins (as used in fitting) are plotted. The exact radio galaxy linear sizes ($D$) falling into each size bin are detailed in corresponding panels. The distributions change for different redshifts as well as for different radio galaxy linear sizes, where larger radio sources on average seem to be also more powerful. Note that these distributions of parent populations depend strongly on the slope $\alpha_{\rm s}$ which for $z>0.3$ is associated large uncertainties. Here, for pictorial purposes, only the best-fits for each redshift range are presented. Colours denoting linear sizes of the radio galaxies are labeled.}
\label{rys:klf-frac-size} 
\end{figure*}

Energy extraction from black hole spin is an attractive way to explain high jet luminosities and efficiencies \cite[e.g.][]{p4, b20, r5, m18}. After \citet{m17} we note that {\mbox{$ {Q \propto j^2 B_{\rm pol}^2 (M_{\rm BH}/M_{\odot})^2}$,}} where $j$ denotes the black hole spin, and $B_{\rm pol}$ is the poloidal magnetic field. In such a case, for a given kinetic luminosity the higher the spin is, the less massive the black hole must be. However, one must bear in mind that the strength of the poloidal magnetic field will have a direct effect on the efficiency of the black hole rotational energy extraction \cite[for discussion on this issue see][]{m17}. A combination of accretion and spin may play a significant role in producing the kinetic luminosities estimated here.  However, our results suggest that also environments and lifetimes of radio galaxies have a direct influence on the radio luminosity of large scale structures of radio galaxies.  It is worth considering, therefore, whether it is possible to reduce the redshift evolution in $Q_{\rm B}$ based on the interplay between the environments and radio galaxy lifetimes in our results. For instance, the  $Q_{\rm B}$ cosmological evolution of $n_{\rm q} = 7.0$ will yield milder evolution of the radio source maximum lifetimes, $n_{\rm t} = -1.0$, and the trend with redshift of the galaxy cluster central densities can be as strong as $n_{\rm r} = 10.5$ (Model S) or even $n_{\rm r}=11.5$ (Model W). With this pattern of cosmological evolution the kinetic luminosity break would reach $\sim 1\times10^{41}$~W at $z=2.0$, which further implies maximum black hole masses of $10^{10}$~M$_{\odot}$ if the sources accrete close to their Eddington luminosity, and jets are predominantly accretion powered. However, this causes $n_{\rm r}$ to be unfeasibly strong. One may get more realistic results while assuming $n_{\rm r}=4.0$, as then our results show that $n_{\rm q}=8.0$ and $n_{\rm t}=-2.5$, which leads to $Q_{\rm B}(z=2.0) \cong 6.6\times10^{41}$~W, $t_{\rm max_{\rm m}}(z=2.0) \cong 3.9 \times10^6$~yr, and $\rho_{\rm m}(z=2.0) \cong 4\times10^{22} \rm kg\, m^{-3}$ (Model S).  The redshift evolution of the central densities may seem to be rather high, since at $n_{\rm r} = 4.0$ it would exceed the effect from the Universe's expansion; however, as discussed in  \S\ref{sec:res-density} such a strong evolution may be realistic if some intrinsic change in the environments, or other effect such as evolution of core radius, take place. Alternatively, radio sources may be allowed to reach older ages than the ones found with the best-fitting parameters (but kept within the confidence intervals as found in this work). To lower the kinetic luminosities the sources would have to be younger or reside in higher density environments  than our best-fitting parameters suggest, or undergo a combination of the two effects. For instance, if one attempts to match the kinetic luminosities and ages estimated by \citet{r1} for 3CRR sources, the lifetimes would have to be allowed to reach $10^8$~yr, and hence sources would have to reside in very high density environments. Finally, there is a possibility that, contrary to  what we assumed so far, the intrinsic source parameters are not independent. Scenarios where $Q$ and $t$, and/or $t$ and $R_{\rm T}$ are correlated are sometimes considered, but are beyond the scope of our current work.

It is necessary to point out that as the cosmological evolution of the source parameters is certainly real and is not an effect of an observational bias due to high flux limit of the observed samples, the location of the kinetic luminosity break can be only constrained in the highest redshift bin. From our results it seems that in the lower redshift bins the $Q_{\rm B}$ must be at the lower end of the luminosities that still contribute to reconstruction of the observed samples, but its exact location is uncertain (Figure~\ref{rys:klf-frac-size}).In addition, one may notice in Figure~\ref{rys:klf-frac-size}, that the independent-$z$ fits predict lower number density of sources than the combined-$z$ fits at intermediate and high redshifts. The reason for this behaviour lies in tying the $\alpha_{\rm s}$ exponent of Eqn.~\ref{eqn:schech-stdn} and Eqn.~\ref{eqn:schech-wlt} in the case of  the combined-$z$ fits, where $\alpha_{\rm s}$ is required to stay the same for all redshifts (unlike for the independent-$z$ fits where the exponent may vary freely between redshift ranges). The best fits of both model S and W of the combined-$z$ fits predict much flatter slopes of the kinetic luminosity distributions than the independent-$z$ fits. This will cause the existence of low radio luminosity sources that are not observed due to the high flux limit of the samples, but are unveiled in the KLFs. Both of these issues this may be verified with larger samples with lower flux limits than are currently being used in this study.

\subsection{Constraints on AGN duty cycles} 
 
The black hole masses of massive elliptical galaxies, which are typical hosts of powerful radio galaxies, are known \cite[e.g.][]{w7,w8,d8,g4}. Hence, once number densities and kinetic luminosities of the sources have been estimated, one can set constraints on the AGN duty cycle which defines the fraction of black holes being active at a particular cosmological epoch.

The  duty cycle, $\delta(M_{\rm BH}, z)$, is defined as 
 
\begin{equation} 
\delta(M_{\rm BH}, z) = \frac{\phi(M_{\rm activeBH}, z)}{\phi(M_{\rm BH}, z)}, 
\end{equation} 
where $\phi(M_{\rm activeBH}, z)$ is the space density of active AGN with mass $M$ and at the redshift $z$, while $\phi(M_{\rm BH}, z)$ analogously defines the space density of all AGN \cite[see also][]{k9}. Since in our work other types of radio loud AGN are excluded, e.g. FR~Is and blazars, we shall refer to the duty cycle of FR~II radio galaxies only in this discussion; this will be denoted by $\delta_{\rm FRII}(M_{\rm BH}, z)$.

A typical FR~II source can be assumed to have a black hole mass of $\sim 10^9$~M$_{\odot}$. If we assume that these are the most typical sources during each epoch (i.e. there is no significant mass evolution), then the typical kinetic luminosity of these black holes is around the value of $Q_{\rm B}$ of the considered cosmological epoch.  
For the lowest redshifts, with $Q_{\rm B} \simeq 10^{39}$~W, the space number density is approximately $\phi(M_{\rm activeBH}, z_1 ) \sim 10^{-7}$~dN~Mpc$^{-3} \Delta \log_{10} (Q)^{-1}$ (Figure~\ref{rys:klf}); $\phi(M_{\rm activeBH}, z)$ is found analogously for the other two redshift bins. These values are compared with the local black hole mass functions compiled by \citet{s17}, which give approximately $\phi(M_{\rm BH}=10^9M_{\odot}, z=0) = 10^{-4}$~dN~Mpc$^{-3}$$\Delta \log_{10} (M_{\rm BH})^{-1}$.  Assuming that, around $10^9$~M$_{\odot}$, the average fractional accretion rate and efficiency of conversion to kinetic power are independent of black hole mass, we can infer that $\Delta \log_{10} (Q)=\Delta \log_{10} (M_{\rm BH})$.  Then, neglecting redshift-evolution of the black hole mass function, we find duty cycles of $\delta_{\rm FRII}(M_{\rm BH}, z_1) \sim10^{-3}$, $\delta_{\rm FRII}(M_{\rm BH}, z_2) \sim10^{-4.5}$, and $\delta_{\rm FRII}(M_{\rm BH}, z_3) \sim10^{-5.5}$. Although in our analysis we use ranges of redshifts, the measurement is instantaneous, and hence can be compared to other results for a single redshift; here our measurement will be an average for the redshifts within each range we consider.

We can compare the estimated duty cycles with the maximum lifetimes of radio galaxies obtained from our fits to determine whether all or just a fraction of $10^{9}$~M$_{\odot}$ black holes are likely to spend some time as a powerful FR~II radio galaxy, where we define `powerful' to mean at the break luminosity, i.e. the most powerful `typical' objects in the observed epochs.  From our fits, we found that the sources live for a maximum of approximately $10^{7.8}$~yr in our lowest redshift range, and since $\Delta z_1 = 10^{9.5}$~yr, then each radio galaxy is on for 1.6 per cent of this epoch, compared to a duty cycle of 0.1 per cent.  If we assume that an AGN goes though the FR~II stage only once, if at all, during its lifetime, then around 10 per cent of `typical' radio galaxy hosts are likely to spend time as a powerful FR~II in the most recent epoch. Moreover, if we assume that these galaxies can go through their FR~II episodes multiple times throughout this epoch, then the fraction will be even smaller. For the higher redshifts the inferred duty cycles are smaller as compared to the local Universe.  From the fits, the radio sources at higher redshifts are found to live for a shorter time; the lifetimes of $10^{7.2} - 10^{7.5}$~yr for redshift range $z_2$ and  $10^{6.6} - 10^{7.2}$~yr for $z_3$ are found, while the time spanned by each epoch remains close to $\Delta z_1$.  However, the reduction in lifetimes is not sufficient to explain the drop in duty cycle at high redshifts: the most powerful FR~IIs become rarer, and a smaller fraction of potential FR~II hosts spend time as powerful FR~IIs (perhaps $<1$~per cent for $z_2$, and $<0.1$~per cent for $z_3$).  

These apparent reductions in duty cycle may result from our assumption of no evolution in the number density of $10^{9}$~M$_{\odot}$ black holes from the present epoch.  In fact, based on the evolution of X-ray luminosity functions, \cite{m5} \cite[see also][]{n2} estimate that the black hole number density of SMBHs declines from $\phi(M_{\rm BH}=10^{9}M_{\odot}, z=0.3) \sim 10^{-4}$ to $\phi(M_{\rm BH}=10^9M_{\odot}, z=2.0) \sim 10^{-5.8}$. This reduction in number density at higher redshifts could account for most of the apparent drop in duty cycle which we infer from our fits.  It is also important to note that since the  break luminosity of the kinetic luminosity function increases by 2 decades from $z_1$ to $z_3$, the active fractions quoted here are tracing significantly different populations in terms of kinetic power, and if lower-power radio galaxies are included (e.g. the FR~I population) the total duty cycle of radio-loud AGN may be much greater than the values given here.

It is interesting to compare our estimates of powerful radio galaxy duty cycles with estimates of AGN duty cycles based on radiative luminosity functions. In the higher redshift ranges, our duty cycle estimates are at least a decade lower than those inferred for $10^{9}$~M$_{\odot}$ black holes by \cite{s17} and the lower bounds estimated by \cite{k9}.  This discrepancy may be understood in terms of the radio loud fraction of AGN, since for the high kinetic powers observed at higher redshifts, the accretion rates are expected to be large fractions of the Eddington rate and hence correspond to the underlying quasar population, which is known to be predominantly radio-quiet.  Hence at these redshifts, the lower duty cycle inferred from kinetic luminosities may reflect the small fraction of high accretion-rate, radio-loud objects.  However, at lower redshifts, the situation turns around, and our inferred duty cycle for $10^{9}$~M$_{\odot}$ may be up to a decade {\it larger} than that estimated from radiative luminosities by \cite{s17}.  This difference may reflect the evolution of the FR~II kinetic break luminosity to lower values, which may correspond to a lower typical black hole mass at low redshifts than the $10^{9}$~M$_{\odot}$ we assume here, for which the duty cycles predicted by \cite{s17} are higher.  An alternative possibility is that the reduction in break luminosity corresponds to a change to a lower accretion rate (as a fraction of the Eddington rate), where a kinetically-dominated accretion mode dominates over radiatively efficient accretion \cite[e.g. see][and references therein]{m5}.

\section{Summary} 
\label{sec:summary} 
 
We perform multidimensional Monte Carlo simulations to investigate fundamental source parameters (such as kinetic luminosity, age and ambient density), and their possible trends with redshift of powerful radio galaxies and radio-loud quasars of FR~II morphology. We present a new method, based on radio luminosity functions, for estimating the parent populations of complete, flux limited radio samples. In our analysis we use the KA/KDA semi-analytical model of FR~II type radio source time evolution, and we stress that other models, such as those of BRW, MK, or the modified models of \cite{b1,b2}, might yield somewhat different results.
 
We find that:

\begin{enumerate}[(i)]

\item The total lifetimes of the radio galaxies are found to be few $\times10^7$~yr at low redshift and decrease for earlier epochs. This is in agreement with independent studies on the 3CRR radio sources, but may be specific to these most powerful radio galaxies and quasars. With the current sample and its strong flux limit we are unable to draw final statements about the evolution of the lifetimes of the sources which is suggested by our current results. 
 
\item Our results suggest cosmological evolution of one or more source parameters. In particular the mean density of the immediate source environments (or the  $\rho_oa_o^{\beta}$ parameter if we allow core radius or $\beta$ to change with redshift) is found to undergo evolution with redshift; the hypothesis that there is no evolution is ruled out with probability of $99$ per cent. 
 
\item The central density of the FR~II environments is found to undergo redshift evolution of approximately $(1+z)^{4}$; evolution stronger than the one expected from the Universe expansion is possible if we consider additional effects from change of environment or core radius evolution. 
 
\item The function describing the initial distribution of kinetic luminosities modelled by the Schechter function (or its modification) or smoothly broken power-law is favoured to simple power-law; however, the hypothesis of an unbroken power-law distribution cannot be ruled out at a confidence level of more than $95$ per cent. 
  
\item The estimated kinetic luminosities are within $10^{38} - 10^{41}$~W for FR~II type sources, which is consistent with previous studies. We observe that the kinetic luminosities depend on the cosmological epoch and the linear size of the source, where larger in linear size as well as higher redshift sources  are more powerful. 
 
\item The FR~IIs' kinetic luminosity function undergoes cosmological evolution of the break luminosities of at least $(1+z)^{3}$ and may be as strong as  $(1+z)^{10}$.  The uncertainty originates from the strong degeneracy between $Q_{\rm B}$, $t_{\rm max}$ and $\rho_o$. Evolution stronger than $(1+z)^9$ is rather unlikely since as the consequence the black hole masses of the most luminous FR~II sources would have to be of  $>10^{11}$~M$_{\odot}$ assuming that there is no strong spin powering of the jets. 

\item Our results suggest that, at least at high redshifts, FR~II sources most probably accrete at moderate/high Eddington ratios and the black hole spin may play a significant role in the jet production, as both effects seem to be necessary to explain the high estimated kinetic luminosities of FR~II sources at higher $z$.

\item We estimated the duty cyles of FR~II radio galaxies at the break in the kinetic luminosity function, finding them to decrease with redshift from $\sim10^{-3}$ for $z_1\leq0.3$, to $\sim10^{-4.5}$ at $0.3 < z_3\leq0.8$, to $\sim10^{-5.5}$ at $0.8 < z_3\leq2.0$.  The decrease in duty cycle at higher redshifts may be explained by a combination of the reduction in the lifetime of FR~II radio galaxies together with evolution in the number density of massive black holes. The shift in kinetic luminosity break to higher values also indicates an intrinsic change in the population of kinetic powers.  Interestingly, at low redshifts the duty cycle of powerful FR~IIs exceeds that estimated for $10^{9}$~M$_{\odot}$ AGN based on radiative luminosities.  This difference can be explained if the typical black hole mass of FR~IIs shifts to lower masses at low $z$. Alternatively, the low-$z$ FR~II population may become dominated by a kinetically dominated, radiatively inefficient mode of accretion.

\end{enumerate}

\section*{Acknowledgments} 

ADK thanks Peter A. Curran for many valuable discussions. The authors thank Judith H. Croston, Thomas J. Maccarone and Andreas Papadopoulos for their useful and insightful comments on the manuscript. The authors thank the referee Paul Wiita for comments which improved the content of the paper. ADK acknowledges the Leverhulme Trust for the financial support. PU acknowledges funding from an STFC Advanced Fellowship. PU and ADK acknowledge funding from the European Community's Seventh Framework Programme (FP7/2007-2013) under grant agreement number ITN 215212 ``Black Hole Universe''.  This work made an extensive use of the Iridis Compute Cluster maintained by the University of Southampton, Southampton, UK, and, in the later stages of the project, of the Sciama High Performance Compute Cluster maintained by the University of Portsmouth and SEPNet (South-Eastern Physics Network) UK.

 

\label{lastpage} 
\bsp

\newpage 
\appendix

\section{ \\ \Huge{Supplementary Material}}

\begin{figure*}
\includegraphics[angle=270,width=126mm]{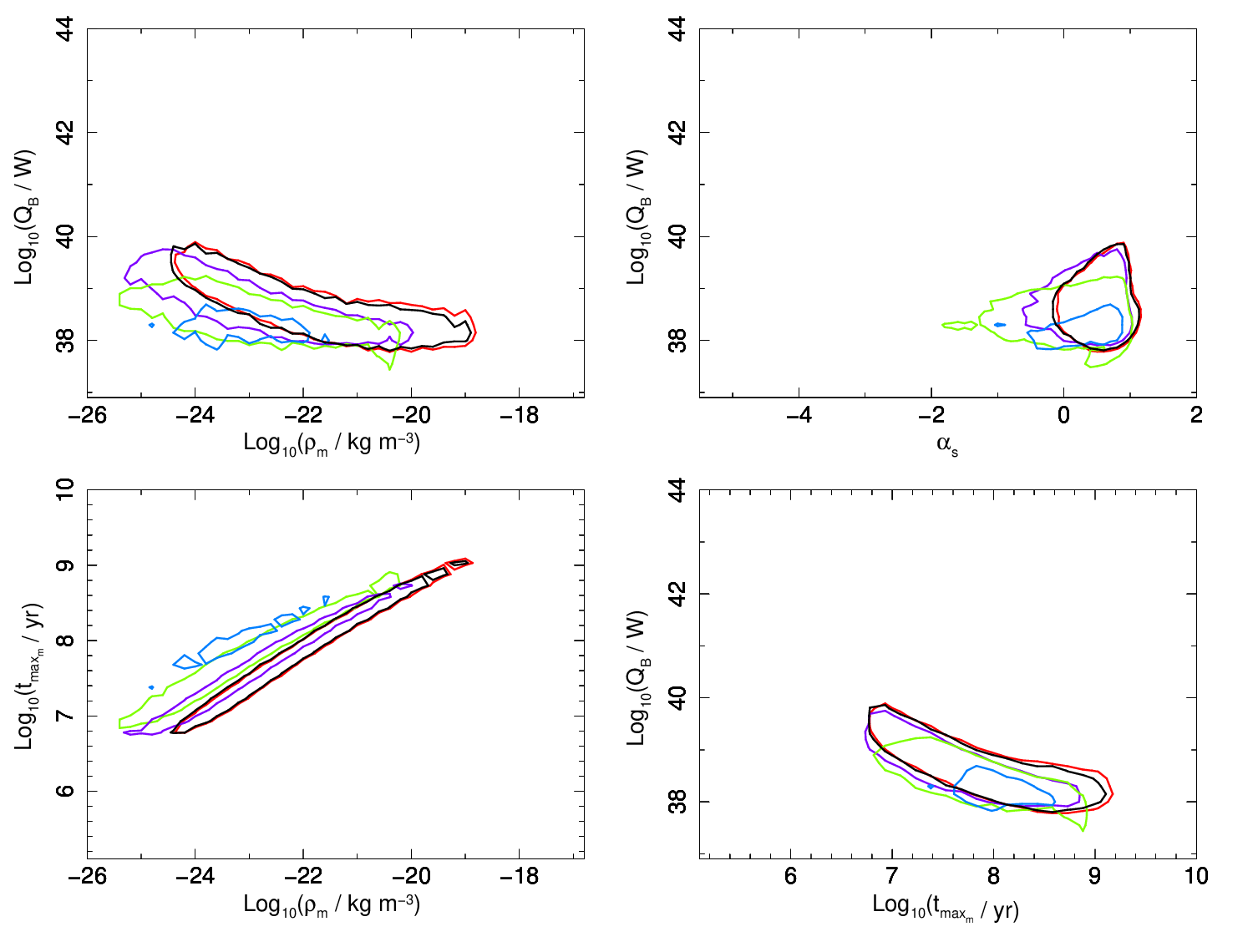}
\caption{90 per cent   joint confidence  intervals for radio galaxy populations  created with different assumptions on the standard deviation of the density log-normal distribution, where  $\rm \sigma_{\rm log_{10}(\rho_0)} = 0.0$ (i.e. $\rho_o$ is a delta function, drawn in red), $\rm \sigma_{\rm log_{10}(\rho_0)} = 0.15$ (black),  $\rm \sigma_{\rm log_{10}(\rho_0)} = 0.50$ (violet),  $\rm \sigma_{\rm log_{10}(\rho_0)} = 0.75$ (green), and  $\rm \sigma_{\rm log_{10}(\rho_0)} = 1.0$ (blue). Results shown for the redshift range $0 < z_1 < 0.3$ and for Model S as described in \S5.4.1 of the main paper. Findings are based on the 3CRR and BRL data fits. The case where the standard deviation is equal to $\rm \sigma_{\rm log_{10}(\rho_0)} = 0.15$ is the initial, standard case analysed in depth in \S5 of the main article.}
\end{figure*}

\begin{figure*}
\includegraphics[angle=270,width=126mm]{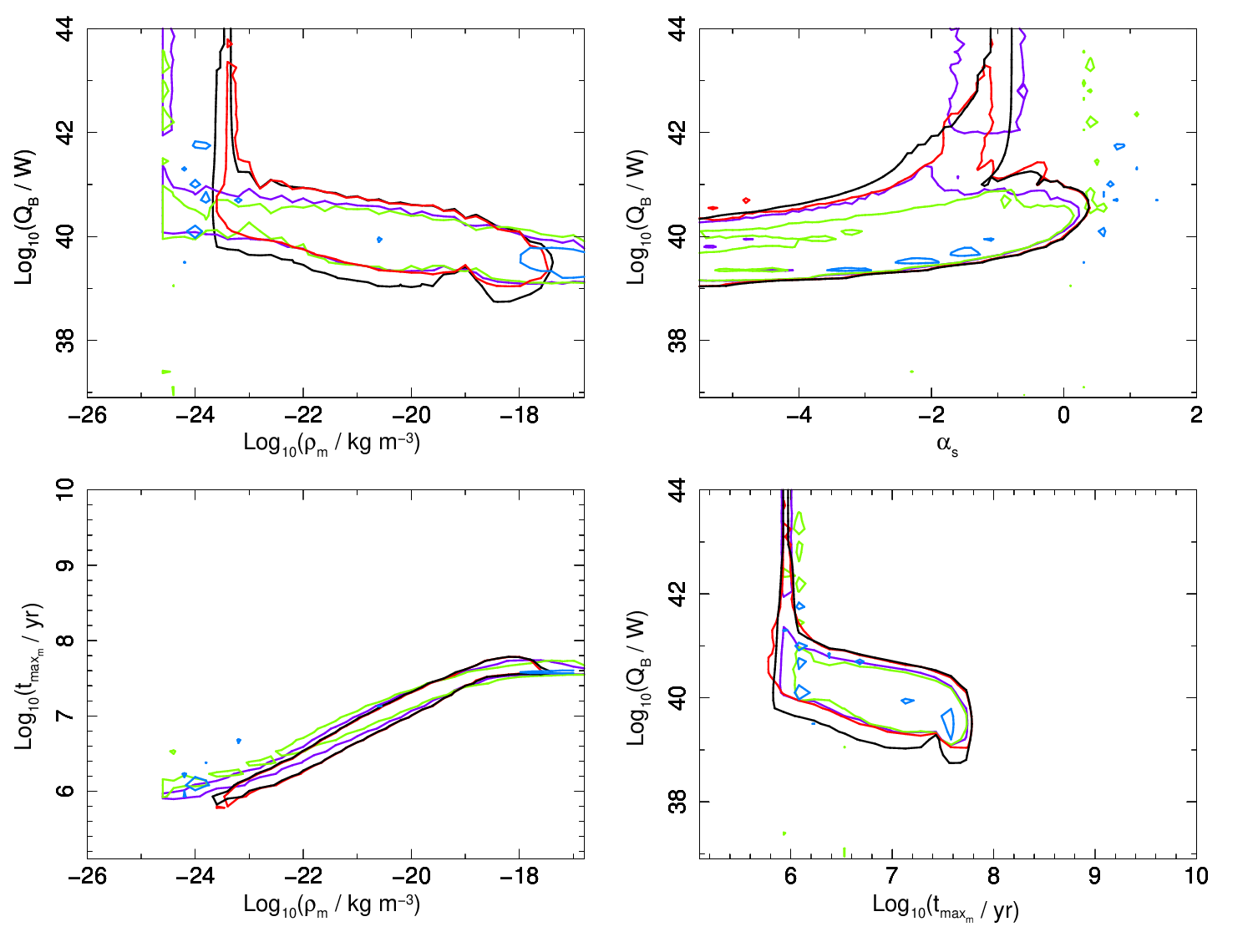}
\caption{90 per cent   joint confidence  intervals for radio galaxy populations  created with different assumptions on the standard deviation of the density log-normal distribution, where  $\rm \sigma_{\rm log_{10}(\rho_0)} = 0.0$ (i.e. $\rho_o$ is a delta function, drawn in red), $\rm \sigma_{\rm log_{10}(\rho_0)} = 0.15$ (black),  $\rm \sigma_{\rm log_{10}(\rho_0)} = 0.50$ (violet),  $\rm \sigma_{\rm log_{10}(\rho_0)} = 0.75$ (green), and  $\rm \sigma_{\rm log_{10}(\rho_0)} = 1.0$ (blue). Results shown for the redshift range $0.8 < z_3 < 2.0$ and for Model S as described in \S5.4.1 of the main paper. Findings are based on the 3CRR and BRL data fits. The case where the standard deviation is equal to $\rm \sigma_{\rm log_{10}(\rho_0)} = 0.15$ is the initial, standard case analysed in depth in \S5 of the main article.}
\end{figure*}

\begin{figure*}
\includegraphics[angle=270,width=126mm]{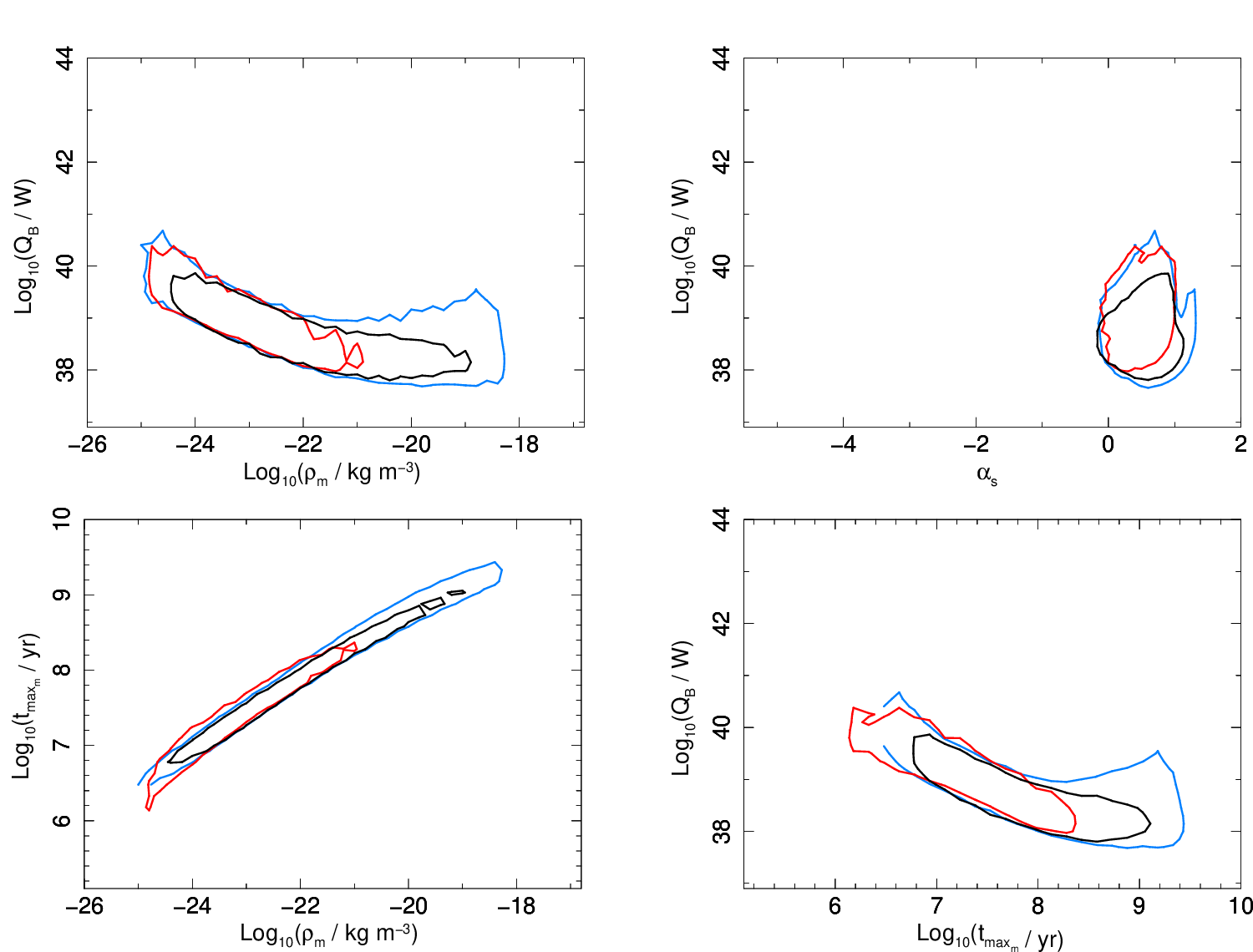}
\caption{90 per cent   joint confidence  intervals for radio galaxy populations  created with different assumptions on the standard deviation of the log-normal maximum lifetimes distribution of the sources, where  $\sigma_{\rm log_{10}(t_{\rm max})} = 0.05$ (black), $\sigma_{\rm log_{10}(t_{\rm max})} = 0.3$ (blue), and  $\sigma_{\rm log_{10}(t_{\rm max})} = 0.6$ (red). Results shown for the redshift range $0 < z_3 < 0.3$ and for Model S as described in \S5.4.2 of the main paper. Findings are based on the 3CRR and BRL data fits. The case where the standard deviation is equal to $\sigma_{\rm log_{10}(t_{\rm max})} = 0.05$ is the initial, standard case analysed in depth in \S5 of the main article.}
\end{figure*}

\begin{figure*}
\includegraphics[angle=270,width=126mm]{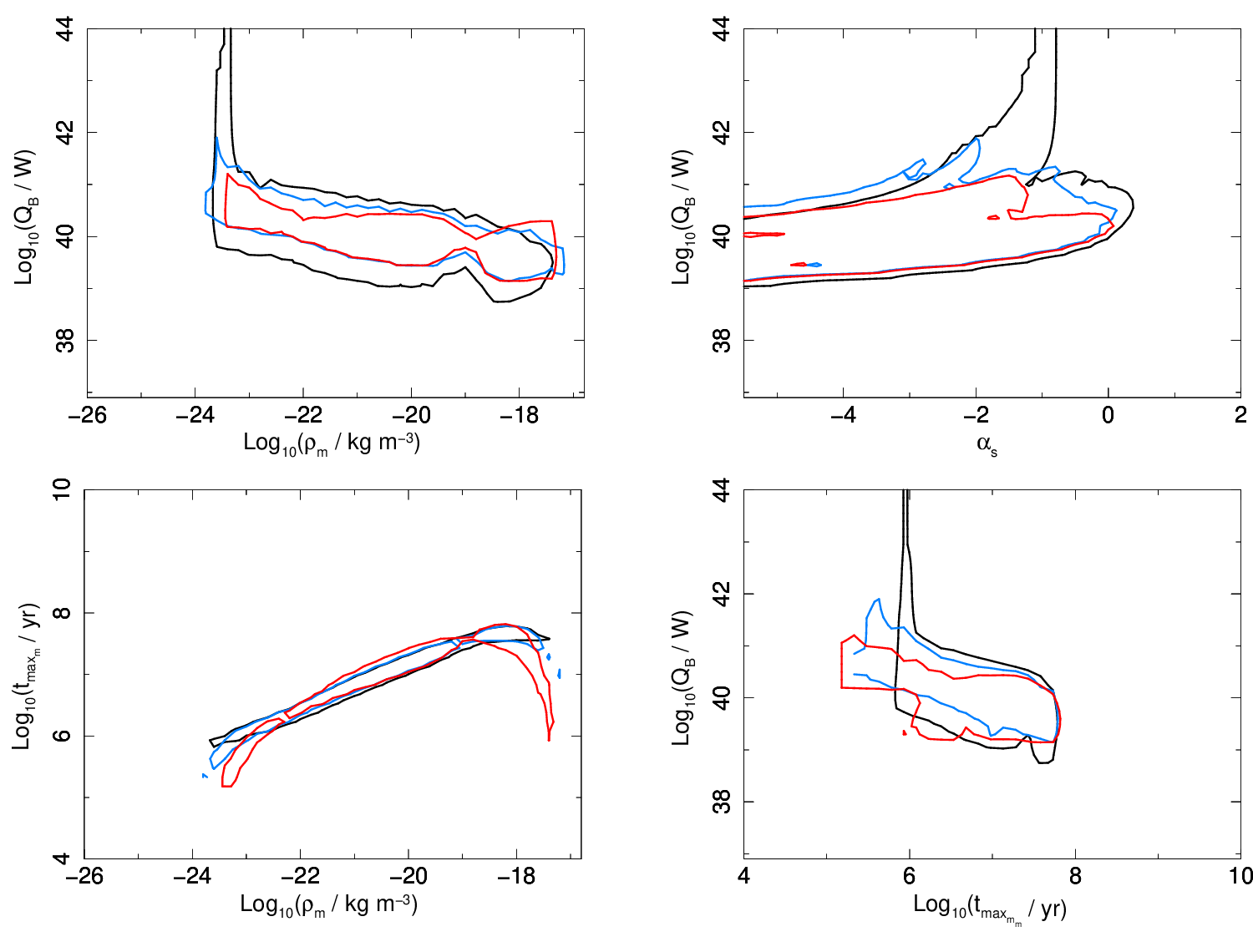}
\caption{90 per cent   joint confidence  intervals for radio galaxy populations  created with different assumptions on the standard deviation of the log-normal maximum lifetimes distribution of the sources, where  $\sigma_{\rm log_{10}(t_{\rm max})} = 0.05$ (black), $\sigma_{\rm log_{10}(t_{\rm max})} = 0.3$ (blue), and  $\sigma_{\rm log_{10}(t_{\rm max})} = 0.6$ (red). Results shown for the redshift range  $08 < z_3 < 2.0$ and for Model S as described in \S5.4.2 of the main paper. Findings are based on the 3CRR and BRL data fits. The case where the standard deviation is equal to $\sigma_{\rm log_{10}(t_{\rm max})} = 0.05$ is the initial, standard case analysed in depth in \S5 of the main article.}
\end{figure*}

\begin{figure*}
\includegraphics[angle=270,width=126mm]{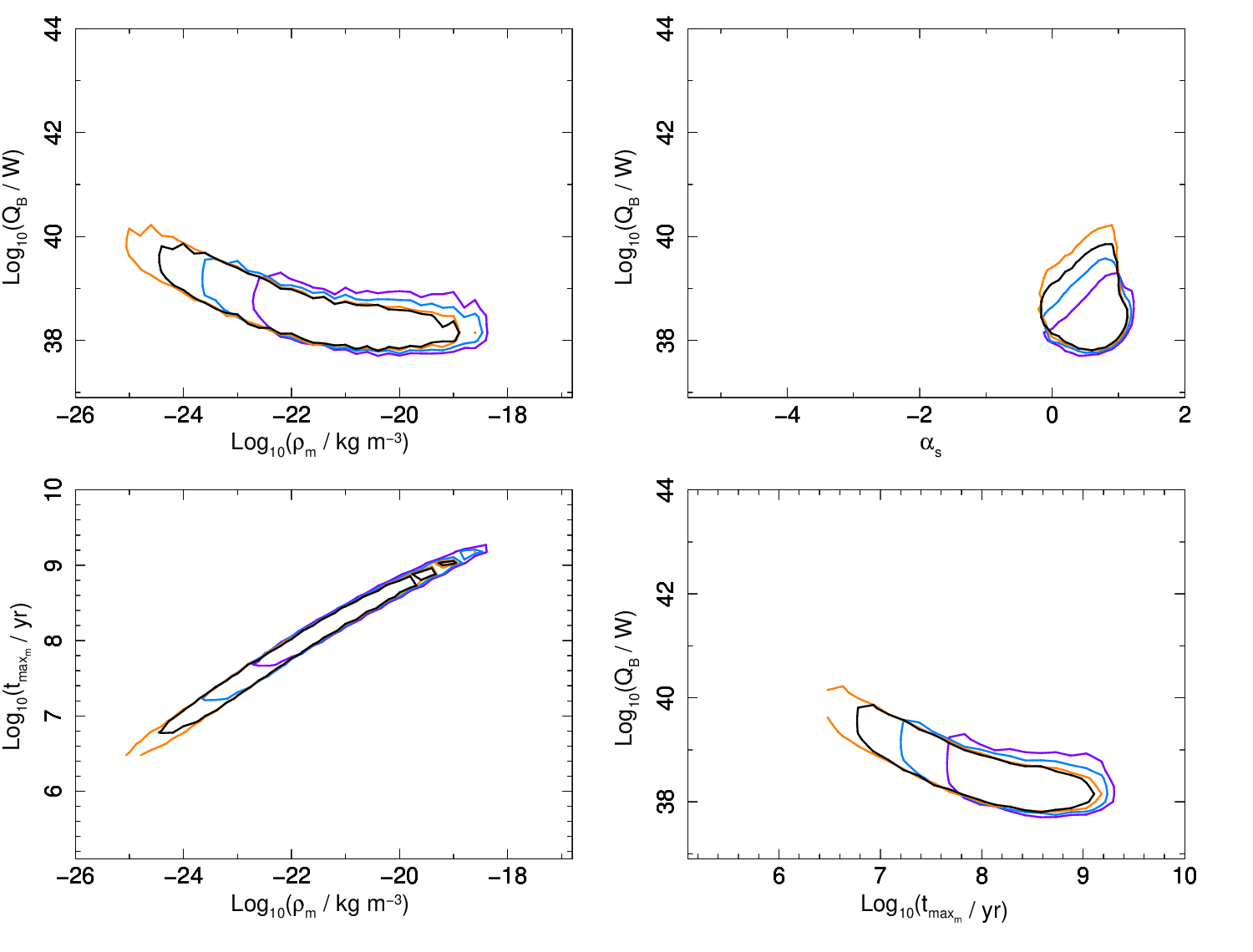}
\caption{90 per cent   joint confidence  intervals for radio galaxy populations  created with different assumptions of the maximum head advance speeds of jets: $0.8c$ (orange), $0.4c$ (black), $0.15c$ (blue), and $0.05c$ (violet) in units of the light speed $c$. Results shown for the redshift range $0 < z_1 < 0.3$ and for Model S as described in \S5.4.3of the main paper. Findings are based on the 3CRR and BRL data fits. The case when the maximum head advance speed is assumed to be $0.4c$ is the initial, standard case analysed in depth in \S5 of the main article. }
\end{figure*}

\begin{figure*}
\includegraphics[angle=270,width=126mm]{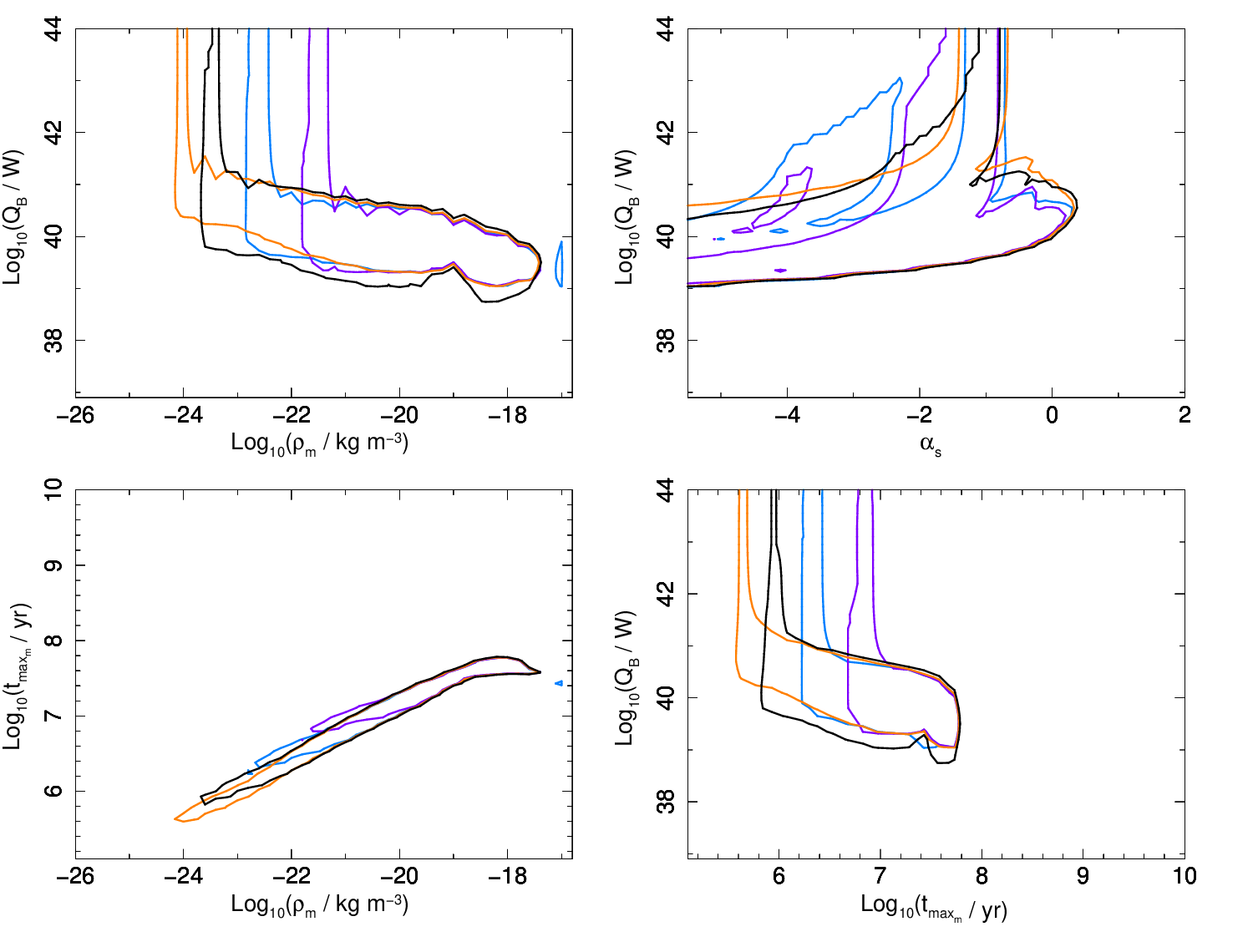}
\caption{90 per cent   joint confidence  intervals for radio galaxy populations  created with different assumptions of the maximum head advance speeds of jets: $0.8c$ (orange), $0.4c$ (black), $0.15c$ (blue), and $0.05c$ (violet) in units of the light speed $c$. Results shown for the redshift range $0.8 < z_3 < 2.0$ and for Model S as described in \S5.4.3 of the main paper. Findings are based on the 3CRR and BRL data fits. The case when the maximum head advance speed is assumed to be $0.4c$ is the initial, standard case analysed in depth in \S5 of the main article. }
\end{figure*}

\begin{figure*}
\includegraphics[angle=270,width=125mm]{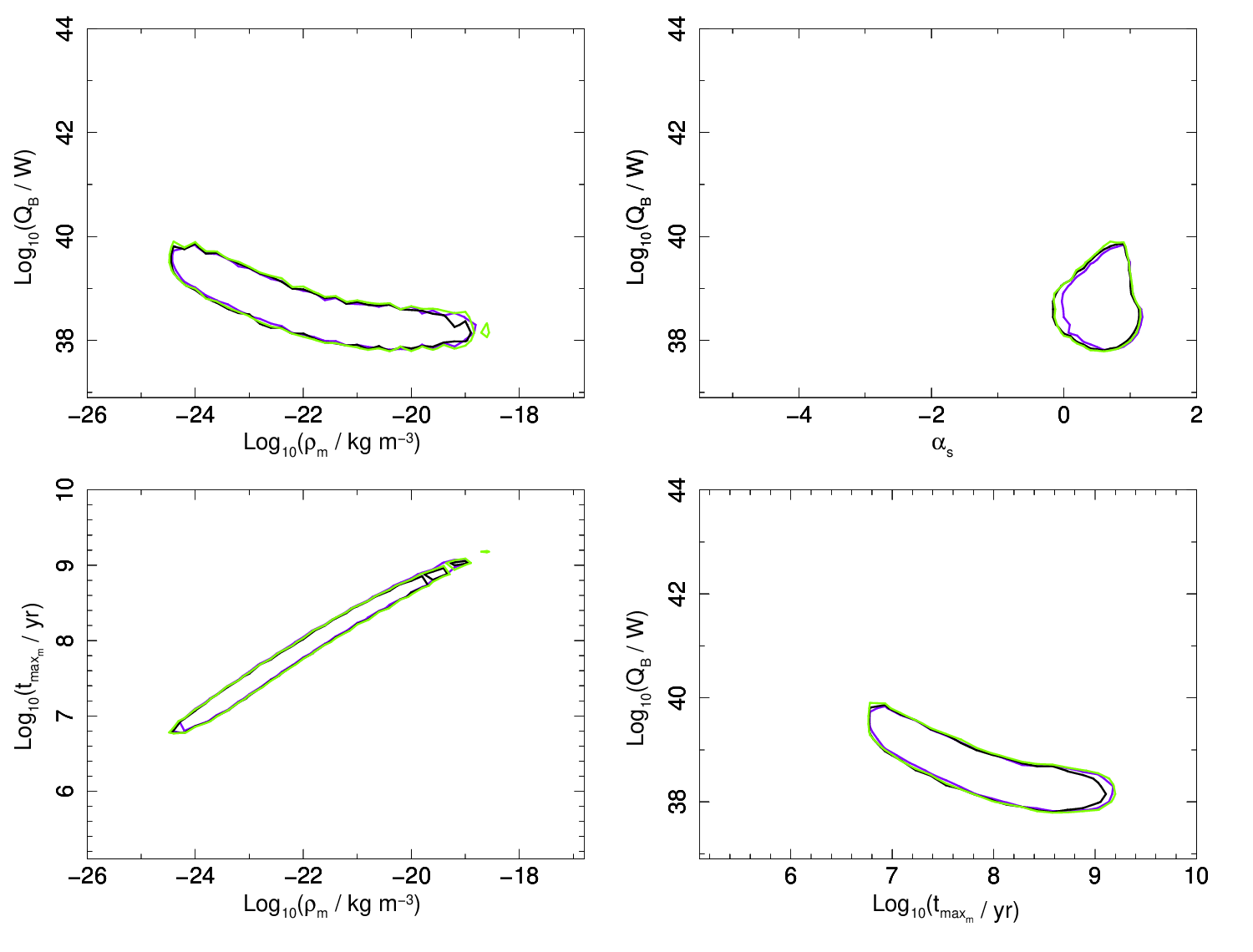}
\caption{90 per cent   joint confidence  intervals for radio galaxy populations  created with different assumptions on the particle injection index, where  (i) $m\in[2;3]$ (uniform distribution, drawn in black), (ii) $m=2.3$ (single value, violet), and (iii) $m_o=2.4$ and $\sigma_{m_o}=0.3$ (green). Results shown for the redshift range $0 < z_1 < 0.3$ and for Model S as described in \S5.4.4 of the main paper. Findings are based on the 3CRR and BRL data fits. The case where the particle injection index is assumed to be a uniform distribution, the case (i) above, is the initial, standard case analysed in depth in \S5 of the main article.}
\end{figure*}

\begin{figure*}
\includegraphics[angle=270,width=122mm]{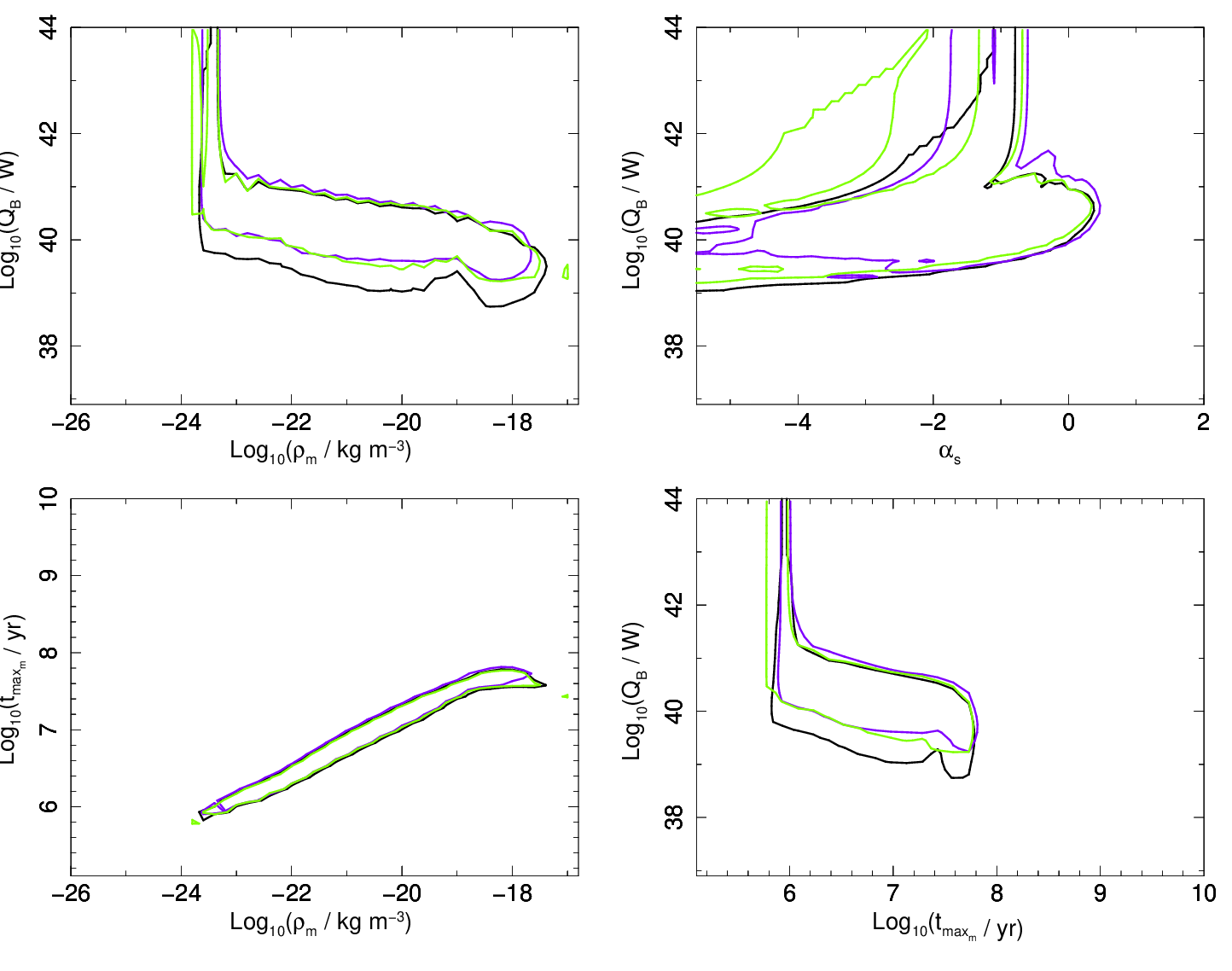}
\caption{90 per cent   joint confidence  intervals for radio galaxy populations  created with different assumptions on the particle injection index, where  (i) $m\in[2;3]$ (uniform distribution, drawn in black), (ii) $m=2.3$ (single value, violet), and (iii) $m_o=2.4$ and $\sigma_{m_o}=0.3$ (green). Results shown for the redshift range $0.8 < z_3 < 2.0$ and for Model S as described in \S5.4.4 of the main paper. Findings are based on the 3CRR and BRL data fits. The case where the particle injection index is assumed to be a uniform distribution, the case (i) above, is the initial, standard case analysed in depth in \S5 of the main article.}
\end{figure*}

\begin{figure*}
\includegraphics[angle=270,width=126mm]{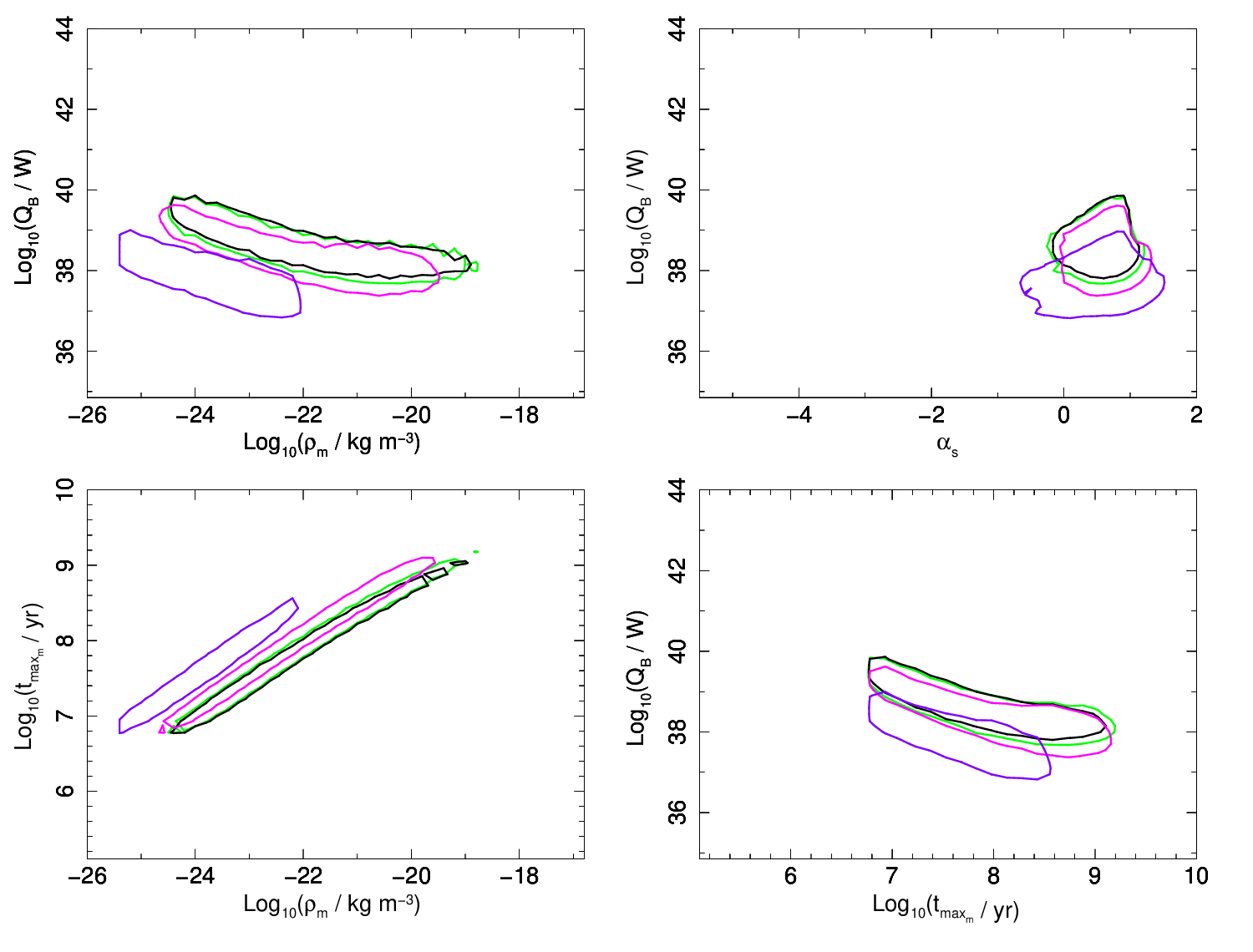}
\caption{90 per cent   joint confidence  intervals for radio galaxy populations  created with different assumptions on the standard deviation of the density log-normal distribution, where  (i) $\gamma_{\rm min}=1$ and $\gamma_{\rm max}=10^{10}$ (drawn in black), (ii) $\gamma_{\rm min}=10^2$ and $\gamma_{\rm max}=10^{10}$ (magenta), (iii) $\gamma_{\rm min}=10^4$ and $\gamma_{\rm max}=10^{10}$ (violet), and(iv) $\gamma_{\rm min}=1$ and $\gamma_{\rm max}=10^{5}$ (green). Results shown for the redshift range $0 < z_1 < 0.3$ and for Model S as described in \S5.4.4 of the main paper. Findings are based on the 3CRR and BRL data fits. The case where the Lorentz factors are set to $\gamma_{\rm min}=1$ and $\gamma_{\rm max}=10^{10}$ is the initial, standard case analysed in depth in \S5 of the main article.}
\label{rys:app:b-z03}
\end{figure*}

\begin{figure*}
\includegraphics[angle=270,width=126mm]{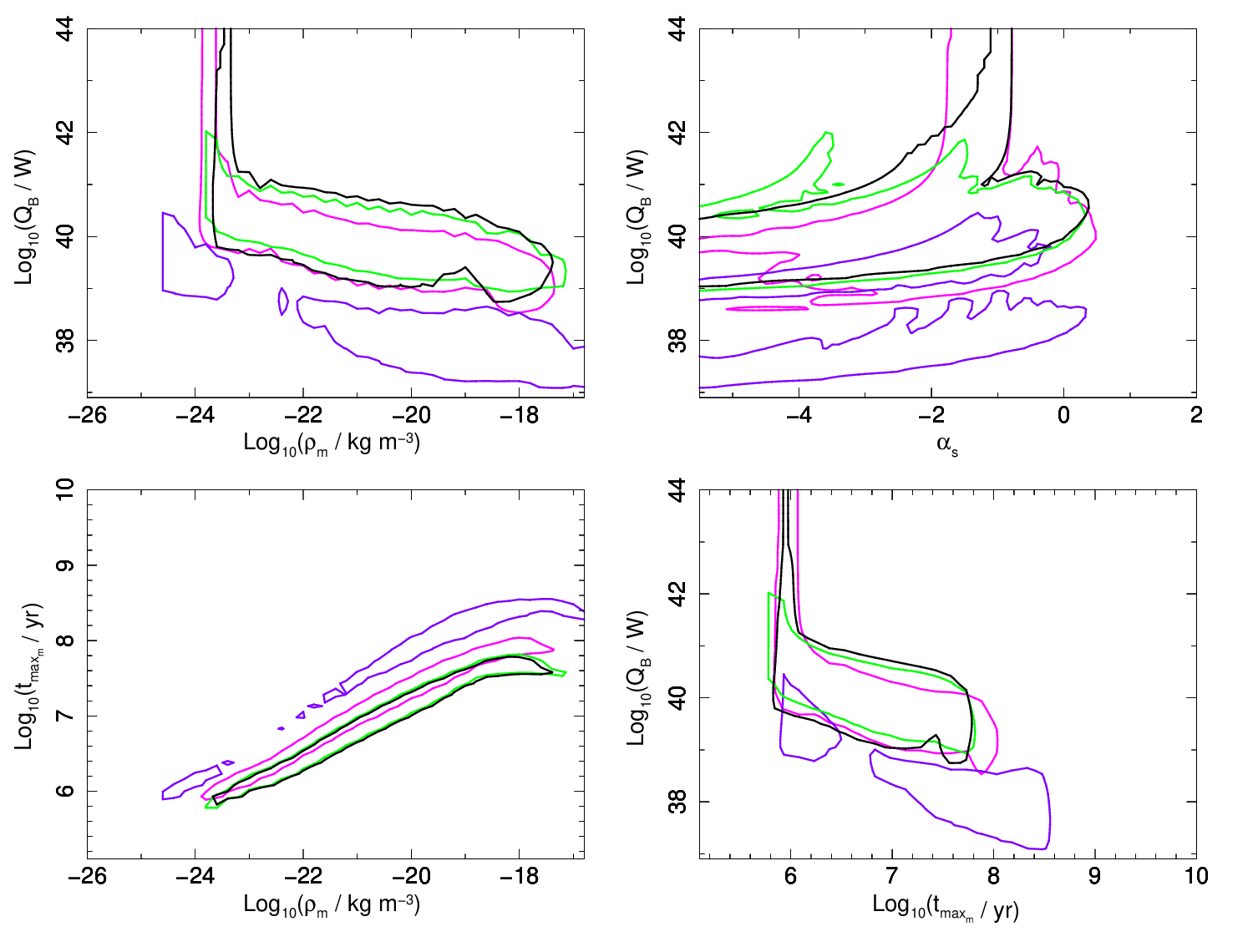}
\caption{90 per cent   joint confidence  intervals for radio galaxy populations  created with different assumptions on the standard deviation of the density log-normal distribution, where  (i) $\gamma_{\rm min}=1$ and $\gamma_{\rm max}=10^{10}$ (drawn in black), (ii) $\gamma_{\rm min}=10^2$ and $\gamma_{\rm max}=10^{10}$ (magenta), (iii) $\gamma_{\rm min}=10^4$ and $\gamma_{\rm max}=10^{10}$ (violet), and(iv) $\gamma_{\rm min}=1$ and $\gamma_{\rm max}=10^{5}$ (green). Results shown for the redshift range $0.8 < z_3 < 2.0$ and for Model S as described in \S5.4.4 of the main paper. Findings are based on the 3CRR and BRL data fits. The case where the Lorentz factors are set to $\gamma_{\rm min}=1$ and $\gamma_{\rm max}=10^{10}$ is the initial, standard case analysed in depth in \S5 of the main article.}
\label{rys:app:b-z82}
\end{figure*}

\begin{figure*}
\includegraphics[angle=270,width=126mm]{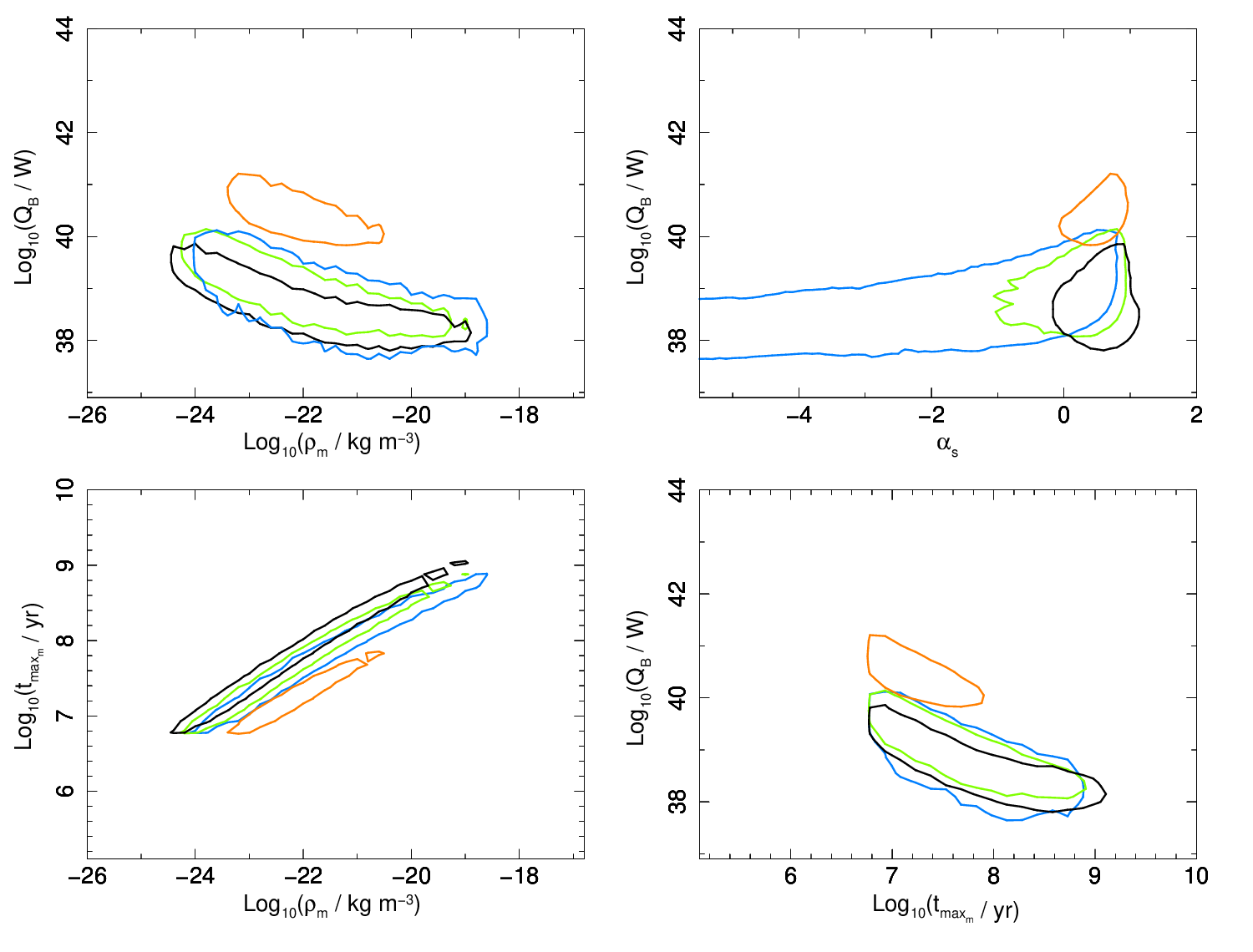}
\caption{90 per cent   joint confidence  intervals for radio galaxy populations  created with different assumptions on the particle content of the jet, where  $k' =0 $ (lightweight jets, black), $k' \in [0; 10]$ (modest inclusion of protons in the jet, light green), $k'\in[0; 100]$ (heavy jets, blue), and $k'=100$ (heavy jets, orange). Results shown for the redshift range $0 < z_1 < 0.3$ and for Model S as described in \S5.4.5 of the main paper. Findings are based on the 3CRR and BRL data fits. The case where $k'=0$ is the initial, standard case analysed in depth in \S5 of the main article.}
\label{rys:app:a-z03}
\end{figure*}

\begin{figure*}
\includegraphics[angle=270,width=126mm]{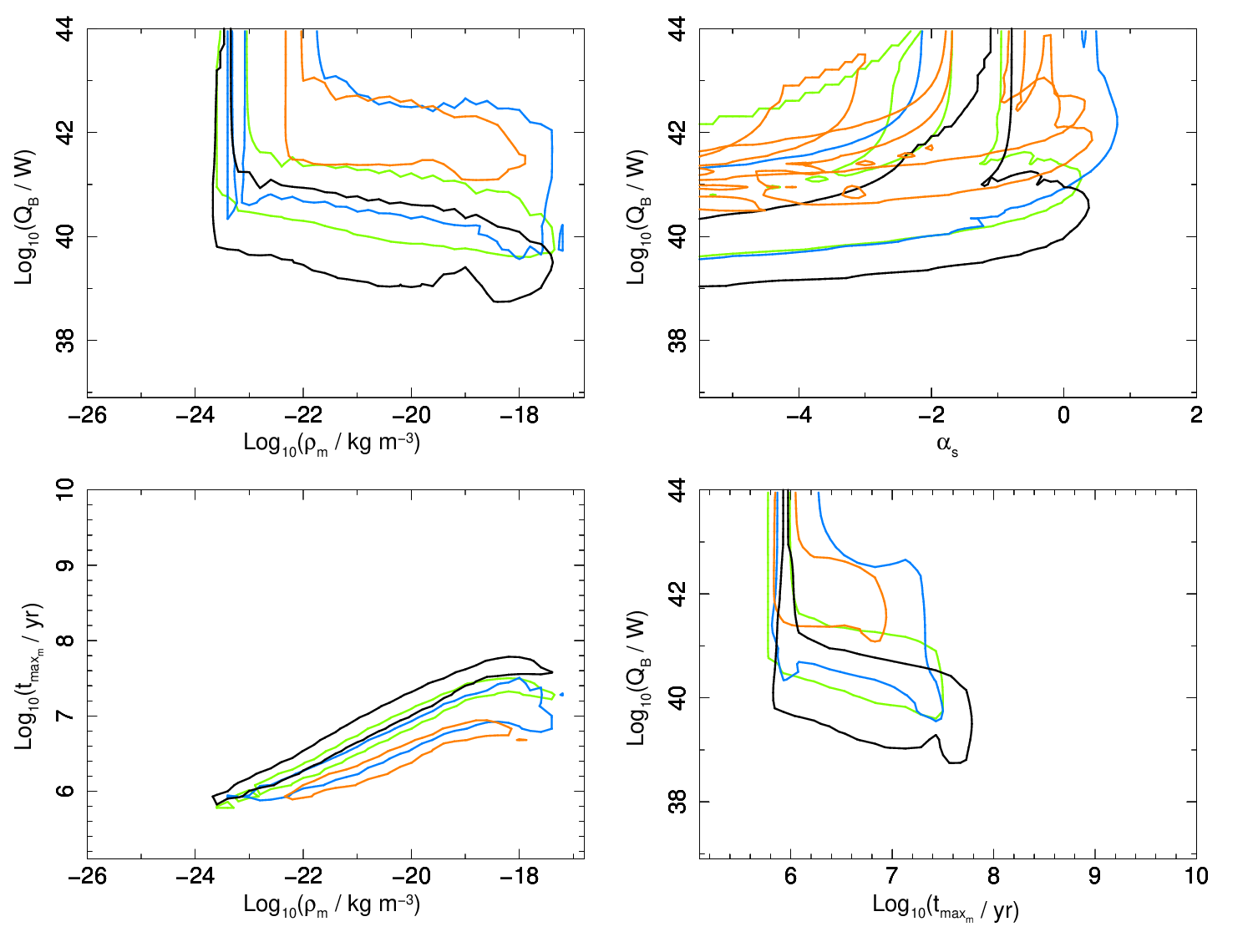}
\caption{90 per cent  joint confidence  intervals for radio galaxy populations  created with different assumptions on the particle content of the jet, where  $k' =0 $ (lightweight jets, black), $k' \in [0; 10]$ (modest inclusion of protons in the jet, light green), $k'\in[0; 100]$ (heavy jets, blue), and $k'=100$ (heavy jets, orange). Results shown for the redshift range  $0.8 < z_1 < 2.0$ and for Model S as described in \S5.4.5 of the main paper. Findings are based on the 3CRR and BRL data fits. The case where $k'=0$ is the initial, standard case analysed in depth in \S5 of the main article.}
\label{rys:app:a-z82}
\end{figure*}

\begin{figure*}
\includegraphics[angle=270,width=126mm]{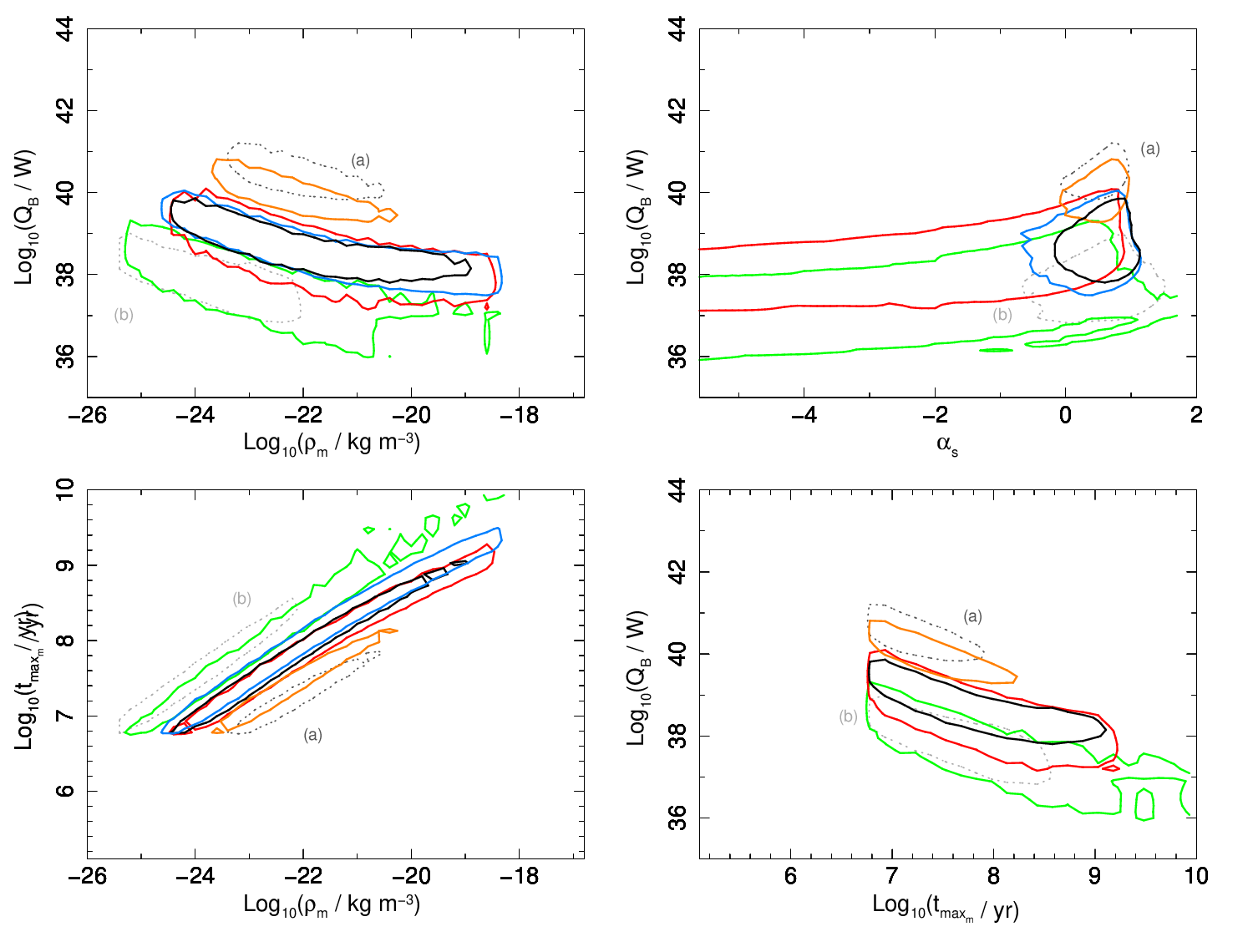}
\caption{90 per cent  joint confidence  intervals for radio galaxy populations created with various assumptions on the minimum energy of the initial particle distribution and the particle content of the jet. The following cases are plotted: 
(i)   $k'=0$, $\gamma_{\rm min}=1$ and $\gamma_{\rm max}=10^{10}$ (solid, black, the default case),
(ii)  $k'\in(0,100)$, $\gamma_{\rm min}=10^2$ and $\gamma_{\rm max}=10^{10}$ (solid, red),
(iii) $k'=100$,       $\gamma_{\rm min}=10^2$ and $\gamma_{\rm max}=10^{10}$ (solid, orange),
(iv)  $k'\in(0,100)$, $\gamma_{\rm min}=10^4$ and $\gamma_{\rm max}=10^{10}$ (solid, green), and 
(v)   $k'=100$,       $\gamma_{\rm min}=10^4$ and $\gamma_{\rm max}=10^{10}$ (solid, blue). 
For reference plotted are the cases where 
(a) $k'=100$, $\gamma_{\rm min}=1$ and $\gamma_{\rm max}=10^{10}$ (dotted, dark grey, see also Figure~\ref{rys:app:a-z03}), 
and (b) $k'=0$, $\gamma_{\rm min}=10^4$ and $\gamma_{\rm max}=10^{10}$ (dotted, light grey, see also Figure~\ref{rys:app:b-z03}).  
Results shown for the redshift range $0 < z_1 < 0.3$ and for Model S as described in \S5.4.5 of the main paper. Findings are based on the 3CRR and BRL data fits. }
\end{figure*}

\begin{figure*}
\includegraphics[angle=270,width=126mm]{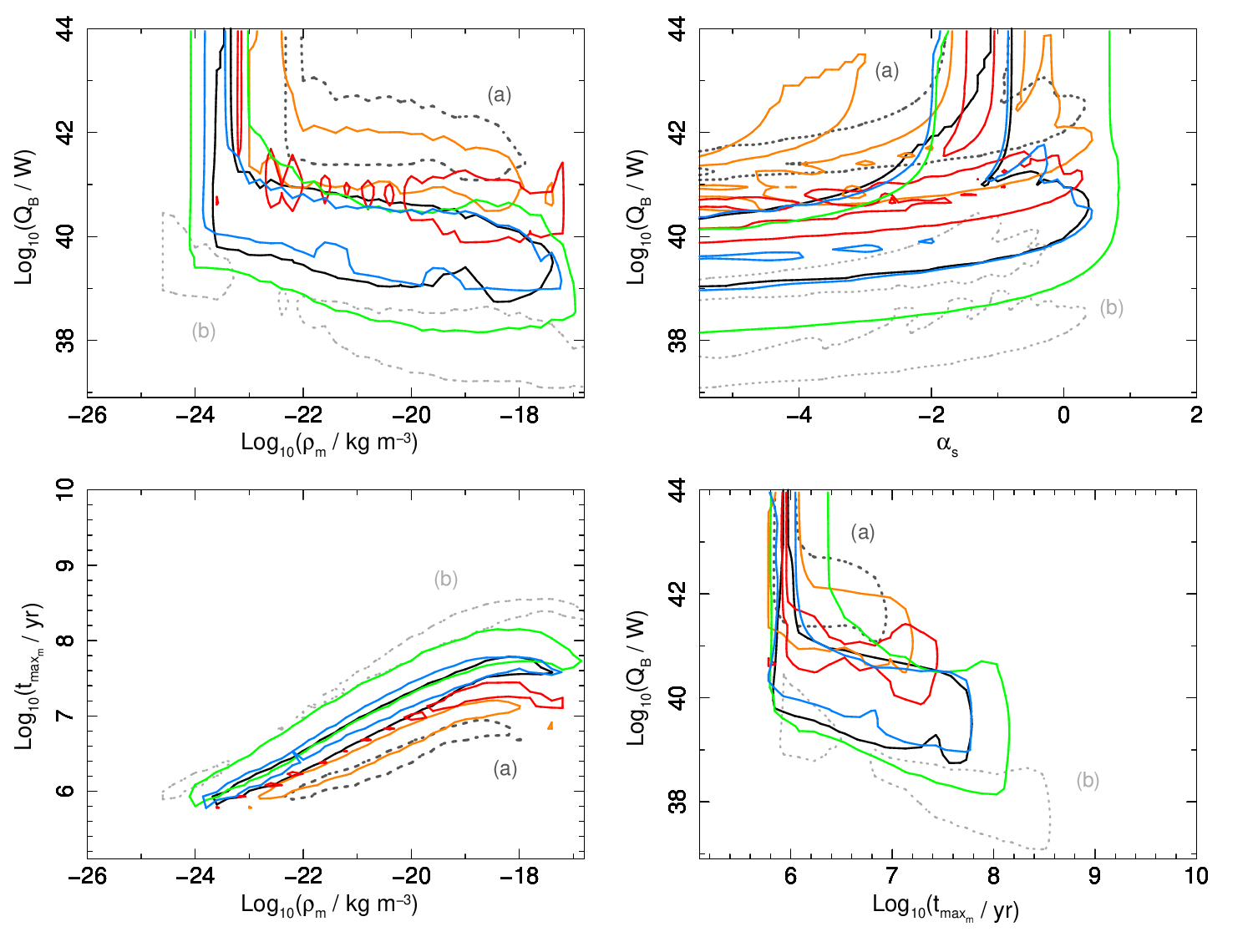}
\caption{90 per cent  joint confidence  intervals for radio galaxy populations created with  various assumptions on the minimum energy of the initial particle distribution and the particle content of the jet. The following cases are plotted:  
(i)   $k'=0$, $\gamma_{\rm min}=1$ and $\gamma_{\rm max}=10^{10}$ (solid, black, the default case),
(ii)  $k'\in(0,100)$, $\gamma_{\rm min}=10^2$ and $\gamma_{\rm max}=10^{10}$ (solid, red),
(iii) $k'=100$,       $\gamma_{\rm min}=10^2$ and $\gamma_{\rm max}=10^{10}$ (solid, orange),
(iv)  $k'\in(0,100)$, $\gamma_{\rm min}=10^4$ and $\gamma_{\rm max}=10^{10}$ (solid, green), and 
(v)   $k'=100$,       $\gamma_{\rm min}=10^4$ and $\gamma_{\rm max}=10^{10}$ (solid, blue). 
For reference plotted are the cases where 
(a) $k'=100$, $\gamma_{\rm min}=1$ and $\gamma_{\rm max}=10^{10}$ (dotted, dark grey, see also Figure~\ref{rys:app:a-z82}), 
and (b) $k'=0$, $\gamma_{\rm min}=10^4$ and $\gamma_{\rm max}=10^{10}$ (dotted, light grey, see also Figure~\ref{rys:app:b-z82}).  Results shown for the redshift range $0.8 < z_3 < 2.0$ and for Model S as described in \S5.4.5 of the main paper. Findings are based on the 3CRR and BRL data fits.}
\end{figure*}
 
\end{document}